\documentclass[sigplan,screen, nonacm]{acmart}
\AtBeginDocument{%
  }
\begin{document}

\title{From Attention to Disaggregation: Tracing the Evolution of LLM Inference}
\author{Srinivasa Rao Aravilli}
\email{srinivasarao.aravilli@capitalone.com}
\affiliation{%
  \institution{Capital One}
  \city{}
  \state{}
  \country{}
}

\author{Mustafa Saify}
\email{mustafa.saify@capitalone.com}
\affiliation{%
  \institution{Capital One}
  \city{}
  \country{}}

\author{Madabattula Rajesh Kumar}
\email{madabattula.rajeshkumar@capitalone.com}
\affiliation{%
  \institution{Capital One}
  \city{}
  \country{}
}

\author{Shashank Srivastava}
\email{shashank.srivastava@capitalone.com}
\affiliation{%
  \institution{Capital One}
  \city{}
  \country{}
}

\pagestyle{plain}
\begin{abstract}
  The 2017 introduction of the Transformer architecture through the paper "Attention Is All You Need" fundamentally transformed sequence modelling by replacing recurrent neural networks with parallelizable self-attention mechanisms, enabling unprecedented scalability. Models such as GPT-1 and BERT showcased the power of large-scale pre-training, while GPT-3, with 175 billion parameters, revealed emergent few-shot and zero-shot learning capabilities. However, as model sizes grew into the hundreds of billions and trillions of parameters, the inference process real-time output generation became the primary bottleneck, constrained by computational throughput, memory bandwidth, and low latency requirements for interactive use cases.
  
Deploying Large Language Models (LLMs) goes far beyond software or model design, it is a complex distributed systems challenge that requires seamless integration of specialized hardware, efficient network scheduling, intelligent resource management, and automated scaling. Fundamentally, LLM inference constitutes a constrained multi-objective optimization problem where latency (e.g., time to first token and total latency) must be minimized, throughput (requests per second) maximized, and cost (dollars per token) reduced. All while adhering to strict service-level agreements and operating within the limits of heterogeneous GPU clusters. This architectural shift, known as "disaggregated inference," is as revolutionary for AI as microservices were for web applications. By decomposing formerly monolithic tasks into modular components that can operate across thousands of GPUs or specialized accelerators, it unlocks significant gains in efficiency, reliability, and agility.

In contrast, traditional monolithic inference pipelines relying on homogeneous GPU hardware encounter bottlenecks caused by competing resource demands between the compute-intensive "prefill" phase and the memory-intensive "decode" phase. These approaches also suffer from inflexible scaling, single points of failure, and energy inefficiencies when handling diverse and dynamic workloads.

This paper explores the evolution toward a disaggregated inference architecture that incorporates distributed systems principles, including service decomposition, resource disaggregation, workload partitioning, and orchestration across heterogeneous hardware. By decoupling prefill and decode phases into independently scalable clusters, this approach mitigates resource contention and enables independent optimization of key performance metrics like Time-to-First-Token (TTFT) and Inter-Token Latency (ITL).

We first analyse six key inference optimizations—Key-Value Caching, Flash Attention, Continuous Batching, Speculative Decoding, Paged Attention, and Radix Attention outlining their mechanisms, historical adoption, and trade-offs in terms of the CAP theorem (Consistency, Availability, Partition Tolerance). We illustrate how these optimizations align with distributed systems concepts such as caching, parallelism, and eventual consistency, while highlighting their constraints in asynchronous environments.

Ultimately, we advocate for a disaggregated paradigm that leverages compute and memory-optimized clusters, dynamic routing, and fault isolation to achieve elastic scalability, reduced latency, and enhanced robustness. This is supported by an in-depth technical survey of three leading frameworks, each representing a distinct archetype: DistServe, a research-first system focused on "goodput" optimization. AIBrix, a cloud-native production orchestration framework and NVIDIA Dynamo, a modular platform for enterprise-scale deployments. Our analysis examines their architectural innovations, implementation strategies, and comparative advantages to guide deployment decisions.

\end{abstract}

\begin{teaserfigure}
    \centering
    \includegraphics[width=1.0\linewidth]{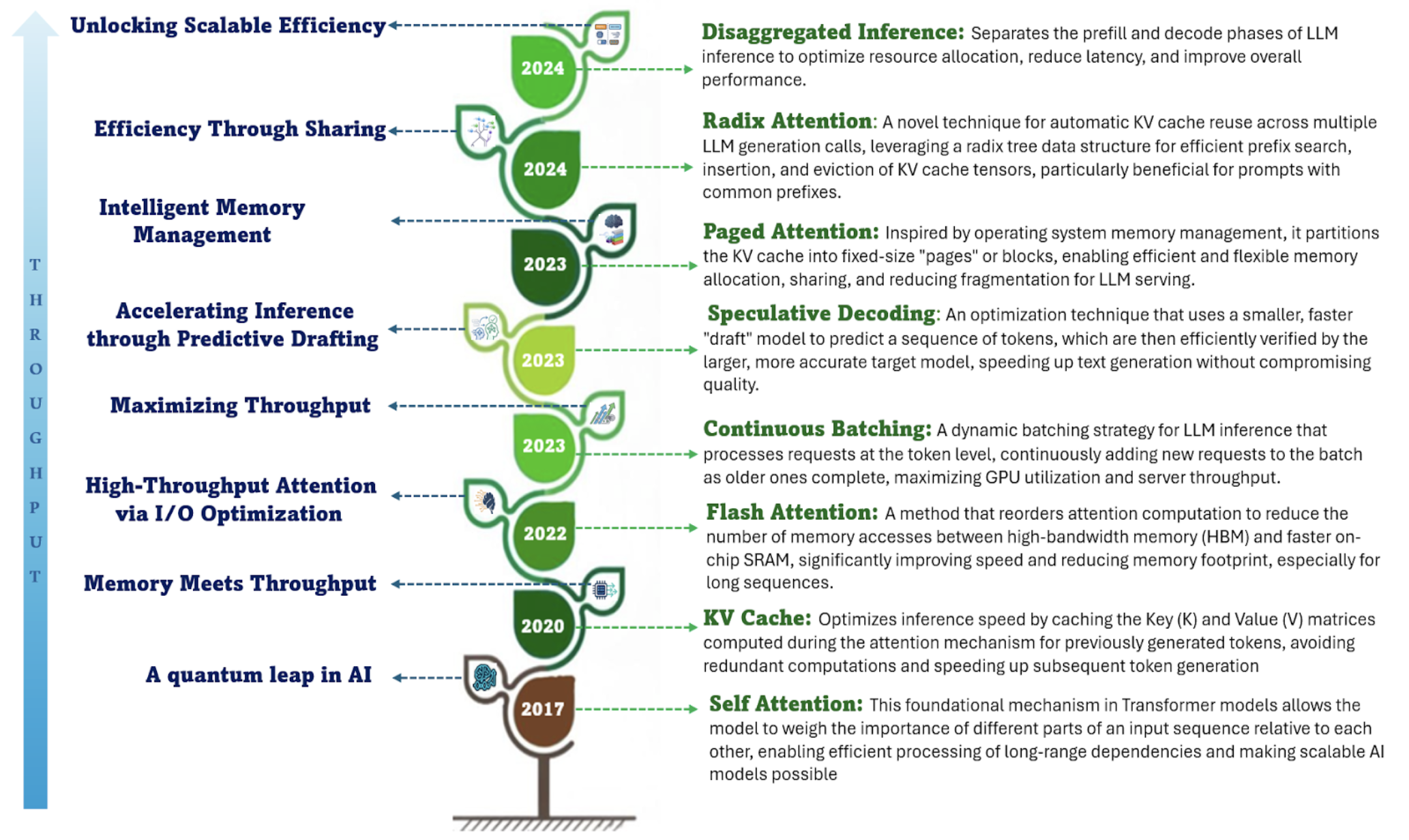}
    \caption{\color{blue} Timeline of LLM Inference Optimization: From Quantum Leaps to Disaggregated Efficiency}
    \label{fig:placeholder}
\end{teaserfigure}
\maketitle

\section{Introduction}

\begin{figure}[h!]
    \centering
    \includegraphics[width=1.0\linewidth]{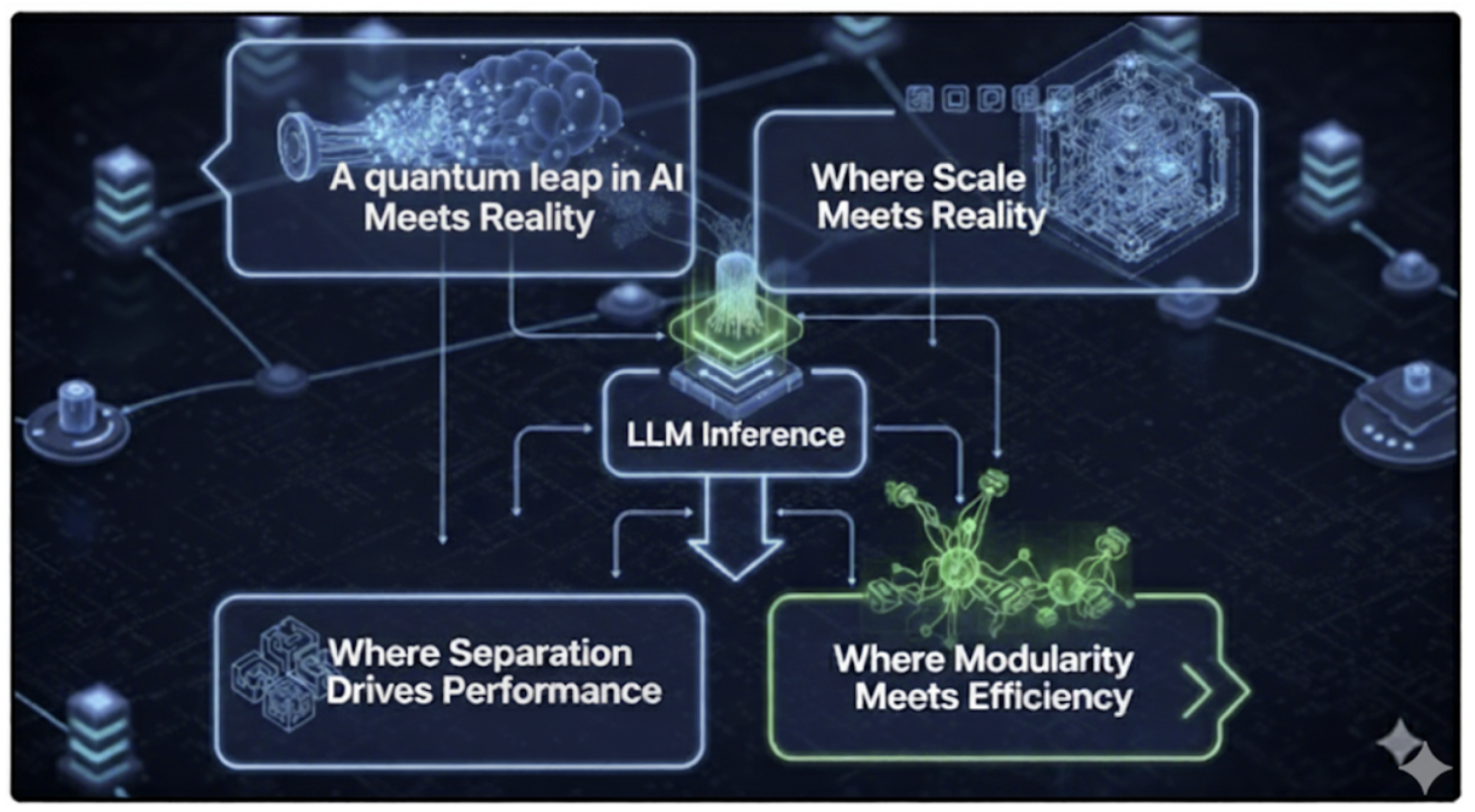}
    \caption{A quantum leap in AI}
    \label{fig:placeholder}
\end{figure}

\subsection{Evolution from Transformers to Large Language Models: \emph{\color{blue} A quantum leap in AI}}
In 2017, the introduction of the Transformer architecture revolutionized natural language processing by replacing sequential recurrent neural networks (RNNs) and long short-term memory (LSTM) units with a fully parallelizable self-attention mechanism capable of handling entire token sequences simultaneously. This breakthrough mitigated vanishing gradient issues, facilitated effective modelling of long-range dependencies, and capitalized on the parallelism inherent in modern GPUs. Subsequent research revealed that scaling parameters, datasets, and computational resources directly cause the emergence of new abilities that are absent in smaller models. This scaling acts as a catalyst, triggering previously unseen capabilities and qualitative shifts in performance that smaller-scale models simply cannot achieve.
This progression is exemplified by the GPT series:

{\bfseries GPT-1} (117 million parameters) established the foundational viability of generative pre-training for natural language understanding. 

{\bfseries GPT-2} (1.5 billion parameters) demonstrated unprecedented zero-shot task transfer capabilities, generating coherent text across diverse domains without task-specific fine-tuning. 

{\bfseries GPT-3} (175 billion parameters) unveiled emergent few-shot and zero-shot reasoning abilities, proving that architectural scale alone could catalyse sophisticated cognitive 
capabilities. 

{\bfseries GPT-3.5} ($ \approx 175 $ billion parameters) introduced conversational refinements and formed the foundation for ChatGPT's November 2022 launch. 

{\bfseries GPT-4} ( $ \approx 1.8 $ trillion parameters) achieved multimodal understanding with text and image processing, released in March 2023.

{\bfseries GPT-5} ($ \approx 635 $ billion parameters for high reasoning) achieved complex tasks, deep analysis and multi-step reasoning, released in August 2025.

\begin{figure}
    \centering
    \includegraphics[width=1.0\linewidth]{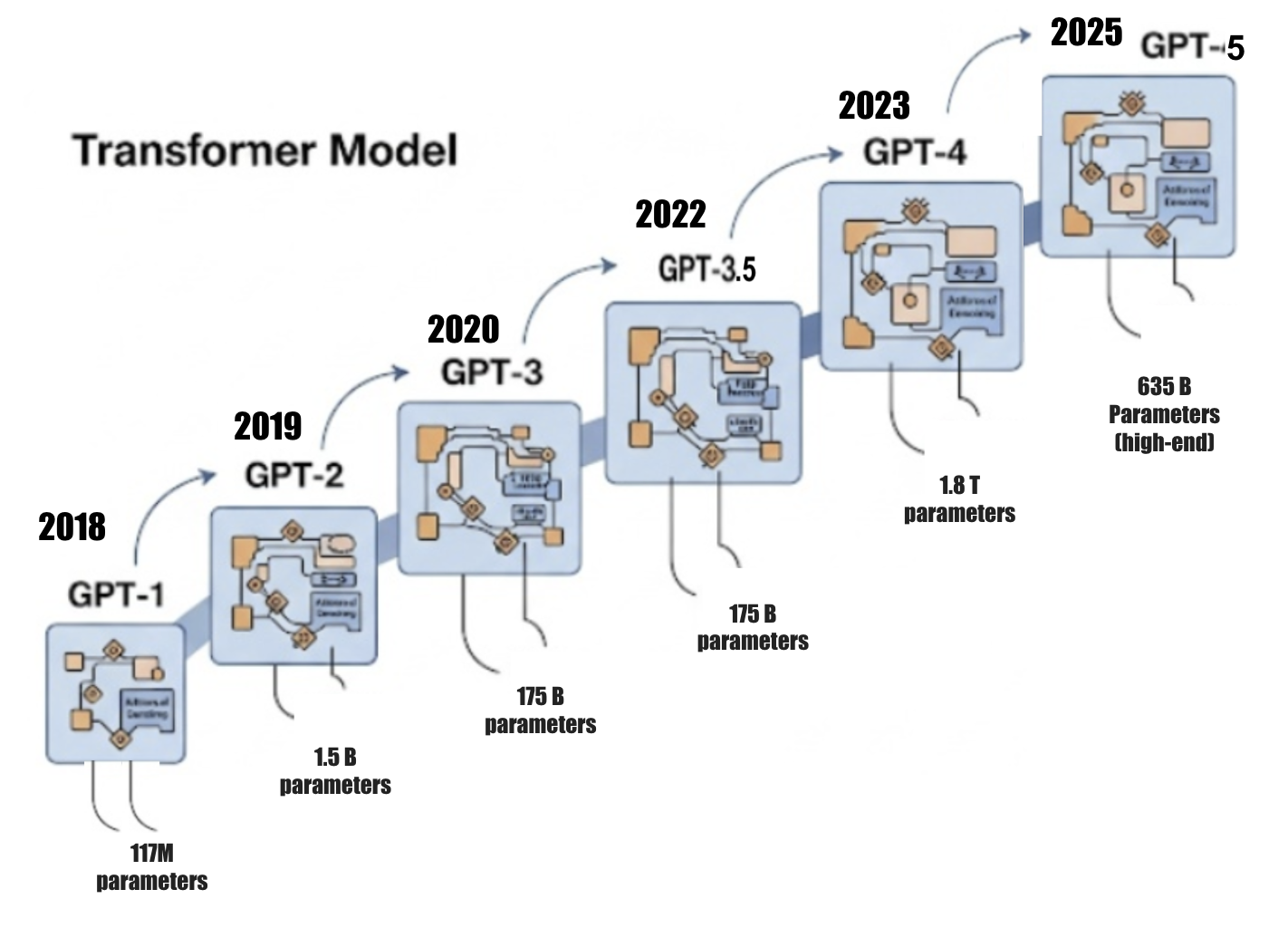}
    \caption{Journey of GPT models}
    \label{fig:placeholder}
\end{figure}

\subsection{The Inference Bottleneck: \emph{\color{blue} Where Scale Meets Reality}}

\emph{\color{blue} The hype around AI has focused on training, but the true technical challenge lies in inference, which is the point where AI systems operate in the real world.}

As models expand to hundreds of billions or even trillions of parameters, inference, the generation of outputs from trained models has emerged as the critical computational challenge. Real-world applications, including conversational agents, code generation tools, and content creation systems, necessitate high throughput (e.g., millions of tokens per second) and low latency (e.g., sub-second time-to-first-token and inter-token delays).

Traditional monolithic inference pipelines, where tokenization, embedding, attention computations, feed-forward networks, and output processing occur within a single integrated runtime on uniform GPUs, are plagued by several issues:

\begin{figure}[h]
    \centering
    \includegraphics[width=1.0\linewidth]{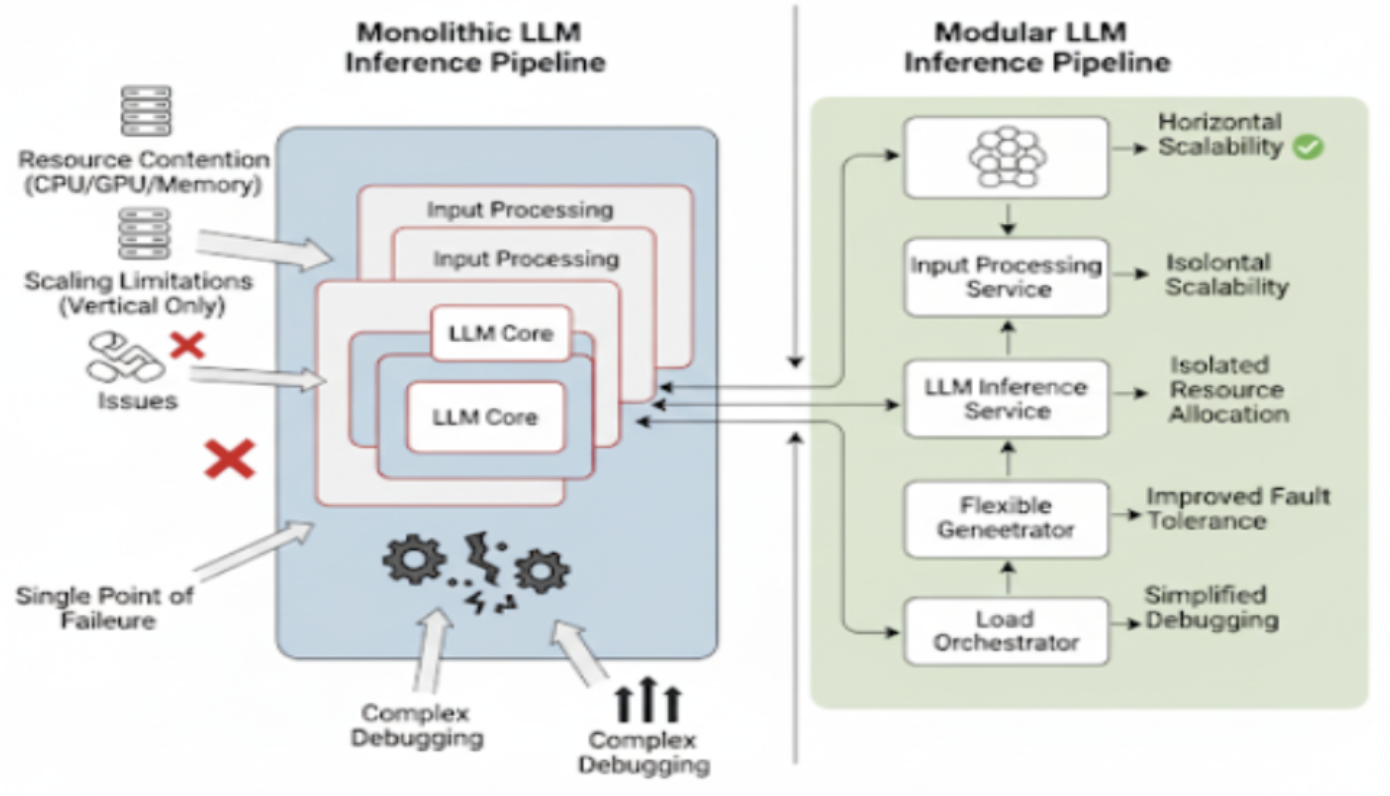}
    \caption{Where Scale Meets Reality}
    \label{fig:placeholder}
\end{figure}

\begin{itemize}
\item {\texttt{\bfseries Resource Contention}: The compute-bound prefill phase and memory-bound decode phase compete for the same GPU cores and memory bandwidth, causing significant underutilization with GPU utilization rates as low as $0.2\%$ in some cases.}
\item {\texttt{\bfseries Rigid Scaling}: Addressing bottlenecks in one phase (e.g., attention) demands replication of the entire pipeline, leading to overprovisioning and inefficient resource allocation.}
\item {\texttt{\bfseries Fault Fragility}: Failures in any component disrupt the whole inference process, creating vulnerable single points of failure that compromise system reliability.}
\item {\texttt{\bfseries Deployment Inflexibility}: Optimizations in kernels or batching strategies require redeploying the complete system, impeding rapid iteration and development cycles.}
\item {\texttt{\bfseries Heterogeneous Workload Inefficiency}: Uniform hardware struggles to optimally support both compute-intensive and memory-intensive tasks, with DRAM bandwidth saturation identified as the primary bottleneck in large-batch inference.}
\end{itemize}

These constraints highlight that conventional inference approaches cannot meet the demands of modern large language models, driving innovation toward specialized hardware architectures, modular pipeline designs, and co-optimized software-hardware solutions to overcome the inference bottleneck.

\subsection{Microservices and Heterogeneous GPU Clusters: \emph {\color{blue} Where Modularity Meets Efficiency}}

\begin{figure}[h]
    \centering
    \includegraphics[width=1.0\linewidth]{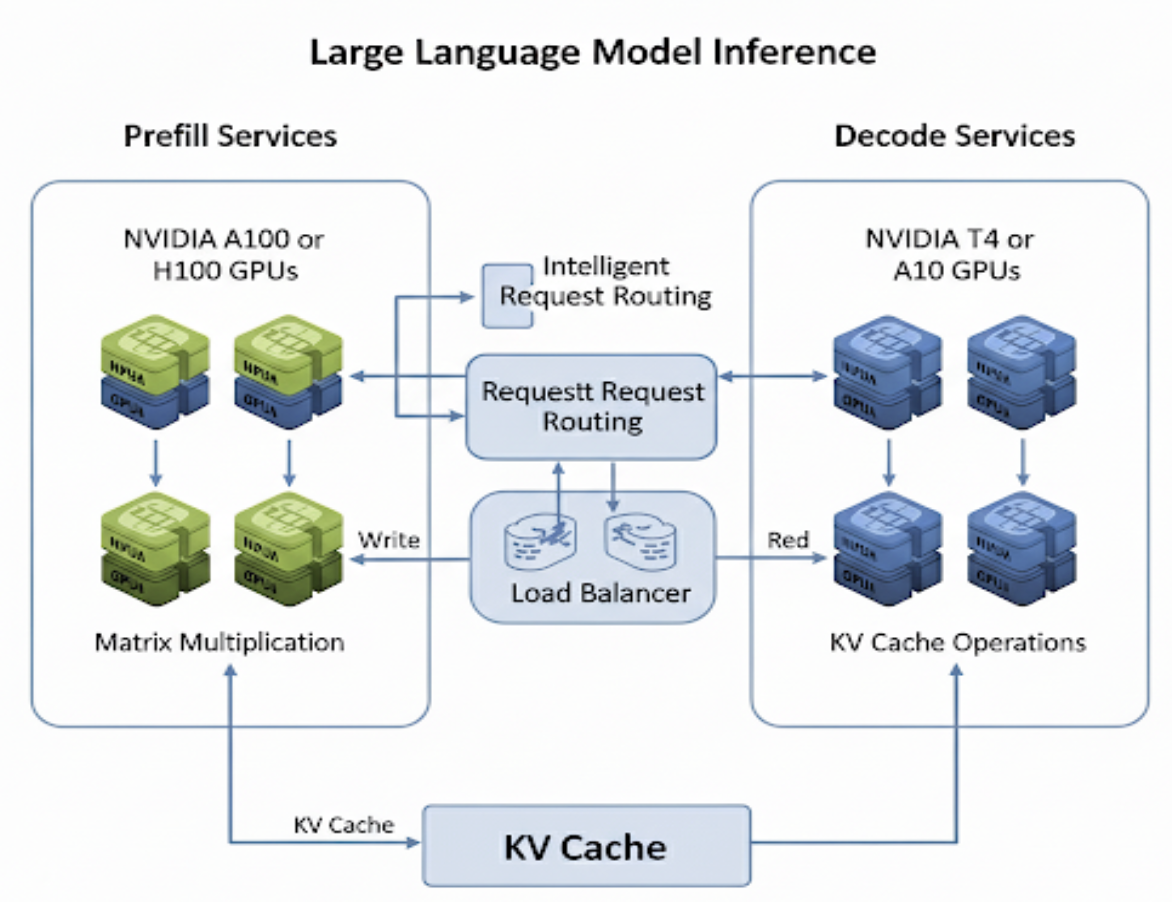}
    \caption{Where Modularity Meets Efficiency}
    \label{fig:placeholder}
\end{figure}

To overcome the limitations of monolithic inference systems, modern AI inference architectures have adopted microservices principles. This involves breaking down the inference pipeline into discrete, modular services such as Tokenization, Embedding, Prefill, Decode, and Post-Processing each independently scalable, monitored, and deployable. Orchestration platforms like Kubernetes enable seamless service discovery, load balancing, and rolling updates, providing resilience and allowing zero-downtime deployments.

A cornerstone of this microservices paradigm is the strategic deployment of heterogeneous GPU clusters tailored to workload characteristics.

Prefill services, which perform extensive matrix-matrix multiplications during initial context processing, are optimally hosted on compute-optimized GPUs such as NVIDIA A100 or H100, equipped with thousands of CUDA cores, specialized Tensor Cores, and multi-TB/s high-bandwidth memory (HBM) for maximum computational throughput. 

Decode services, focused on memory-bound sequential token generation through key-value (KV) cache operations—benefit significantly from memory-optimized GPUs like NVIDIA T4 or A10, which provide larger on-chip caches, more efficient memory hierarchies, and reduced power consumption for sustained inference workloads. This hardware-software co-optimization approach enables intelligent request routing that enhances GPU utilization rates, minimizes energy expenditure, and optimizes cost-performance ratios across the entire inference pipeline.

\begin{table*}[h!]

\begin{tabular}{|l|p{6cm}|p{9cm}|} 
    \hline
    \textbf{Principle} & \textbf{Monolithic Inference}  & \textbf{Microservice/Dis-aggregated Inference} \\
    \hline
    \textbf{Failure domain} & Whole model crashes & Isolated stateless "prefill" and stateful "decode" pools \\
    \hline
    \textbf{Elasticity} & Scale only by cloning the entire model & Independently scale hot paths (prefill GPU fleet) vs. cold paths (decode GPU fleet) \\
    \hline
    \textbf{Heterogeneity} & One accelerator SKU & Mix of A100/H100 for prefill, L40S/L4 for decode \\
    \hline
    \textbf{Consistency model} & Single address space & Explicit state transfer (KV cache, paged attention) across services \\
    \hline
\end{tabular}
\caption{Comparison of Monolithic and Microservice/Dis-aggregated Inference}
    \label{tab:inference_comparison}
\end{table*}

The microservices architecture delivers measurable performance improvements, with some implementations achieving up to $70\%$ reduction in processing times through independent service scaling and specialized hardware allocation. This modular approach ensures that system failures are isolated to individual services rather than cascading through the entire pipeline, maintaining partial functionality even during component outages.

\subsection{The Need for Disaggregated Inference: \emph {\color{blue} Where Separation Drives Performance}}

\begin{figure}[h]
    \centering
    \includegraphics[width=1.0\linewidth]{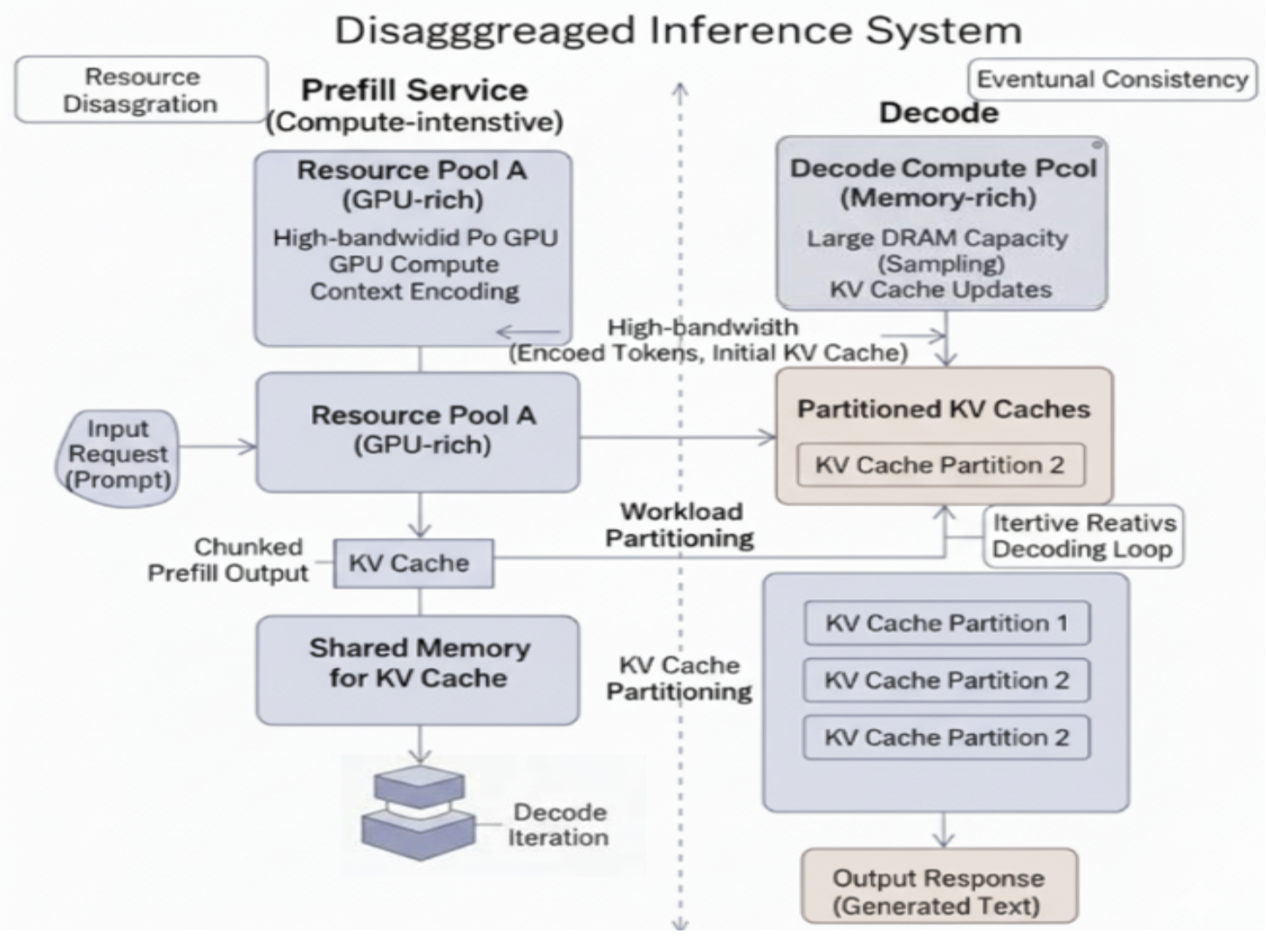}
    \caption{Where Separation Drives Performance}
    \label{fig:placeholder}
\end{figure}

Despite the advantages of microservices and heterogeneous hardware, managing asynchronous and concurrent requests poses additional hurdles, including cross-service compute and memory contention, priority inversion, cache hotspots, and intricate scheduling. Disaggregated inference mitigates these by segregating prefill and decode into autonomous services or clusters, enabling independent scaling and evolution.

This architecture draws from core distributed systems principles
\begin{itemize}
\item {\texttt{\bfseries Resource Disaggregation}: Decouples compute (prefill) from memory (decode), analogous to cloud-based separation of compute and storage.}
\item {\texttt{\bfseries Workload Partitioning}: Distributes tasks across nodes for optimal load balancing and fault isolation.}
\item {\texttt{\bfseries Eventual Consistency}:  Allows transient KV cache inconsistencies to maintain high availability during network partitions.}
\item {\texttt{\bfseries Partitioning for Scalability}: Shards KV caches and model parameters to eliminate centralized bottlenecks.}
\end{itemize}
In the context of the CAP theorem, disaggregated inference favours Availability and Partition Tolerance over strict Consistency, tolerating brief cache inconsistencies to maintain throughput amid network or hardware partitions. By directing requests to operational nodes and containing failures, disaggregation ensures uninterrupted service and facilitates modular advancement of inference functionalities.

In the subsequent sections, we delineate six essential enhancements Key-Value Caching, Flash Attention, Continuous Batching, Speculative Decoding, Paged Attention, and Radix Attention—each rooted in distributed systems principles and tailored to optimize inference for sequential workloads. We explore their mechanisms, adoption histories, benefits, and CAP trade-offs, laying the groundwork for a disaggregated inference framework that fulfils the requirements of contemporary, large-scale LLM deployments.

\section{Core Algorithmic and Architectural Enhancements in Large Language Model Inference}

\subsection{KV Cache: \emph {\color{blue} Memory Meets Throughput}}
\begin{figure}[h]
    \centering
    \includegraphics[width=1.0\linewidth]{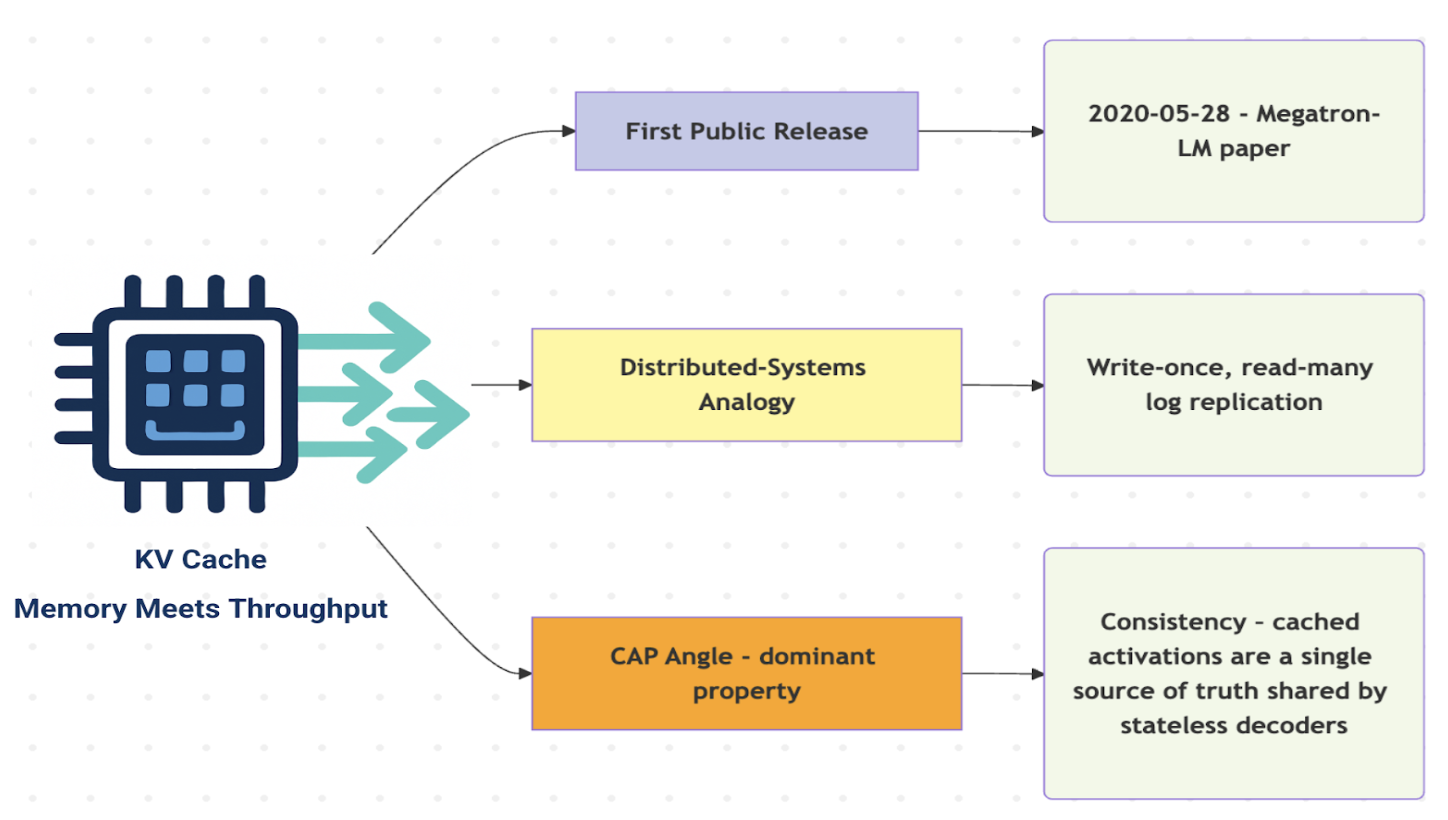}
    \caption{Memory Meets Throughput}
    \label{fig:placeholder}
\end{figure}
LLM inference faces inefficiency due to its autoregressive nature, where each new token generation requires re-processing all previous tokens, causing redundant computation. In Transformers, Key (K) and Value (V) vectors for earlier tokens are recalculated unnecessarily, leading to quadratic scaling of computation with sequence length and increased overhead in distributed setups.

The KV Cache solves this by storing pre-computed K and V vectors for all previous tokens. During decoding, only the new token’s vectors are calculated while reusing cached K and V pairs, reducing complexity from O(N²) to O(N) per token and significantly speeding up inference.

Introduced between 2018 and 2020 with models like GPT-2, KV caching incurs linear memory growth with sequence length and model size, consuming several gigabytes in large models.

\begin{figure}[h]
    \centering
    \includegraphics[width=1.0\linewidth]{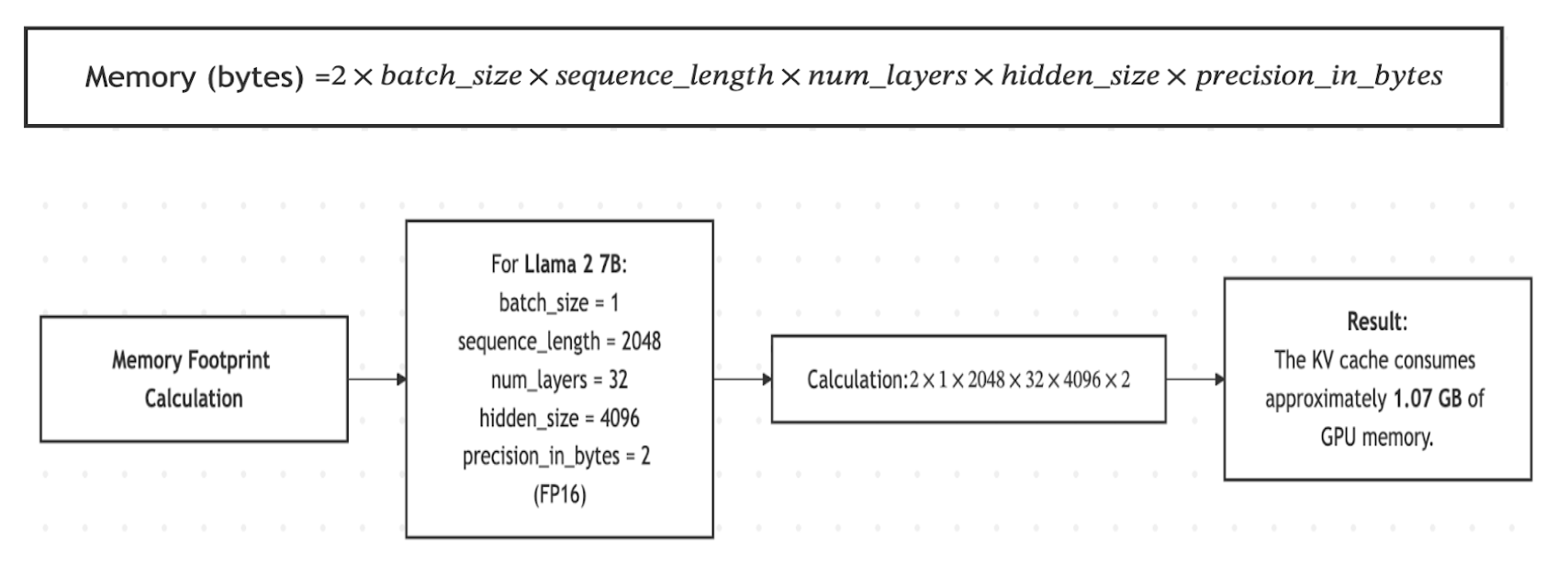}
    \caption{KV cache memory usage calculation}
    \label{fig:placeholder}
\end{figure}

From a distributed systems perspective, the KV cache mirrors caching mechanisms in distributed databases or content delivery networks (CDNs), storing intermediate attention states to avoid redundant computation. Applying the CAP theorem to distributed LLM inference, the KV cache balances Consistency (coherent cache state across nodes), Availability (fast access to cached data), and Partition Tolerance (resilience to node communication failures) to enable scalable and fault-tolerant inference in GPU clusters.

\begin{figure}[h]
    \centering
    \includegraphics[width=1.0\linewidth]{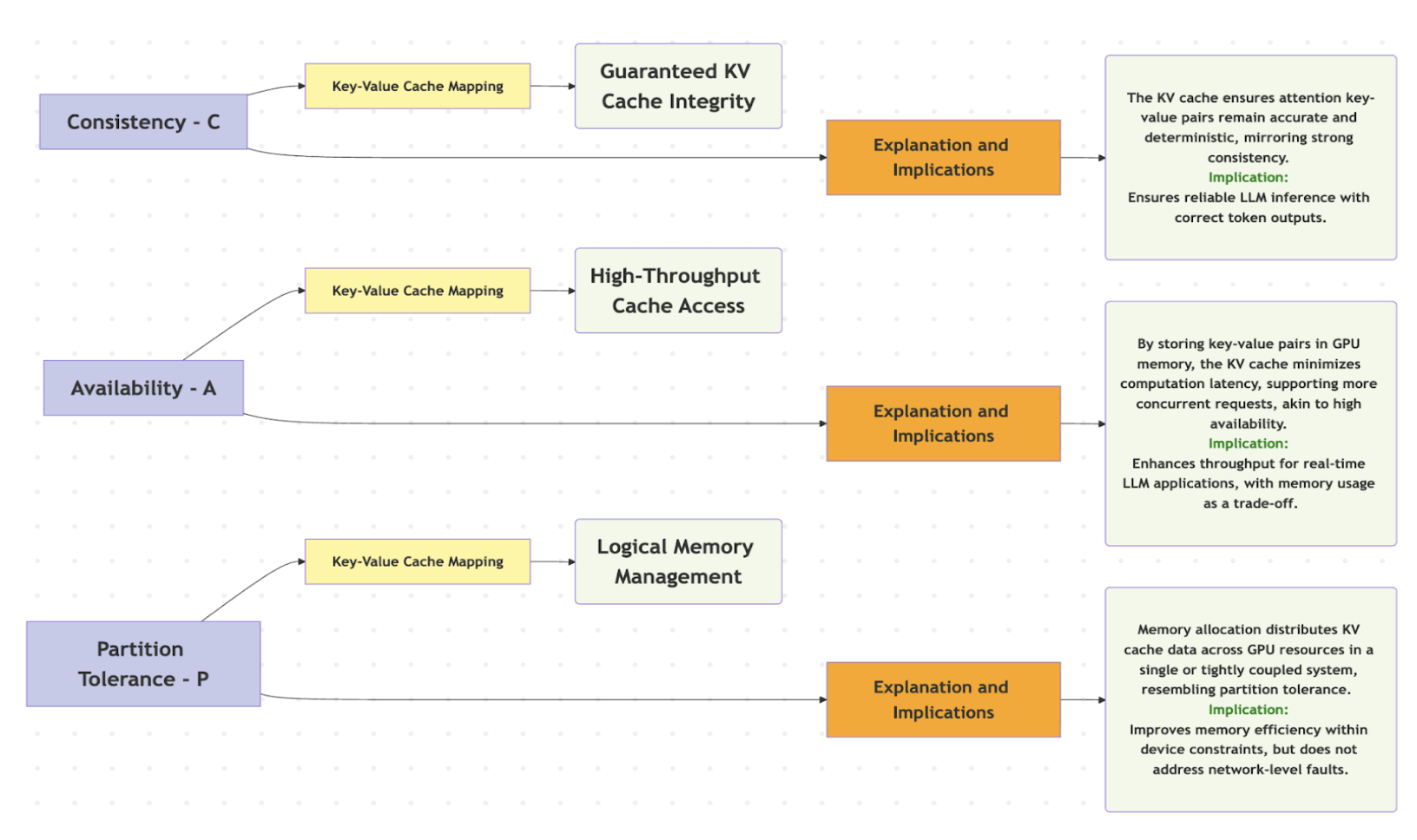}
    \caption{{\bfseries Note}: The CAP theorem is applied metaphorically, treating memory management as logical resource allocation within a single or tightly coupled system, not networked nodes.
}
    \label{fig:placeholder}
\end{figure}

\subsection{Flash Attention: \emph {\color{blue} High-Throughput Attention via I/O Optimization}}

While the KV cache drastically reduced computational redundancy, the self-attention mechanism's performance was still gated by a different bottleneck: the inefficient data movement between the GPU's compute cores and its high-bandwidth memory, rather than the raw FLOPs of the computation itself.

\begin{figure}[h!]
    \centering
    \includegraphics[width=1.0\linewidth]{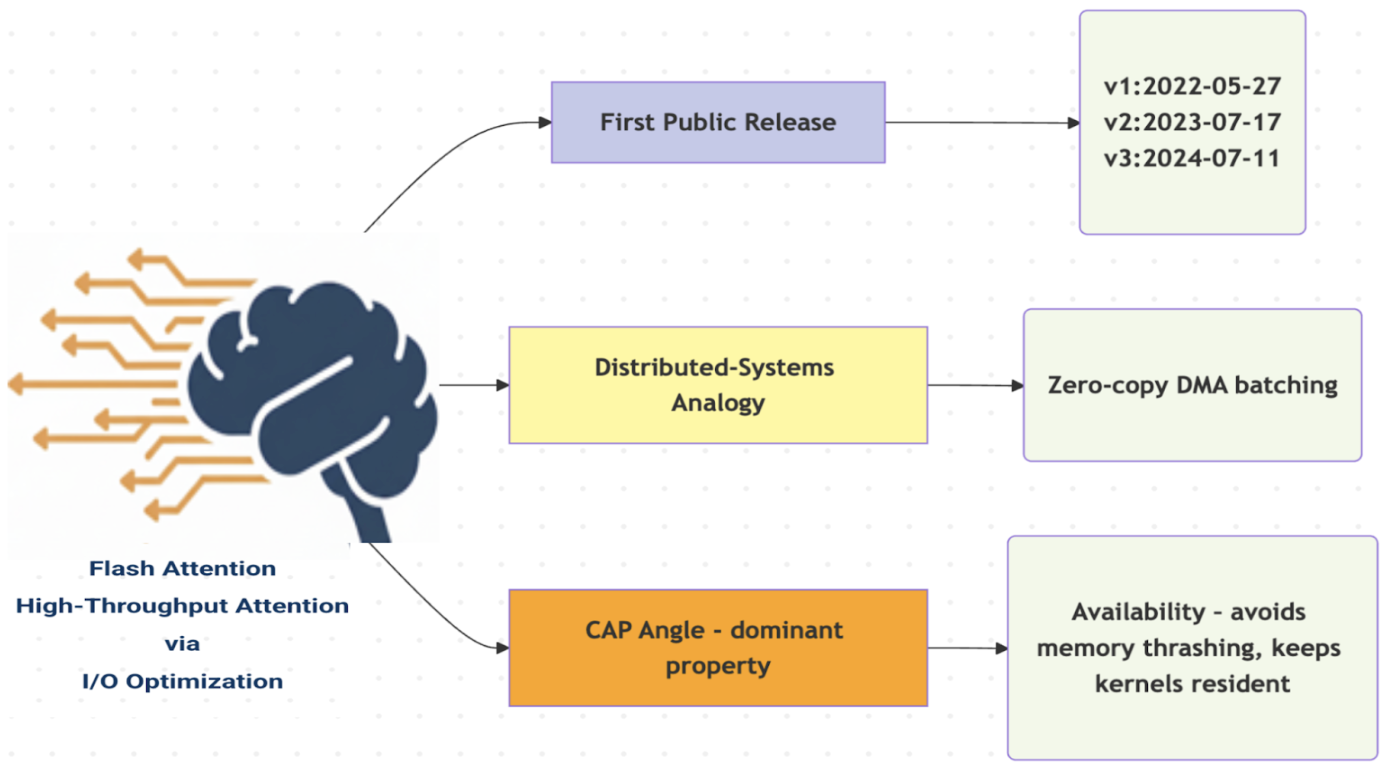}
    \caption{High-Throughput Attention via I/O Optimization}
    \label{fig:placeholder}
\end{figure}

To understand this, one must understand the GPU's memory architecture, which is a classic memory hierarchy:

{\bfseries The GPU Memory Hierarchy}
\begin{figure}[h]
    \centering
    \includegraphics[width=1.0\linewidth]{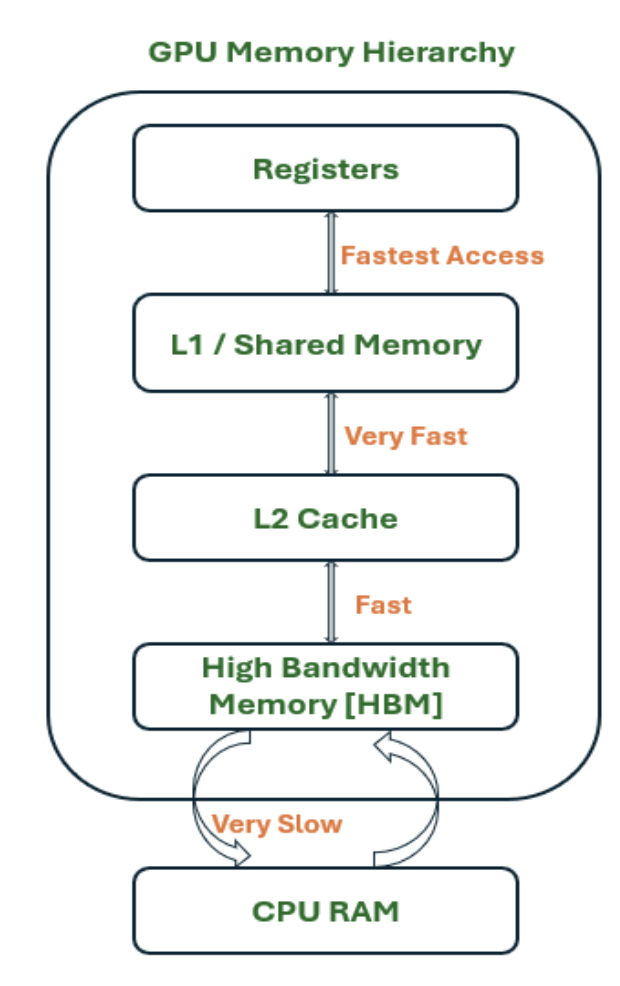}
    \caption{GPU Memory Hierarchy}
    \label{fig:placeholder}
\end{figure}

\begin{enumerate}
    \item \textbf{Registers}
    \begin{itemize}
        \item Located within each streaming multiprocessor (SM).
        \item Fastest access but extremely limited in size.
        \item Hold per-thread variables during kernel execution.
    \end{itemize}
    \item \textbf{L1 / Shared Memory (SRAM)}
    \begin{itemize}
        \item On-chip scratchpad memory accessible by threads within a block.
        \item Slightly larger than registers yet still very small.
        \item Very fast access (nanoseconds), ideal for reuse across operations.
    \end{itemize}
    \item \textbf{L2 Cache}
    \begin{itemize}
        \item Shared across all SMs on the GPU die.
        \item Larger capacity than L1 but with slower access times.
        \item Caches data transferred between on-chip memory and HBM.
    \end{itemize}
    \item \textbf{High-Bandwidth Memory (HBM)}
    \begin{itemize}
        \item The GPU’s main memory (VRAM), with tens of gigabytes of capacity.
        \item Fast bandwidth but higher latency compared to on-chip caches.
        \item Main storage for large data structures (model weights, feature maps).
    \end{itemize}
    \item \textbf{CPU RAM}
    \begin{itemize}
        \item System memory on the host CPU.
        \item Very slow relative to GPU memory layers.
        \item Data must traverse the PCIe or NVLink bus to reach the GPU, introducing significant latency.
    \end{itemize}
\end{enumerate}

\begin{table*}[h!]

\begin{tabular}{|l|p{3cm}|p{3cm}|p{3cm}|p{3cm}|} 
    \hline
    \textbf{Memory Level} & \textbf{Capacity}  & \textbf{Access Latency} & \textbf{Bandwidth} & \textbf{Scope}\\
    \hline
    \textbf{Registers} & ~256KB per SM & 1 cycle & Fastest & Per-thread\\
    \hline
    \textbf{Shared Memory} & 128KB per SM & 5-10 cycles & Very Fast & Per-block\\
    \hline
    \textbf{L2 Cache} & 6-50MB & 50-100 cycles & Fast & Global\\
    \hline
    \textbf{HBM/Global Memory} & 32-80GB & 300-500 cycles & 1-3 TB/s & Global\\
    \hline
\end{tabular}
\caption{Key GPU Memory Hierarchy Characteristics Performance Trade-offs}
    \label{tab:inference_comparison}
\end{table*}

A standard attention implementation requires multiple passes over the input data. The large intermediate attention matrix ($S = Q * K^{\mathsf{T}}$) must be written to and read back from HBM, creating a severe memory I/O bottleneck. The SMs spend more time waiting for data from HBM than they do actually computing.

Flash Attention is a groundbreaking algorithm that reorders the attention computation to avoid this I/O bottleneck. It uses techniques like tiling and fused kernels.

\begin{itemize}
\item It breaks the large attention matrix into smaller blocks or "tiles" that can fit into the fast on-chip SRAM (L1/L2 cache).
\item It performs the entire attention computation (matrix multiplication and SoftMax) for a block in one go within SRAM without ever writing the full intermediate matrix to HBM.
\end{itemize}

Flash Attention specifically optimizes this hierarchy by:
\begin{itemize}
\item {\texttt{\bfseries Minimizing HBM Access}: Keeping attention computation in fast SRAM (shared memory/L1)}
\item {\texttt{\bfseries Blocking Strategy}: Processing attention in blocks that fit entirely in shared memory}
\item {\texttt{\bfseries Fused Operations}: Combining multiple operations to reduce intermediate HBM writes}
\item {\texttt{\bfseries Asynchronous Processing}: Overlapping computation with memory transfers between levels}
\end{itemize}

This memory hierarchy understanding is fundamental to why Flash Attention achieves such significant performance improvements by working with the GPU's memory architecture rather than against it, minimizing expensive data movement between slow HBM and fast on-chip memory.

\begin{figure}[h]
    \centering
    \includegraphics[width=1.0\linewidth]{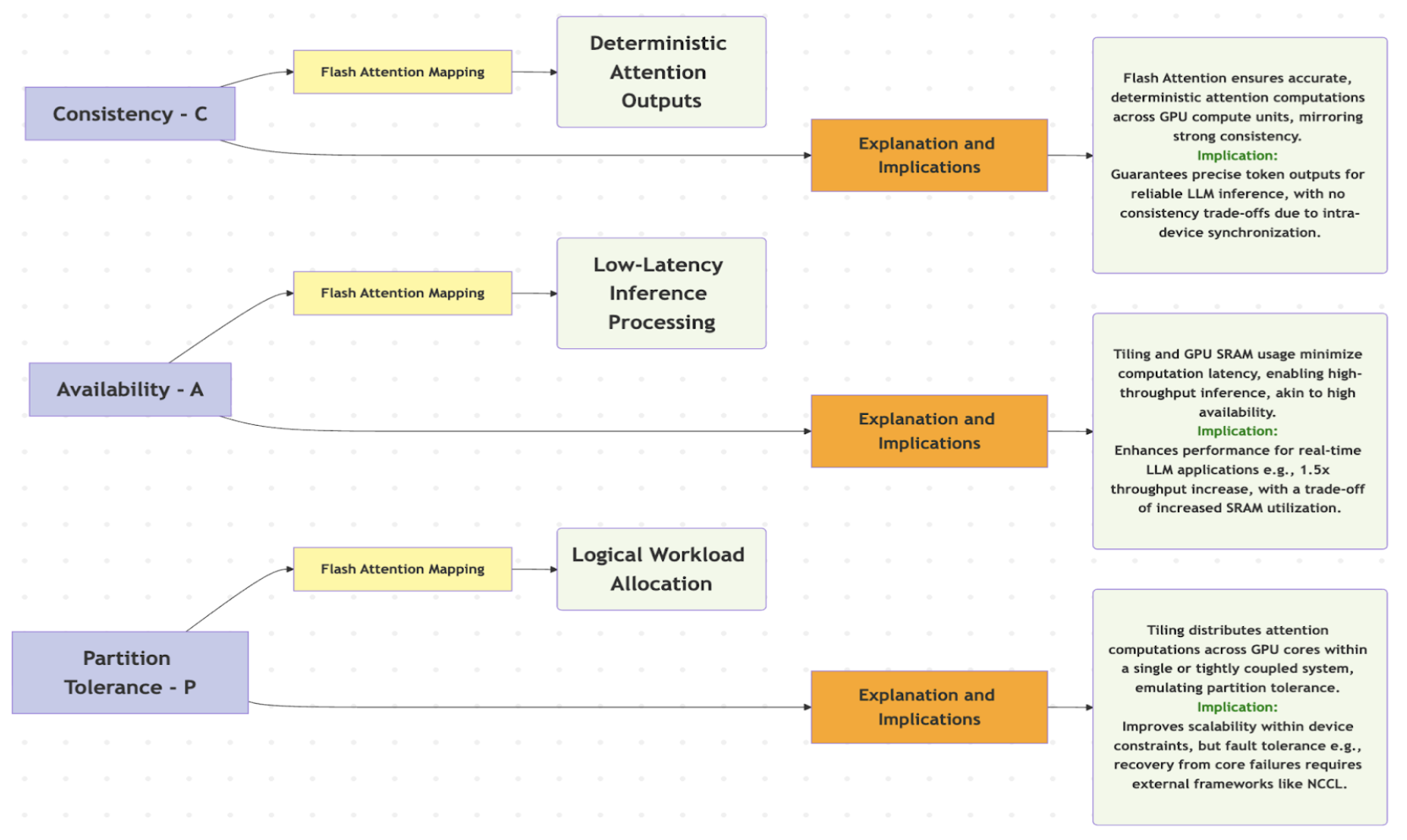}
      \caption{{\bfseries Note}: The CAP theorem is applied metaphorically, with workload allocation treated as logical resource management within a single or tightly coupled system, not networked nodes.
}
    \label{fig:placeholder}
\end{figure}

From a distributed systems perspective, Flash Attention illustrates both parallelism (tiling, akin to MapReduce) and locality (using SRAM as a near cache). Interpreted through the CAP theorem, it emphasizes Availability (rapid computation) and Partition Tolerance (operation across independent nodes), while providing consistency within its defined computational window rather than 'eventual consistency' in the traditional distributed systems sense.

\subsection{Speculative Decoding: \emph {\color{blue} Accelerating Inference through Predictive Drafting}}

Speculative decoding accelerates inference by using a small, fast "draft" model to generate a chunk of several tokens at once. This chunk is then passed to the main, large language model for verification in a single, parallel pass. If the draft model's predictions are correct, the entire chunk is accepted, allowing the system to generate multiple tokens in the time it would normally take to generate one. If any tokens are incorrect, the system accepts the correct portion and continues from there.

\begin{figure}[h]
    \centering
    \includegraphics[width=1.0\linewidth]{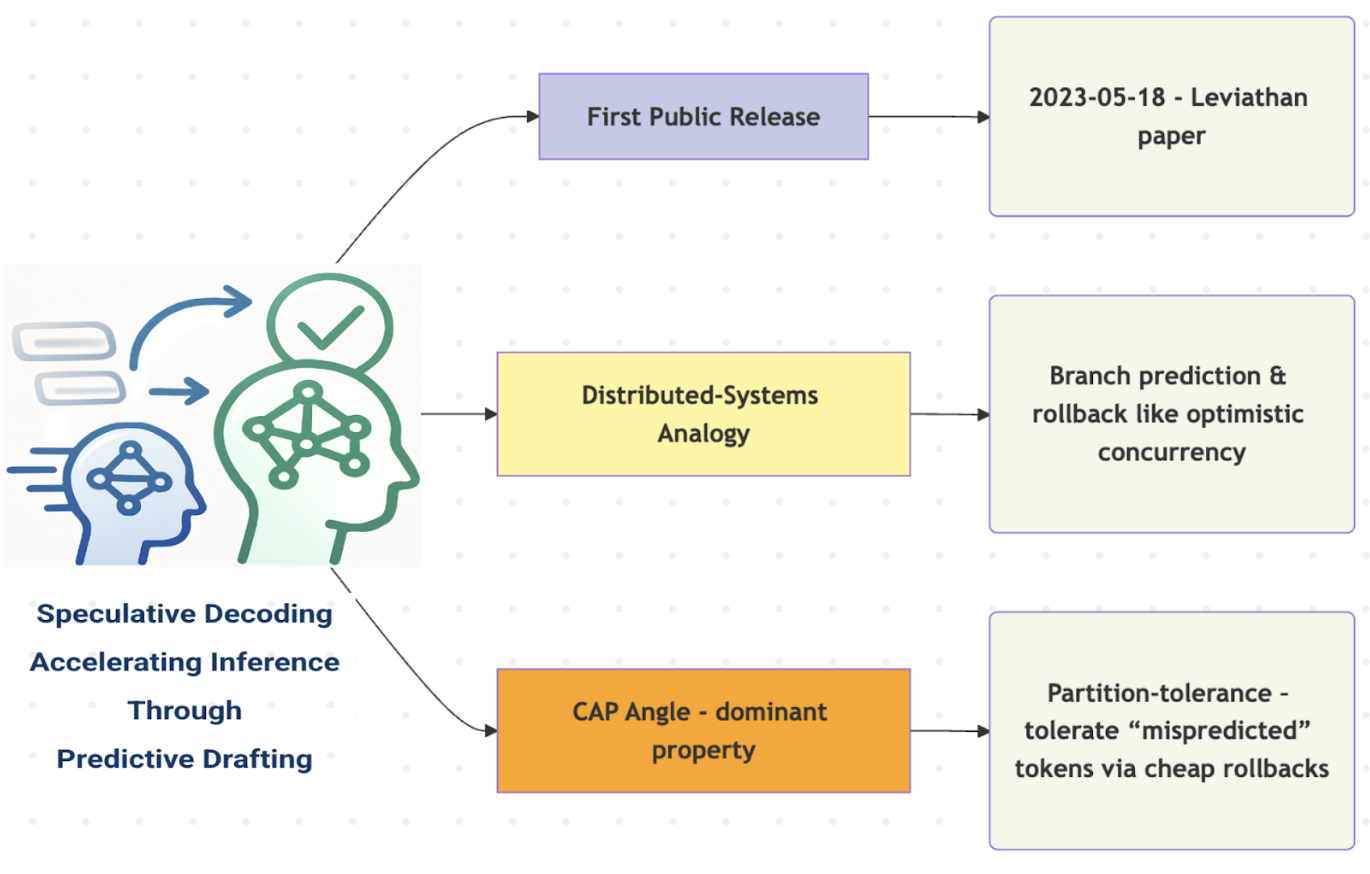}
    \caption{Accelerating Inference through Predictive Drafting}
    \label{fig:placeholder}
\end{figure}

\begin{figure}[h]
    \centering
    \includegraphics[width=1.0\linewidth]{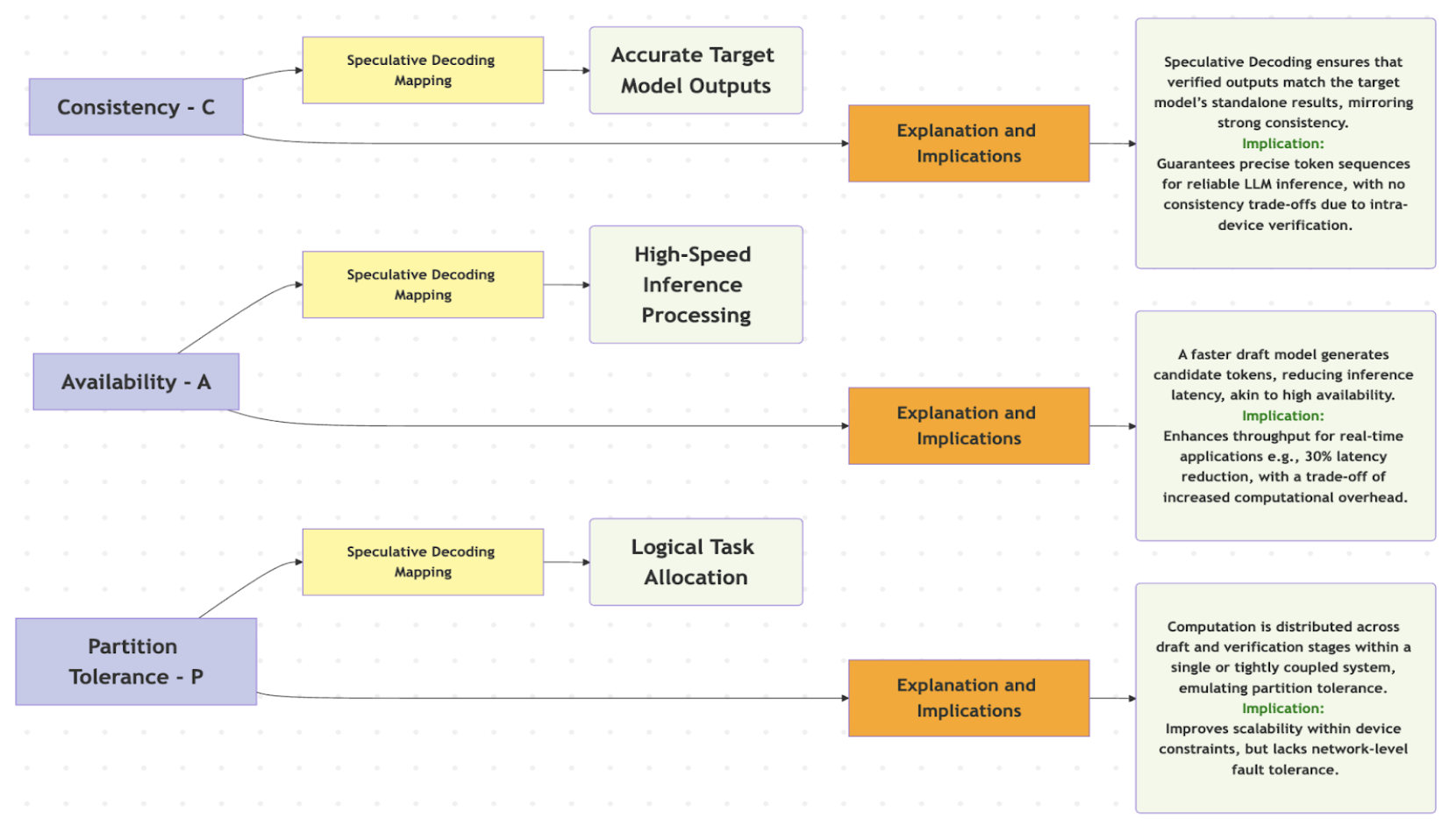}
     \caption{{\bfseries Note}: The CAP theorem is applied metaphorically, treating draft and verification stages as logical components within a single or tightly coupled system, not networked nodes.
}
    \label{fig:placeholder}
\end{figure}

From a distributed systems perspective, this is a form of speculative execution or optimistic concurrency control. The system optimistically performs work (generates a token chunk) and then validates it, a common pattern in high-performance distributed databases to improve latency. Interpreted through the CAP principle, Speculative decoding is a brilliant example of improving Availability (responsiveness) while strictly preserving Consistency. The verification step ensures the final output is always identical to what the large model would have generated on its own.

\begin{figure}[h]
    \centering
    \includegraphics[width=1.0\linewidth]{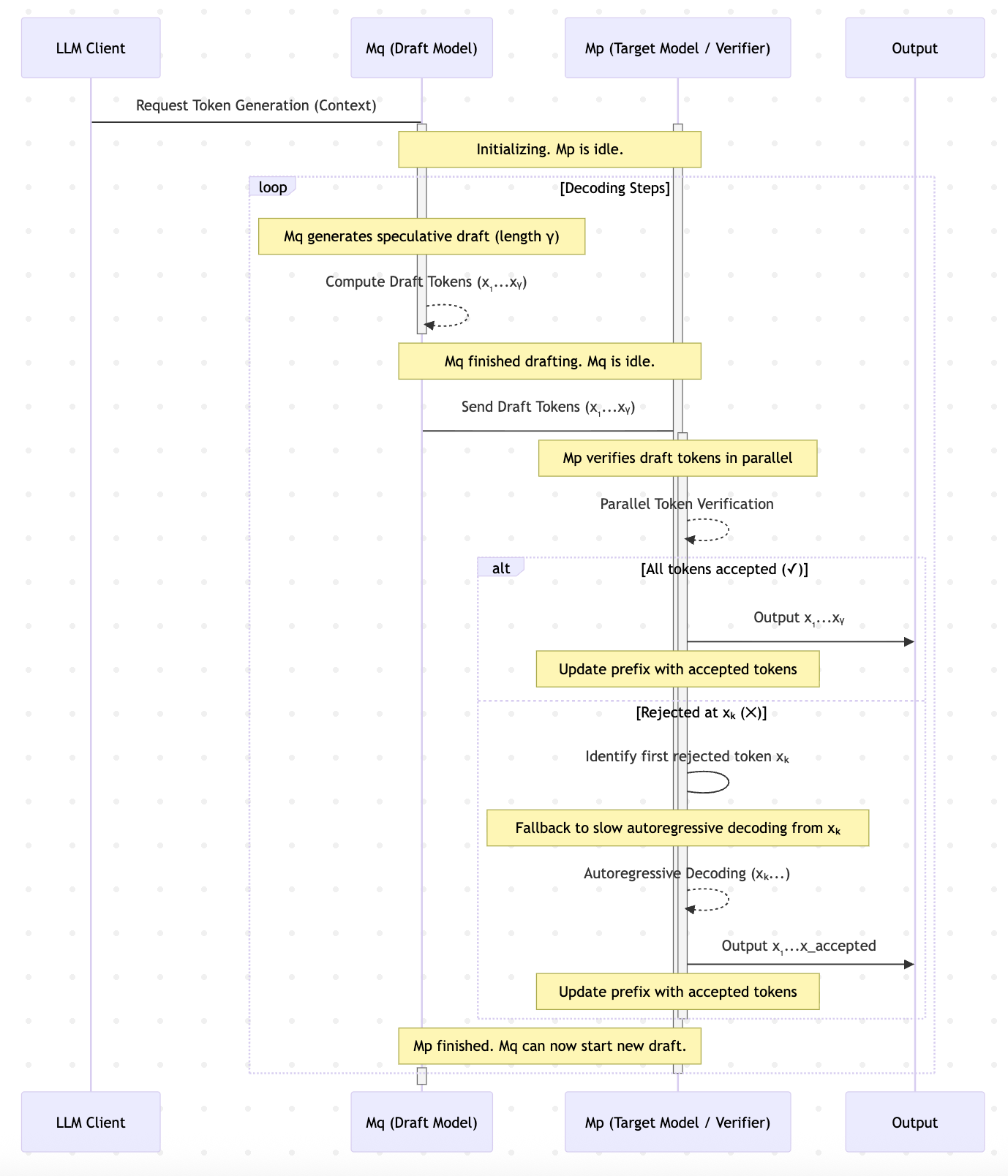}
    \caption{Speculative Decoding process}
    \label{fig:placeholder}
\end{figure}

\subsubsection{Parallel Speculative Decoding : \emph {\color{blue} Accelerating Inference through Parallel Adaptive Drafting}}

Parallel Speculative Decoding (PEARL) is an advanced inference acceleration framework designed to maximize concurrency between the token generation (drafting) and verification phases during Large LanguageModel (LLM) decoding. Traditional Speculative Decoding (SD) employs two cooperating models: a lightweight draft model (Mq)  that predicts multiple candidate tokens ahead, and a larger target model (Mp) that validates these candidates. However, SD operates sequentially the target model (Mp) must wait for the draft model (Mq) to finish its proposals before beginning verification, resulting in mutual idle phases and suboptimal hardware utilization.

PEARL overcomes this mutual waiting bottleneck by enabling true parallelism between Mq and Mp. Both models execute concurrently in a pipelined decoding loop, where continuously proposes speculative tokens while incrementally verifies incoming tokens in parallel. This overlapping execution eliminates synchronization stalls, minimizes idle computation, and ensures continuous token flow throughout the decoding process. The result is a steady, high throughput decoding pipeline that achieves superior inference efficiency without sacrificing prediction accuracy.

To further optimize decoding dynamics, PEARL introduces adaptive drafting strategies through two complementary mechanisms

\begin{enumerate}
    \item \textbf{Pre-Verify Strategy:}
    While the draft model (Mq) is generating a token sequence, the target model (Mp) immediately begins verifying the earliest proposed token.
    \begin{itemize}
	\item {\texttt If this token is rejected, the remaining unverified draft tokens are discarded early, effectively shortening the draft length.}
	\item {\texttt This adaptive truncation prevents wasted computation on low-confidence or high-entropy text segments.}
	\end{itemize}
	
    \item \textbf{Post-Verify Strategy:}
    During the verification of the current draft by Mp, the draft model (Mq) continues speculatively generating additional tokens, assuming acceptance.
    \begin{itemize}
    \item {\texttt If the current draft is fully validated, these extra tokens are seamlessly merged into the next cycle, effectively extending the draft length.}
	\item {\texttt This adaptive extension benefits predictable, low-entropy regions, improving throughput.}
	\end{itemize}

\end{enumerate}

\begin{figure}[h]
    \centering
    \includegraphics[width=1.0\linewidth]{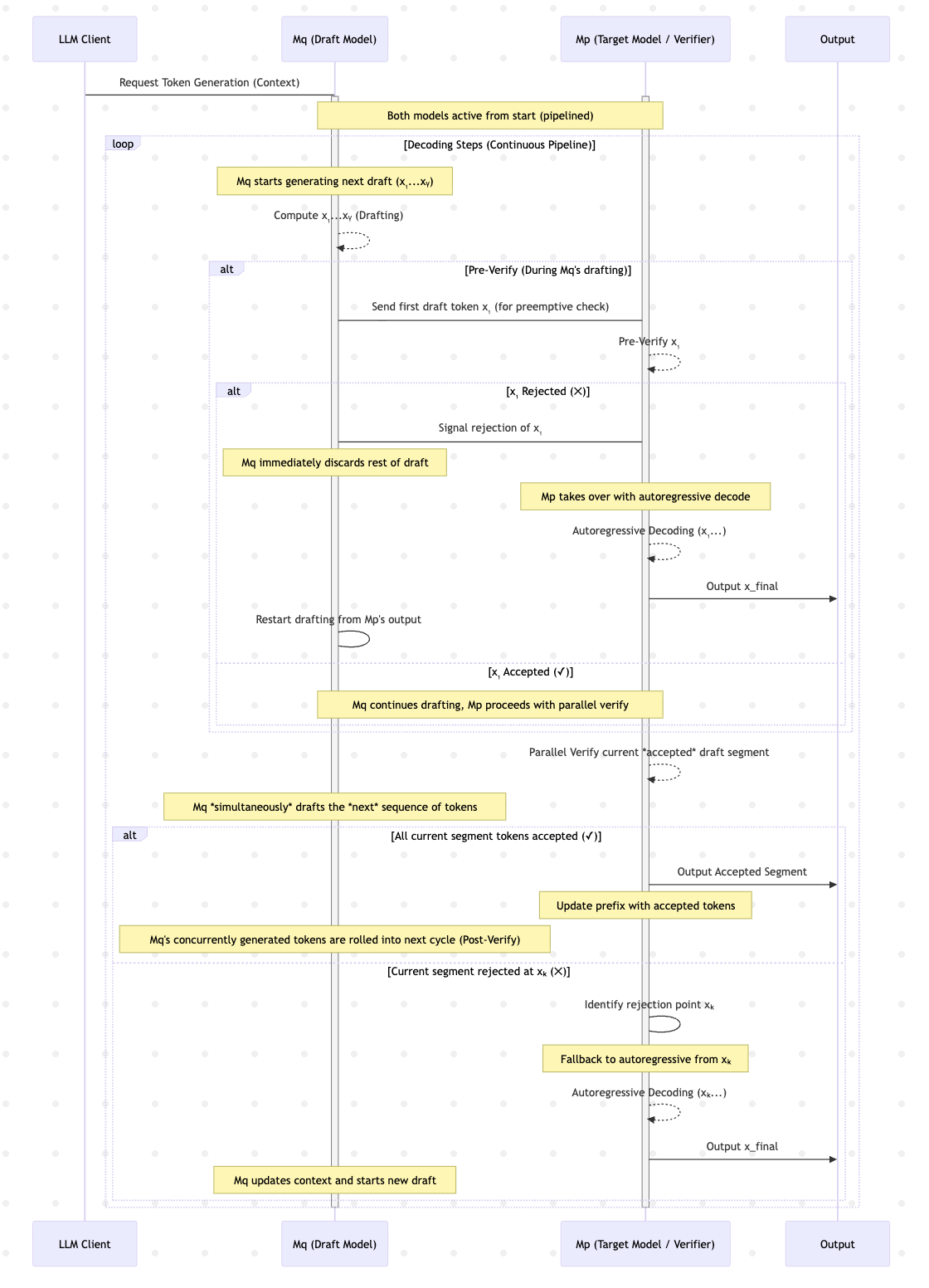}
    \caption{Parallel Speculative Decoding process}
    \label{fig:placeholder}
\end{figure}

\subsection{Continuous Batching: \emph {\color{blue} Maximizing Throughput}}

Traditional static batching groups several requests and processes them together. The entire batch, however, must wait for the longest sequence to finish before the next batch can start. This leads to significant GPU idle time, as resources are stalled waiting for a single request.

\begin{figure}[h]
    \centering
    \includegraphics[width=1.0\linewidth]{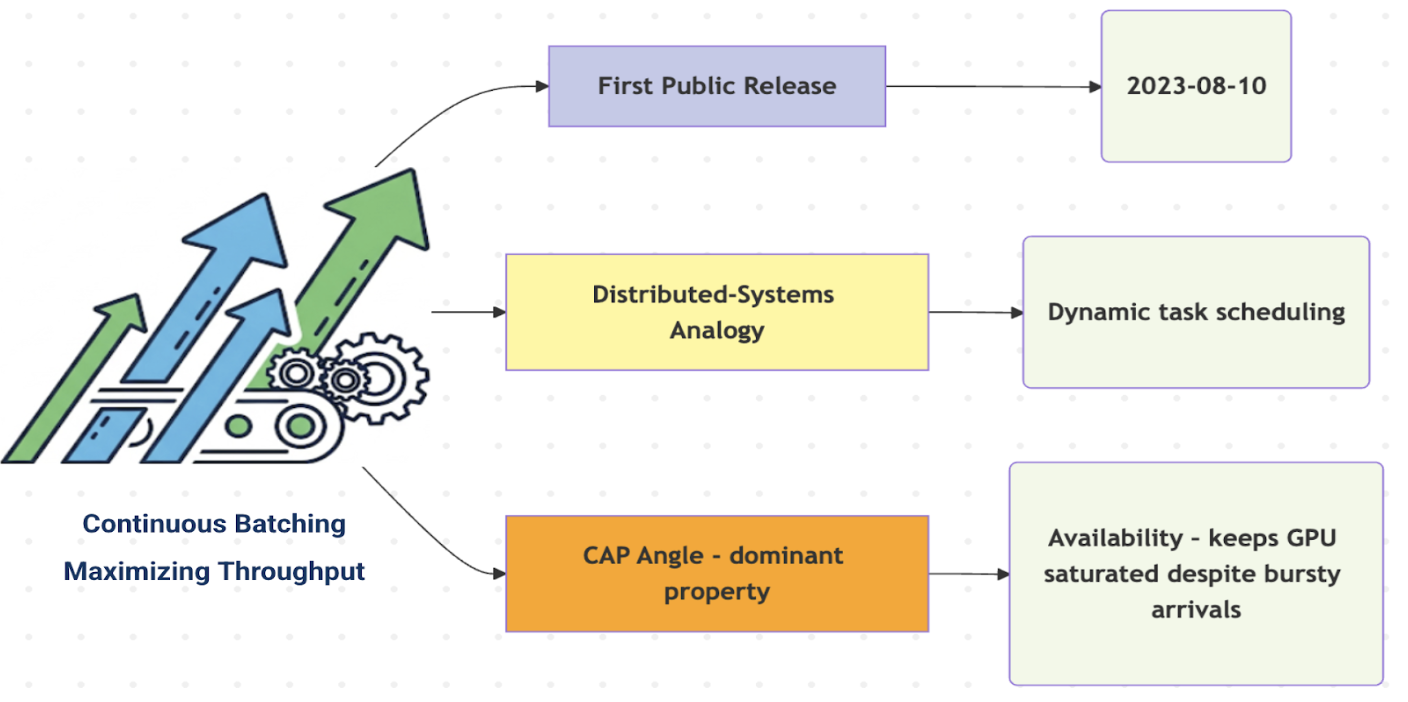}
    \caption{Maximizing Throughput}
    \label{fig:placeholder}
\end{figure}

Continuous batching solves this by operating as a continuous process. As soon as any single sequence in the batch finishes, a new sequence from the queue is immediately added. This ensures the GPU is always processing a full batch of work, maximizing its utilization and dramatically increasing overall server throughput.

From the standpoint of distributed systems, this mechanism embodies a direct application of dynamic scheduling and resource management. It operates analogously to an advanced task scheduler (such as Kubernetes), ensuring that computational resources remain fully utilized to optimize both efficiency and throughput. Considered through the CAP theorem, the design emphasizes Availability. By driving high GPU utilization, the system can support greater concurrency and minimize user wait times, thereby improving responsiveness, while leaving the Consistency of each individual generation intact.

\begin{figure}
    \centering
    \includegraphics[width=1.0\linewidth]{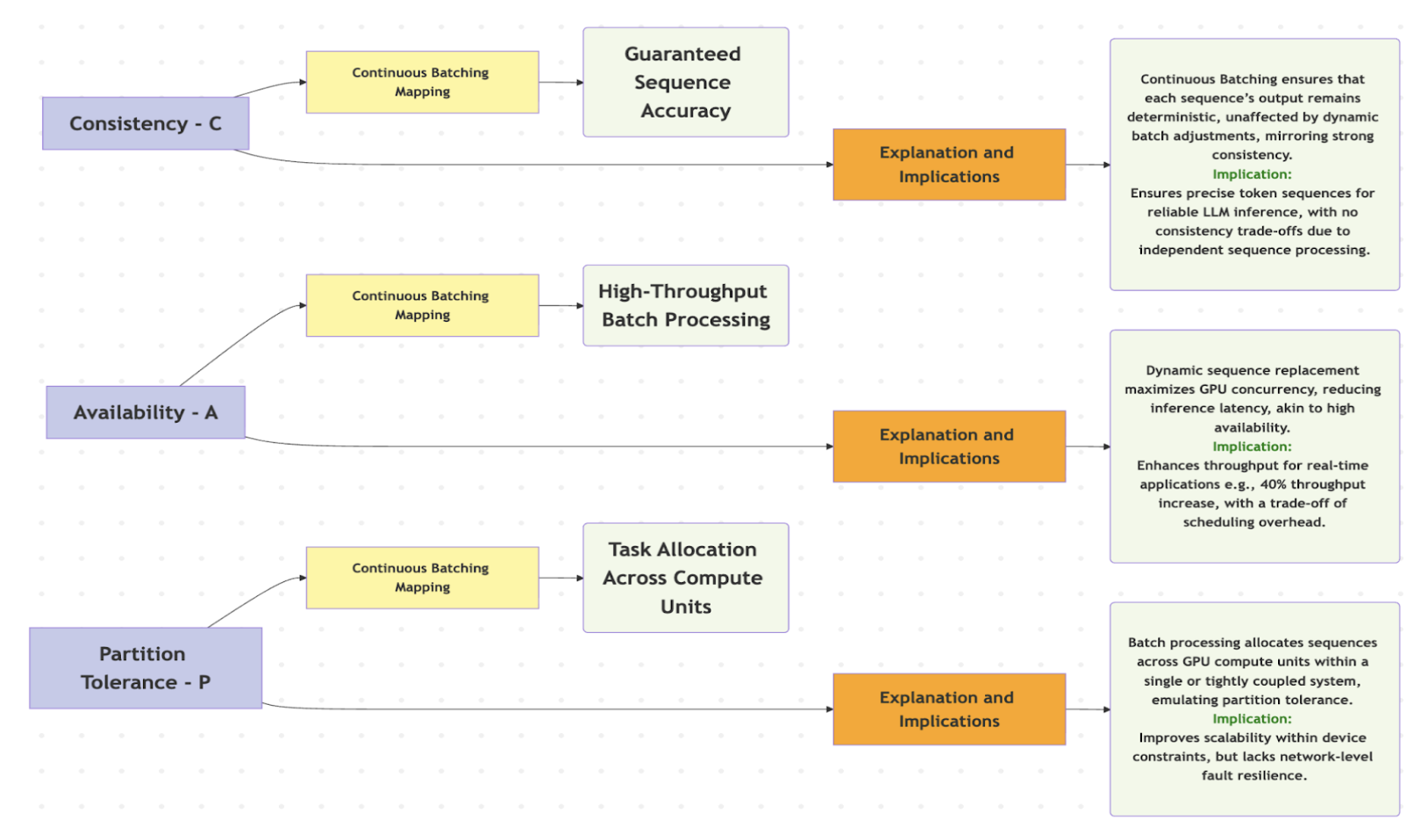}
   \caption{{\bfseries Note}: The CAP theorem is applied metaphorically, treating batch processing as logical task allocation within a single or tightly coupled system, not networked nodes.
}
    \label{fig:placeholder}
\end{figure}

\subsection{Paged Attention: \emph {\color{blue} Intelligent Memory Management}}

The KV cache, while crucial, creates a memory management nightmare. Traditionally, the cache for each sequence requires a single, contiguous block of memory. This leads to:

\begin{itemize}
\item {\texttt{\bfseries Internal Fragmentation}: Wasted space within an allocated block.}
\item {\texttt{\bfseries External Fragmentation}: The memory is so broken up into small, non-contiguous pieces that there isn't a large enough single block to start a new request, even if the total free memory is sufficient.}
External Fragmentation: The memory is so broken up into small, non-contiguous pieces that there isn't a large enough single block to start a new request, even if the total free memory is sufficient.
\end{itemize}

\begin{figure}[h]
    \centering
    \includegraphics[width=1.0\linewidth]{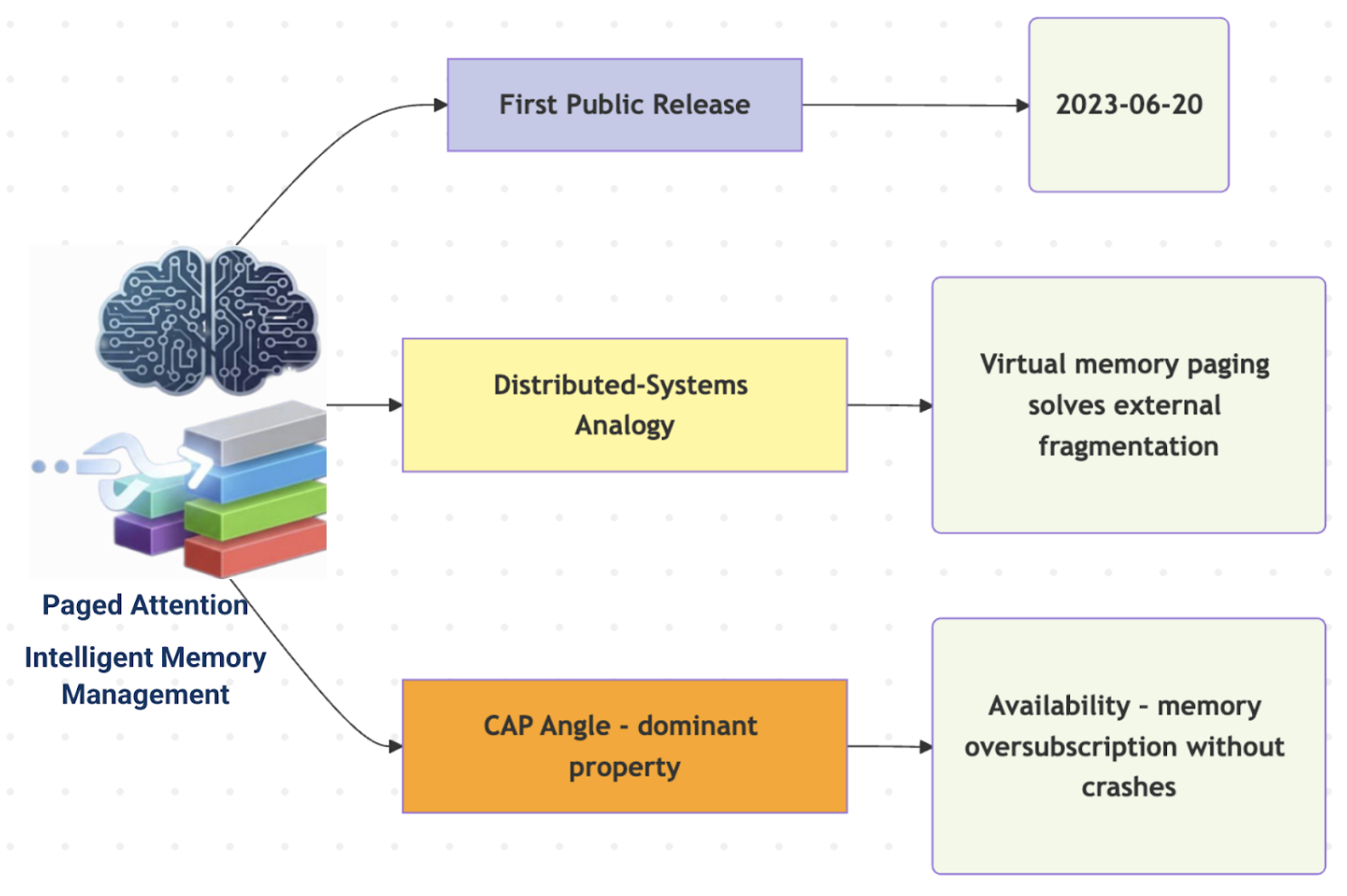}
    \caption{Intelligent Memory Management}
    \label{fig:placeholder}
\end{figure}

Paged Attention solves this by borrowing the concept of virtual memory from operating systems. It partitions the KV cache into fixed-size blocks, or "pages." A sequence's cache can now be stored in non-contiguous pages scattered throughout memory. A "page table" tracks which pages belong to which sequence. This approach eliminates fragmentation, enabling up to $96\%$ memory efficiency.

From a distributed systems perspective, this mechanism is a clear instance of virtualization. By abstracting the underlying physical memory layout, Paged Attention introduces a more flexible and efficient resource management layer, much like hypervisors that virtualize physical servers in cloud infrastructure.
Analysed through the CAP theorem, the focus here is largely on Availability. By eliminating memory fragmentation, Paged Attention enables the system to sustain significantly higher loads without exhausting memory resources, thereby keeping the service accessible to a larger user base. At the same time, it preserves strict Consistency of the key-value (KV) cache data.

\begin{figure}[h]
    \centering
    \includegraphics[width=1.0\linewidth]{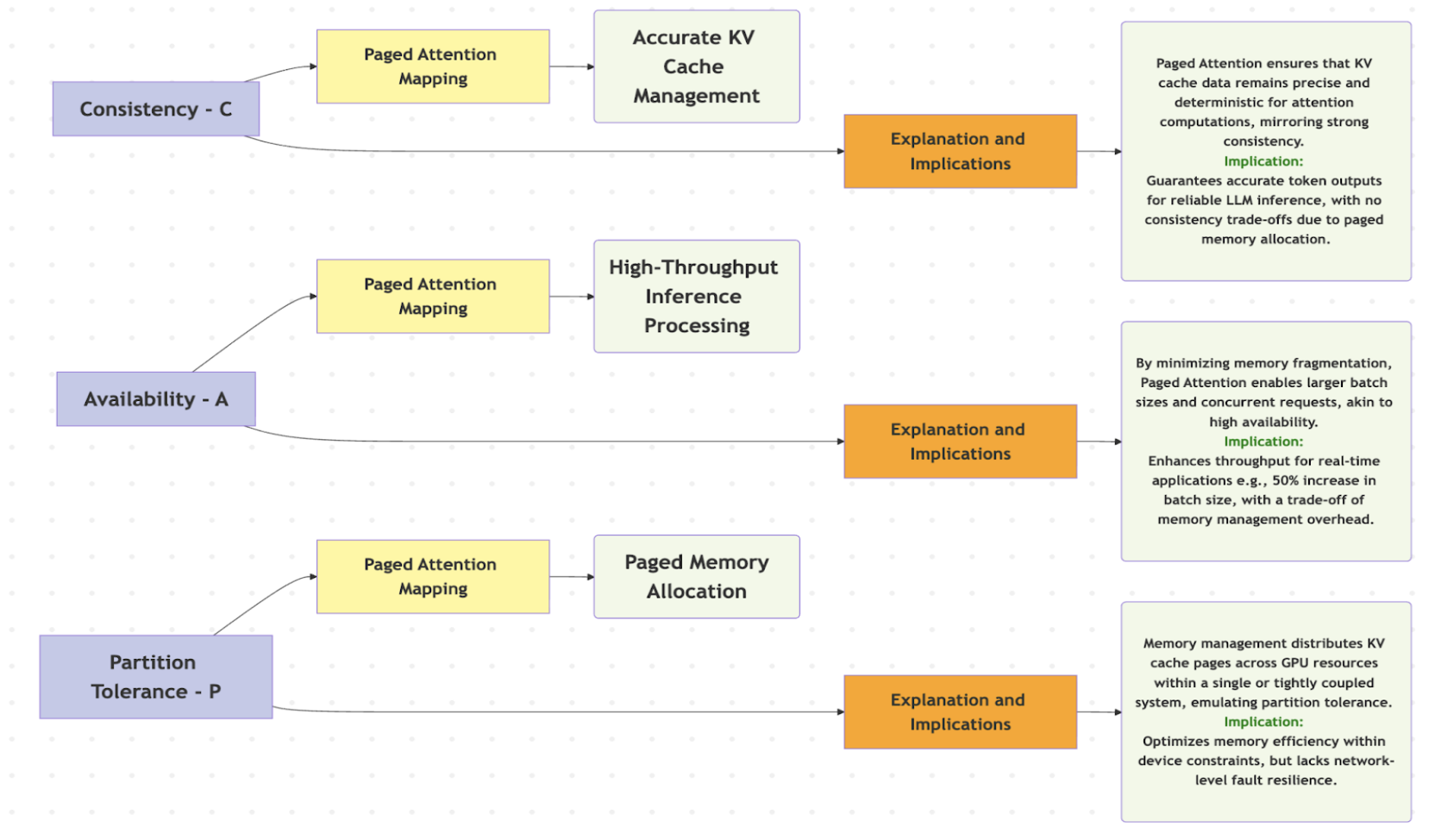}
 \caption{{\bfseries Note}: The CAP theorem is applied metaphorically, treating memory management as logical resource allocation within a single or tightly coupled system, not networked nodes.
}
    \label{fig:placeholder}
\end{figure}

\subsection{Radix Attention : \emph {\color{blue} Efficiency Through Sharing}}

Building on Paged Attention, Radix Attention introduces a mechanism for sharing KV cache memory between different requests. It organizes the KV cache pages into a radix tree (trie), a data structure that excels at prefix matching. If multiple users' prompts share a common prefix (e.g., the same system prompt or introductory question), Radix Attention stores the KV cache for that prefix only once and shares it across all relevant requests.

\begin{figure}[h]
    \centering
    \includegraphics[width=1.0\linewidth]{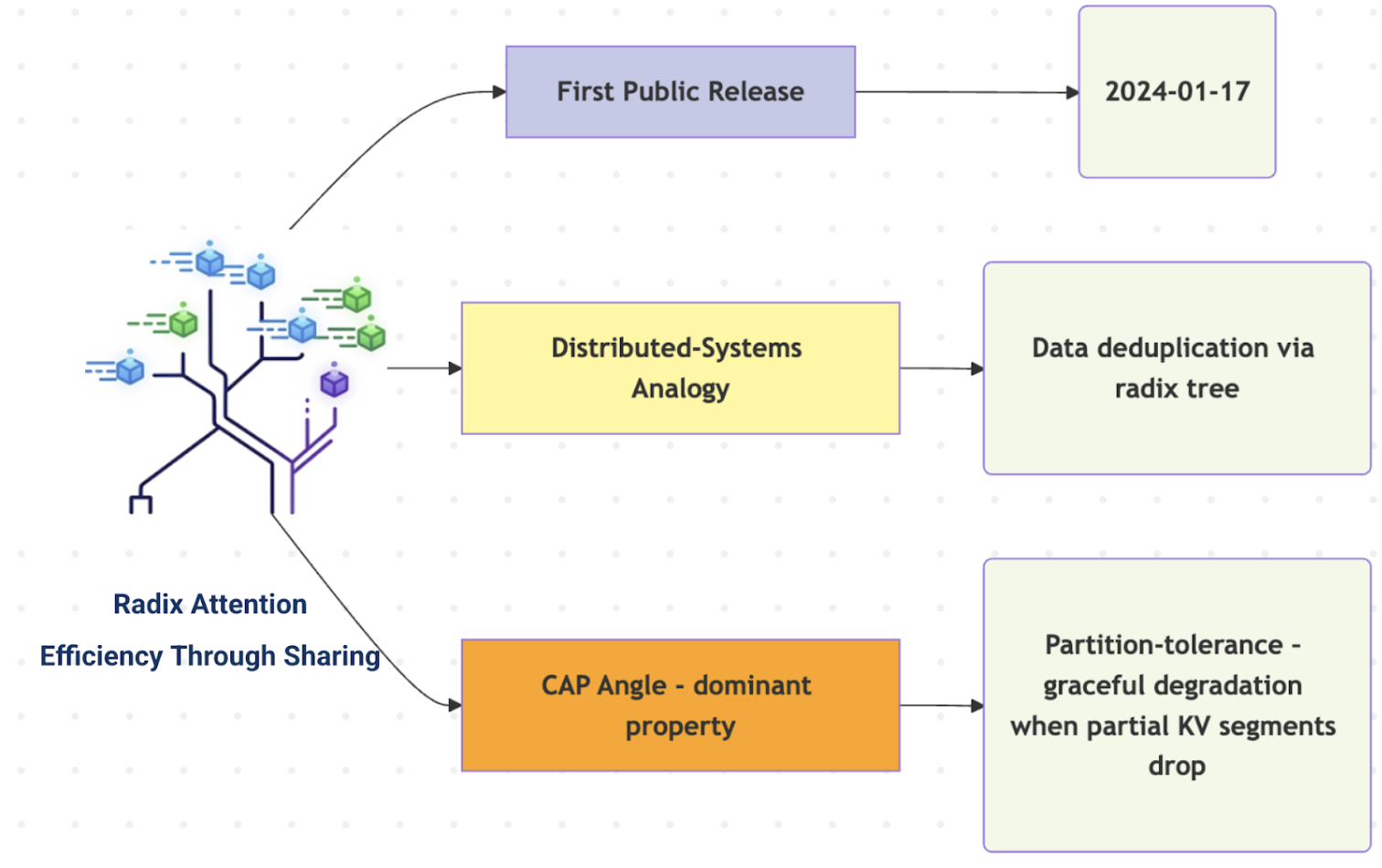}
    \caption{Efficiency Through Sharing}
    \label{fig:placeholder}
\end{figure}

From the perspective of distributed systems, this mechanism represents a form of data deduplication, a technique commonly employed in distributed storage systems to conserve space by storing identical data blocks only once. Interpreted through the CAP theorem, Radix Attention improves Availability by optimizing memory utilization, thereby enabling the system to serve a greater number of concurrent requests that share overlapping context. Crucially, this optimization does not compromise Consistency, since the shared cache produces results identical to those of a fully separate, non-shared cache.

\begin{figure}[h]
    \centering
    \includegraphics[width=1.0\linewidth]{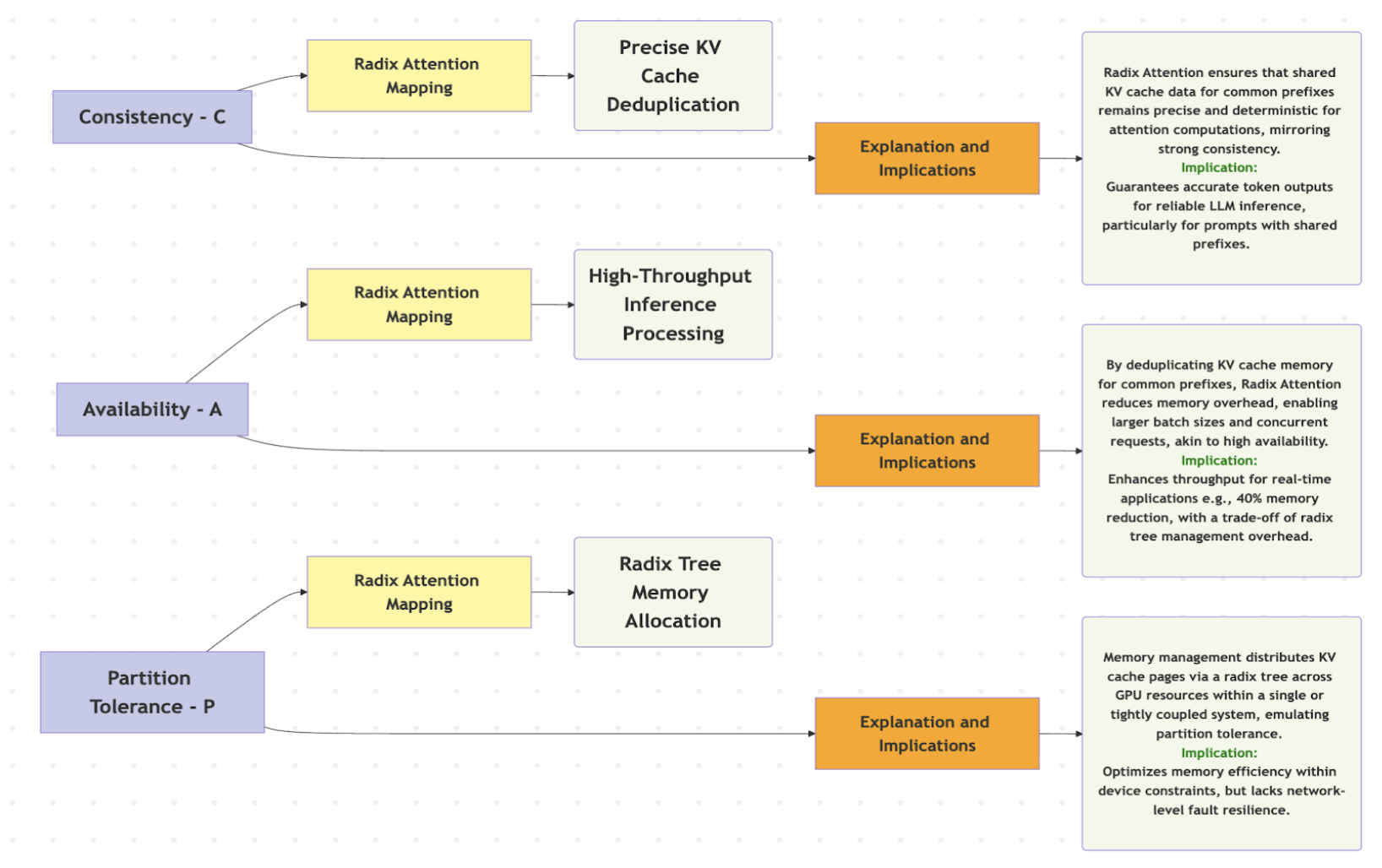}
 \caption{{\bfseries Note}: The CAP theorem is applied metaphorically, treating memory management as logical resource allocation within a single or tightly coupled system, not networked nodes.
 }
    \label{fig:placeholder}
\end{figure}

\section{The Fundamental Challenges of LLM Inference Serving}

The immense scale and computational demands of Large Language Models (LLMs) have introduced significant challenges for traditional serving architectures. A primary bottleneck arises from the bimodal nature of LLM inference, a two-stage process with fundamentally different resource requirements. Understanding this duality is paramount to appreciating why disaggregation has emerged as a crucial architectural pattern for achieving high performance at scale.

\subsection{Anatomy of LLM Inference : \emph {\color{blue} The Prefill and Decode Duality}}

Inference in decoder only large language models (LLMs), such as GPT-style architectures, comprises two distinct phases prefill and decode each with unique computational demands and resource profiles.

{\bfseries Prefill Phase}

The prefill phase processes the entire tokenized input prompt in parallel to generate the Transformer's intermediate representations and construct the key/value (KV) cache. The steps include:

\begin{itemize}
\item {\texttt{\bfseries Embedding Lookup}:Mapping input tokens to dense vector embeddings.}
\item {\texttt{\bfseries Self-Attention Computation}:Generating queries (Q), keys (K), and values (V) for each token through dense matrix multiplications.}
\item {\texttt{\bfseries Transformer Layers}:Propagating Q/K/V tensors via multi-headed attention and feed-forward sublayers to yield updated representations.}
\item {\texttt{\bfseries KV Cache Construction}:Storing K and V tensors for each layer and token in GPU memory for subsequent decoding.}
\end{itemize}

Given the parallel processing of all tokens, the prefill phase exploits high parallelism and maximizes GPU compute and tensor core utilization. It is compute-bound, with performance governed by floating-point operations per second (FLOPS) and tensor-core throughput rather than memory bandwidth. The primary metric is time-to-first-token (TTFT), the delay from request receipt to the first output token. Prefill performs well for fixed-length inputs but may bottleneck under high concurrency, where multiple prompts compete for compute resources.

{\bfseries  Decode Phase }

The decode phase autoregressively generates output tokens one at a time, leveraging the precomputed KV cache and prior tokens:

\begin{itemize}
\item {\texttt{\bfseries Cache Access}:Retrieving all KV vectors for the current position from the cache.}
\item {\texttt{\bfseries Query Computation}:Calculating the query vector (Q) for the latest token.}
\item {\texttt{\bfseries Attention Score Computation}:Performing $$QK^T$$ through matrix vector multiplication.}
\item {\texttt{\bfseries Weighted Sum}:Applying attention weights to the V matrix for the output representation.}
\item {\texttt{\bfseries Token Generation}:Executing feed-forward layers and SoftMax to select the next token.}
\end{itemize}

Due to the sequential dependency on prior tokens' cached states, decoding exhibits low parallelism. It is memory-bound, with performance constrained by frequent accesses to large KV matrices in HBM, leaving GPU compute resources underutilized. The key metric is inter-token latency (ITL), the interval between successive tokens. For extended outputs or high-throughput scenarios, decode frequently dominates overall latency.

\subsection{Key Performance Metrics for LLM Serving}

Evaluating the performance of LLM serving systems requires a nuanced understanding of key metrics that capture both raw efficiency and quality of service.

\begin{figure}
    \centering
    \includegraphics[width=1.0\linewidth]{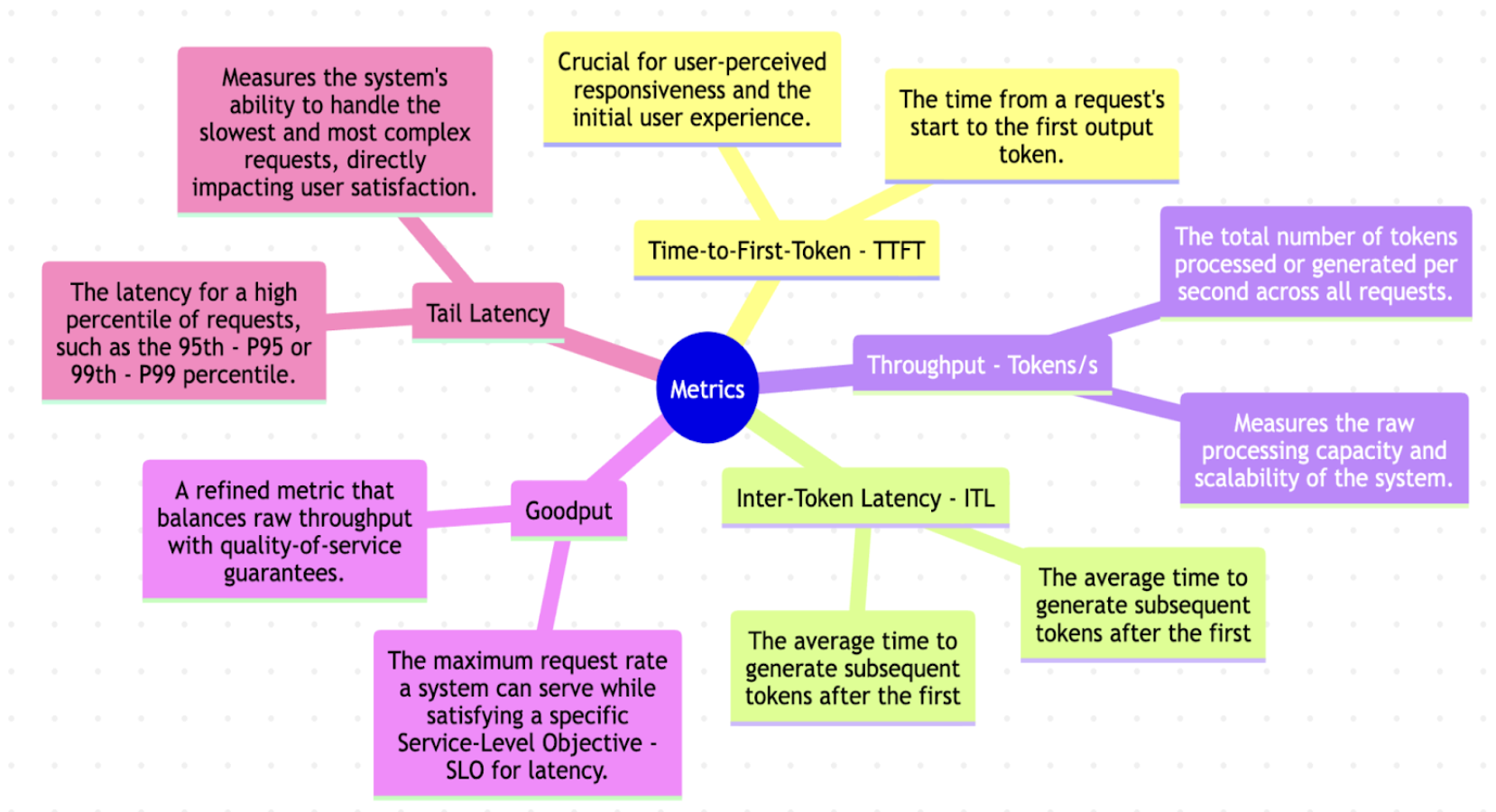}
    \caption{Key Performance Metrics for LLM Serving}
    \label{fig:placeholder}
\end{figure}

\subsection{Congestion under Asynchronous Requests}

The prefill and decode phases function optimally for sequential requests. In asynchronous, high-volume settings, typical of distributed, multi-tenant LLM services—congestion arises:

\begin{itemize}
\item {\texttt{\bfseries Prefill Overload}:Concurrent prompts initiate multiple parallel prefills, overwhelming GPU compute clusters and inflating TTFT.}
\item {\texttt{\bfseries Decode Queuing}:Sequential decodes queue for memory bandwidth, elevating ITL and causing variable tail latencies.}
\item {\texttt{\bfseries Resource Contention}:Shared HBM and PCIe access creates hotspots, diminishing throughput.}
\end{itemize}

\begin{figure*}[h]
    \centering
    \includegraphics[width=1.0\linewidth]{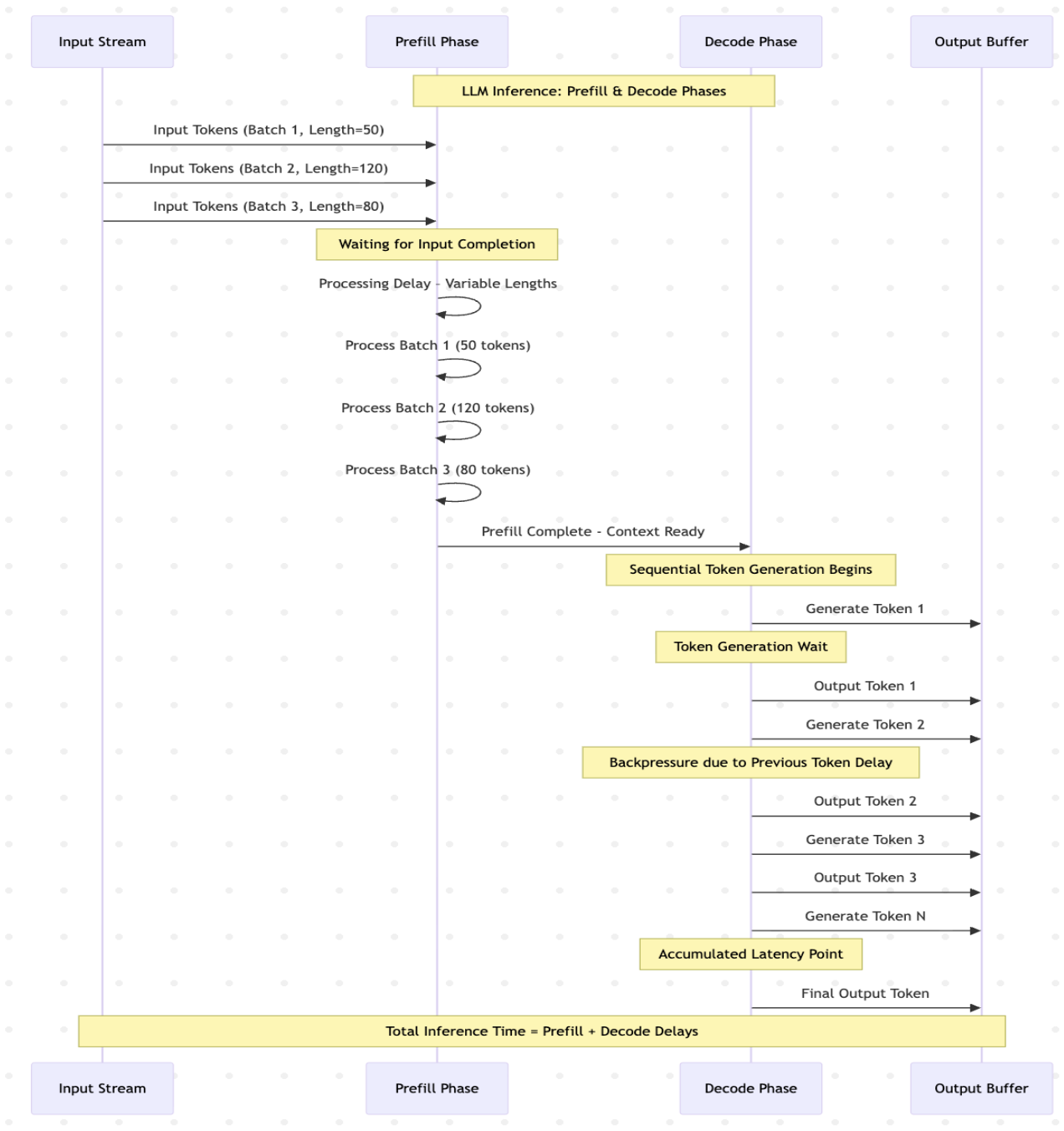}
    \caption{Understanding LLM Inference Latency: Prefill, Decode, and Delays in Token Generation}
    \label{fig:placeholder}
\end{figure*}

In the prefill phase, as shown in the above diagram, input tokens arrive asynchronously with variable lengths, creating irregular processing patterns.When token batches of different sizes (e.g., 50, 120, and 80 tokens) are received sequentially, the system must wait for input completion before proceeding with processing. This waiting period introduces substantial processing delays, particularly when handling variable-length inputs that cannot be efficiently batched together.

This sequential dependency creates a point of accumulated latency that significantly impacts overall inference time.Each batch must be processed individually—for example, Batch 1 (50 tokens), Batch 2 (120 tokens), and Batch 3 (80 tokens) are handled in sequence rather than concurrently.This approach results in underutilized computational resources and extended prefill completion times.

The decode phase further exacerbates latency issues through its inherently sequential token generation process. Each token must be generated individually, creating a cascade of delays. Token generation wait times accumulate as the system produces tokens sequentially, with each subsequent token experiencing backpressure from previous token delays. This sequential dependency creates an accumulated latency point that significantly impacts overall inference time.

The total inference time in traditional architectures represents the sum of prefill delays and decode delays, resulting in suboptimal performance for applications requiring low-latency responses.

\subsection{Why Disaggregation Matters : \emph {\color{blue} Addressing Resource Conflicts}}

In traditional, co-located serving architectures, the prefill and decode phases run on the same GPU. This creates a fundamental and often overlooked problem of resource contention: the compute-bound prefill phase and the memory-bound decode phase compete for the same resources, leading to inefficiencies and making it exceedingly difficult to optimize both TTFT and ITL simultaneously. A single GPU configuration, for instance, cannot be optimally tuned for both parallel matrix operations and high-speed memory access, resulting in resource underutilization during one phase or the other.   

Disaggregation resolves this by physically separating the two phases onto different GPUs or nodes. This architectural shift enables each phase to be independently optimized with dedicated hardware and parallelism strategies tailored to its unique characteristics. For example, prefill can be routed to GPUs with high compute throughput, while decode can be assigned to resources optimized for memory bandwidth. This separation mitigates interference, improves resource utilization, and enables more predictable performance, especially with regards to improving tail latency for the slowest requests. 

\begin{figure*}[h]
    \centering
    \includegraphics[width=1.0\linewidth]{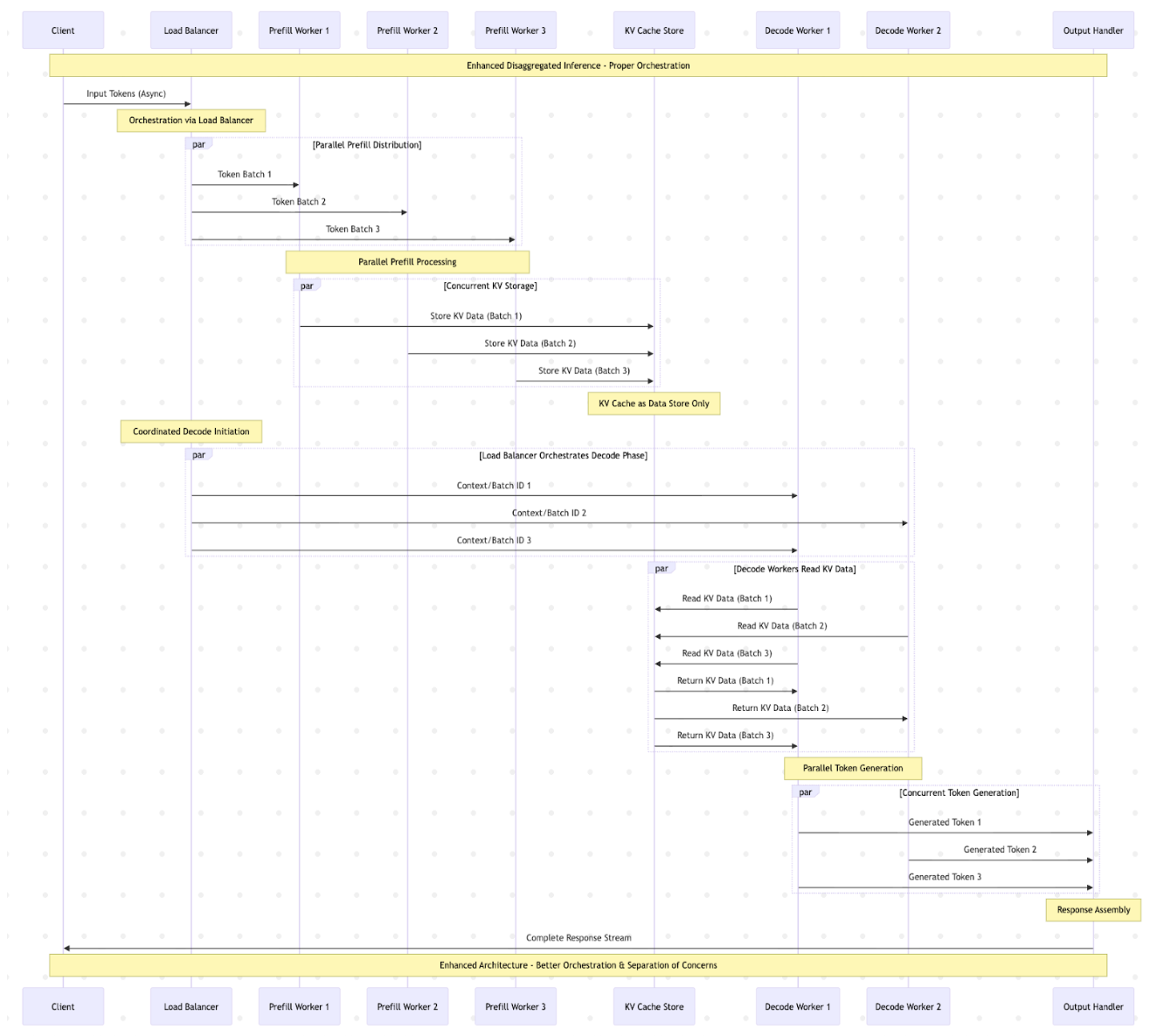}
    \caption{Disaggregated LLM Inference Workflow: Prefill, KV Storage, and Decode Stages}
    \label{fig:placeholder}
\end{figure*}

A disaggregated inference architecture fundamentally transforms LLM processing by distributing computational workloads across specialized worker nodes. This approach enables the parallel execution of both prefill and decode operations. As illustrated in the diagram, the architecture utilizes multiple prefill workers that can process distinct token batches simultaneously. When asynchronous input tokens arrive, a load balancer distributes them in parallel, rather than sequentially, across Prefill Worker 1, Prefill Worker 2, and Prefill Worker 3. This parallel prefill processing dramatically reduces input waiting times and minimizes the processing delays found in traditional architectures. 

A centralized KV Cache Store is a key component of this disaggregated design, acting as an intermediary between prefill and decode operations. Each prefill worker concurrently stores its processed key-value (KV) cache states, which are then immediately available for decode operations without requiring full prefill completion.

The load balancer then coordinates decode initiation by sending context or batch identifiers to Decode Worker 1 and Decode Worker 2. These workers retrieve the corresponding KV data from the KV cache store and generate tokens concurrently. Decoder one worker could generate tokens 1 and 3, while a second decoder worker generates token 3. This parallel process significantly reduces token generation wait times and backpressure effects.

By transforming sequential dependencies into orchestrated parallel operations with a clear separation of concerns, this architecture achieves substantial latency reduction, improved resource utilization, and higher throughput, making it ideal for real-time LLM applications.

However, this approach is not without its trade-offs. The separation of prefill and decode introduces a new challenge: the need to rapidly and reliably transfer the KV cache from the prefill worker to the decode worker. This data transfer introduces a communication overhead that can, in certain scenarios, negate the performance gains of disaggregation. A system's success therefore hinges on whether the performance improvements from parallelization outweigh this communication latency

\section{Architectural Archetypes of Disaggregated Frameworks}

The landscape of disaggregated inference is populated by distinct architectural archetypes, each with a unique philosophy and purpose. This survey paper provides a comparative analysis of three prominent frameworks: DistServe, AIBrix, and NVIDIA Dynamo. These frameworks have been selected to represent different strategic approaches to serving large language models (LLMs). Specifically, DistServe represents a research-oriented framework that explores novel serving strategies like the disaggregation of prefill and decoding to improve goodput, AIBrix embodies a cloud-native mindset emphasizing scalability and operational efficiency, and NVIDIA Dynamo exemplifies an enterprise-scale solution optimized for high performance and deployment in large GPU fleets.

\subsection{Architectural Paradigms and Design Principles}

The "Architectural Paradigms and Design Principles" section outlines the fundamental frameworks and innovative strategies employed in developing disaggregated systems. This includes optimizing Resource Allocation and Scheduling through techniques like dynamic resource provisioning for compute and memory, aiming to maximize hardware utilization and adapt to variable workload demands in cloud-native or HPC environments. Furthermore, Key Innovations encompass sophisticated Communication Strategy and Data Transfer Mechanisms, such as cache-aware data placement and high-bandwidth interconnects (e.g., NVIDIA NVLink or RDMA), crucial for minimizing inter-node latency and enabling efficient data exchange within the disaggregated architecture

\begin{figure*}[h]
    \centering
    \includegraphics[width=1.0\linewidth]{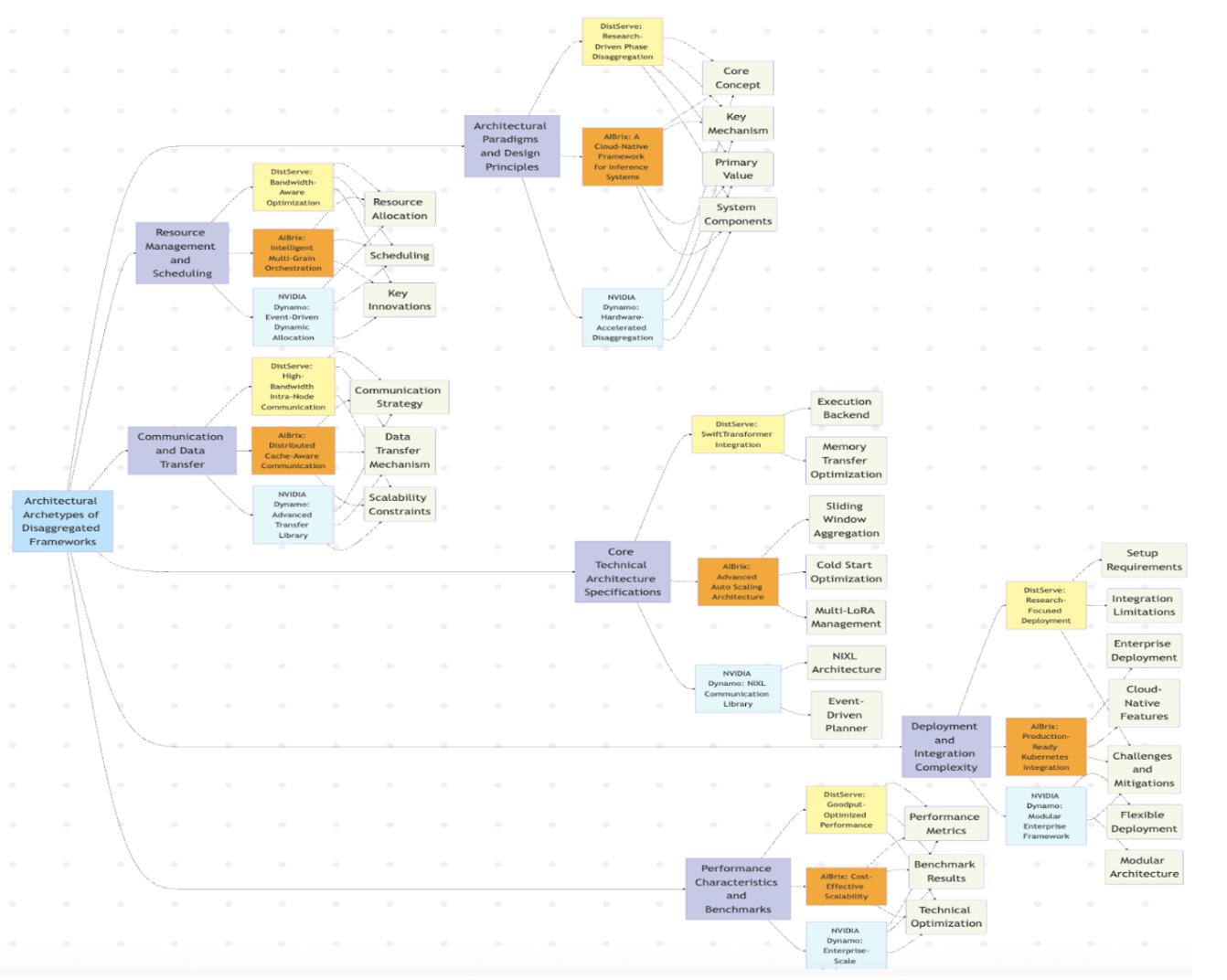}
    \caption{Architectural Paradigms and Design Principles}
    \label{fig:placeholder}
\end{figure*}

\subsubsection{DistServe : \emph {\color{blue} Research-Driven Phase Disaggregation}}

{\bfseries  Core Concept }

DistServe is an academic research project that pioneers fine-grained phase-level disaggregation for LLM inference. It fundamentally decouples the compute-bound prefill phase from the memory-bound decode phase into separate, independently scalable services.

{\bfseries  Key Mechanism}

It employs a simulation-driven scheduler to intelligently route requests between dedicated prefill and decode worker pools, optimizing for "goodput" (the number of requests meeting strict latency SLOs).

{\bfseries  Primary Value}
 
This approach eliminates interference between phases, enabling superior performance predictability and higher throughput while guaranteeing tight Time-to-First-Token (TTFT) and Time-Per-Output-Token (TPOT) deadlines.

{\bfseries  Archetype}

It represents the research-first archetype, prioritizing algorithmic innovation and serving as a blueprint for future production systems.

{\bfseries System Components}
\begin{enumerate}
    \item \textbf{LLMEngine: \emph {\color{blue} Central Inference Orchestrator}}

    The LLMEngine acts as the central coordinator, managing the entire inference workflow and ensuring seamless operation between the disaggregated prefill and decoding stages. It orchestrates task execution, manages the flow of data (including the transfer of key-value cache via a bridge queue), and supports both online and offline inference through its high-level interfaces. As the backbone of the system, it guarantees the architecture operates as a cohesive unit by automatically handling memory management and inter-stage communication.
    
    \item \textbf{Context Stage Engine: \emph {\color{blue} High-Throughput Prefill Processor}}
    
    This component is dedicated to the compute-intensive prefill phase, processing the input prompt to generate the initial key-value (KV) cache. It utilizes independent parallelism strategies tailored to its demands and operates on dedicated GPU resources to prevent interference with decoding. This separation is crucial for optimizing the time to first token (TTFT) by allowing the phase to be scaled and optimized independently for high computational throughput.

    \item \textbf{Decoding Stage Engine: \emph {\color{blue} Latency-Optimized Token Generator}}
    
    Managing the autoregressive decoding phase, this engine incrementally generates output tokens using the KV cache supplied by the prefill stage. It employs distinct parallelism strategies optimized for its iterative nature and runs on isolated GPU resources. This isolation is fundamental to optimizing time per output token (TPOT), enabling efficient token generation without contention from prefill workloads.

    \item \textbf{Bridge Queue: \emph {\color{blue} Bandwidth-Optimized KV Transfer}}
    
    The bridge queue is a critical data structure that facilitates the efficient transfer of KV cache data between the Context and Decoding Stage Engines. It acts as an intermediary, ensuring minimal communication overhead by leveraging a bandwidth-aware placement algorithm to optimize data transfer across the cluster. This component is essential for maintaining the performance of the disaggregated architecture.

    \item \textbf{ParaWorker: \emph {\color{blue} Parallelism-Driven GPU Executor}}
    
    The bridge queue is a critical data structure that facilitates the efficient transfer of KV cache data between the Context and Decoding Stage Engines. It acts as an intermediary, ensuring minimal communication overhead by leveraging a bandwidth-aware placement algorithm to optimize data transfer across the cluster. This component is essential for maintaining the performance of the disaggregated architecture.

    \item \textbf{AsyncLLM Interface: \emph {\color{blue}  Asynchronous Inference Gateway}}
    
    The AsyncLLM Interface is a high-level API designed for real-time inference, serving applications with strict latency requirements such as chatbots and interactive systems. It provides a streamlined interface for submitting queries and receiving low-latency responses by integrating directly with the LLMEngine. This ensures efficient management of the disaggregated prefill and decode processes, while features like continuous batching allow it to dynamically handle incoming request streams. Its design is critical for enabling DistServe to deliver optimal Time to First Token (TTFT) and Time Per Output Token (TPOT), making it ideal for real-time user interactions.

    \item \textbf{OfflineLLM Interface: \emph {\color{blue}  Batch Throughput Maximizer}}
    
    The OfflineLLM Interface supports batch processing workloads where high throughput is prioritized over low latency. It is tailored for large-scale inference tasks such as summarization and code completion, managing batch jobs by processing multiple inputs concurrently. By leveraging DistServe’s disaggregated architecture, it optimizes resource allocation specifically for offline workloads and coordinates with the LLMEngine to efficiently execute prefill and decoding stages. This interface enhances DistServe’s versatility, enabling cost-effective, high-throughput inference for non-interactive use cases.
    
    \begin{figure}[h]
        \centering
        \includegraphics[width=1.0\linewidth]{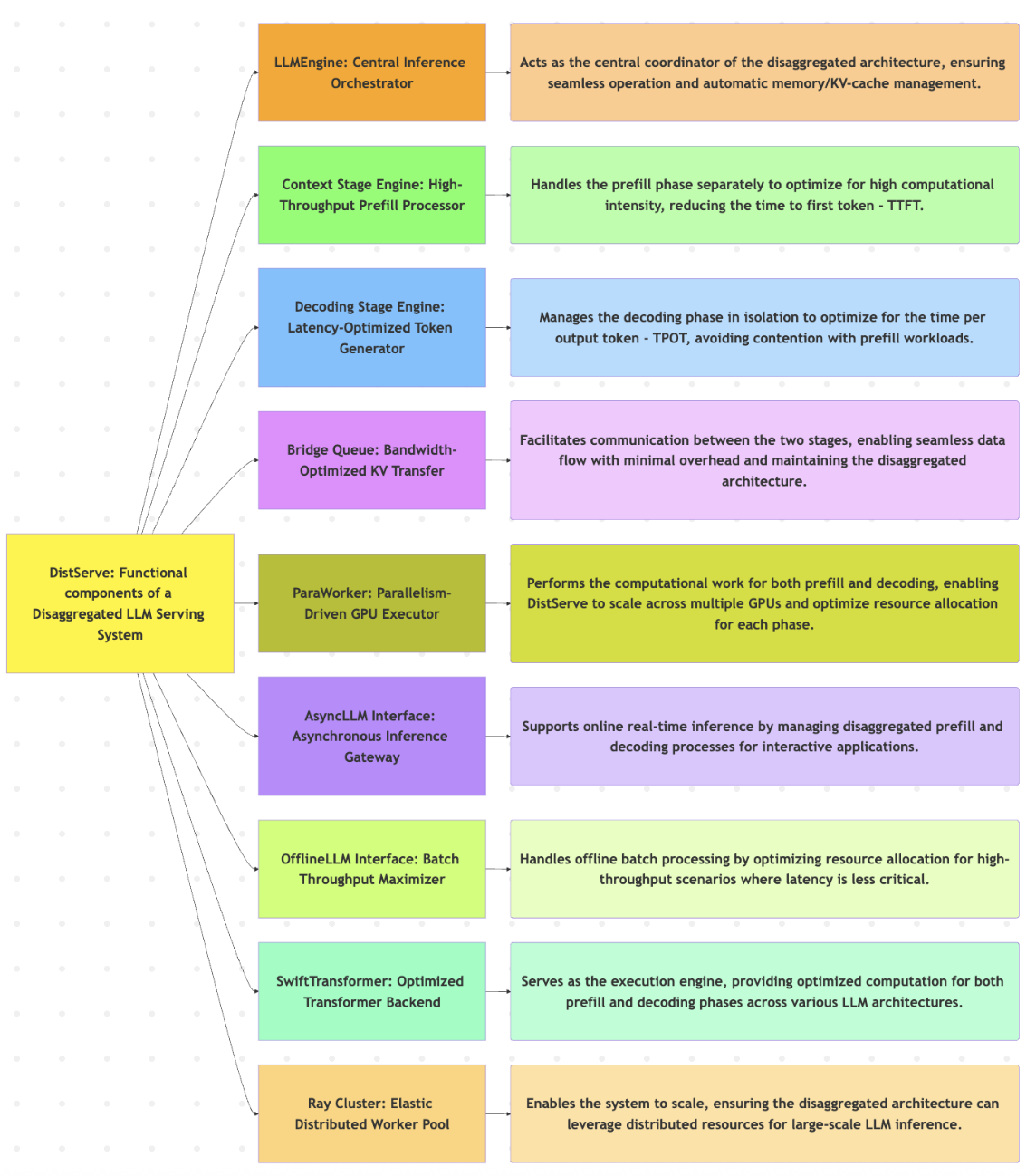}
        \caption{Role in Disaggregated Serving}
        \label{fig:placeholder}
    \end{figure}

    \item \textbf{SwiftTransformer: \emph {\color{blue} Optimized Transformer Backend}}
    
    Serving as the high-performance execution backend, SwiftTransformer is a C++ library that implements core inference operations. It provides optimized computation for supported architectures (e.g., GPT-2, LLaMA2) through features like model parallelism, FlashAttention, and PagedAttention. By integrating with the ParaWorker, it enhances DistServe’s performance by providing a highly efficient backend for executing all computational tasks.

    \item \textbf{Ray Cluster: \emph {\color{blue} Elastic Distributed Worker Pool}}
    
    DistServe leverages a Ray cluster to manage its distributed workers across multiple GPUs or nodes. This framework facilitates the distribution of tasks, automatically initializing a cluster on a single node or allowing manual configuration for multi-node scaling. The Ray cluster provides the foundational elastic infrastructure that enables the disaggregated architecture to leverage distributed resources for large-scale inference.
    
\end{enumerate}

\subsubsection{AIBrix : \emph {\color{blue} A Cloud-Native Framework for Inference Systems}}

{\bfseries  Core Concept }

AIBrix is a comprehensive control plane and orchestration layer designed to manage large-scale LLM inference infrastructure on Kubernetes. It treats the entire GPU cluster as a disaggregated pool of resources for dynamic allocation.

{\bfseries  Key Mechanism }

It employs a hybrid Kubernetes-Ray architecture for granular management, combining Kubernetes' robustness for coarse-grained orchestration with Ray's flexibility for fine-grained, high-performance computing tasks.

{\bfseries  Primary Value }

AIBrix provides end-to-end automation for production LLM serving, featuring LLM-specific autoscaling, a distributed Key-Value (KV) cache, dynamic LoRA adapter management, and proactive health monitoring. This maximizes throughput and cost-efficiency while ensuring reliability.

{\bfseries  Archetype }

It represents the cloud-native, production-ready archetype, prioritizing enterprise-grade operability, multi-tenancy, and integration with existing cloud ecosystems over purely theoretical optimizations.

{\bfseries System Components}

{\color{blue} \bfseries Control Plane Components}
\begin{enumerate}
    \item \textbf{Model Adapter (LoRA) Controller: \emph {\color{blue} Dynamic Adapter Orchestration}}

    This component manages the dynamic loading and unloading of LoRA (Low-Rank Adaptation) adapters, enabling efficient multi-LoRA deployments on a single pod. It handles adapter scheduling and leverages Kubernetes' Service and EndpointSlice mechanisms for optimized service discovery and placement. By significantly reducing the memory footprint for serving multiple fine-tuned versions of a base model, it drastically improves resource utilization and reduces inference costs—particularly beneficial for "long-tail" workloads requiring numerous specialized models.

    \item \textbf{LLM-Specific Autoscaler: \emph {\color{blue} Tailored Dynamic Autoscaling}}

    Designed for the unique resource patterns of LLM inference, this autoscaler dynamically adjusts GPU allocation and inference pod counts based on real-time demand. It uses specialized metrics like KV cache utilization and inference-aware indicators to enable second-level scaling and supports heterogeneous GPU serving to balance cost-efficiency and performance. This ensures efficient handling of fluctuating workloads, preventing both over-provisioning (reducing costs) and under-provisioning (maintaining low latency).

    \item \textbf{GPU Optimizer: \emph {\color{blue} Heterogeneous Resource Allocator}}

    Working alongside the autoscaler, this profiler-based optimizer maximizes cost-efficiency in heterogeneous GPU environments. It analyzes workload patterns and GPU profiles to determine the most cost-effective GPU mix—such as leveraging older or less expensive GPUs for lighter workloads—while still meeting Service Level Objectives (SLOs). This significantly reduces operational costs without compromising performance.

    \item \textbf{Unified AI Runtime: \emph {\color{blue} Vendor-Neutral Inference Sidecar}}
    
    Acting as a lightweight sidecar container, this component provides a management layer between the AIBrix Control Plane and inference engine pods. It abstracts vendor-specific inference engines, offering a unified interface for model downloading, configuration, and metric standardization. By enabling dynamic, cloud-native resource management, it ensures a pluggable architecture that supports multiple inference engines (e.g., vLLM, SGLang, Dynamo) and avoids technology lock-in.

    \item \textbf{Cold Start Manager: \emph {\color{blue} Latency-Optimized Preloader}}

    This component mitigates the "cold start" problem in serverless or auto-scaled environments by proactively managing inference pod lifecycles. It pre-warms or pre-loads models onto idle workers and predicts traffic patterns to ensure ready instances are available, minimizing latency for initial requests. This is critical for maintaining low latency in interactive applications, especially during traffic spikes after low-usage periods

    \item \textbf{Accelerator Diagnose Tools: \emph {\color{blue} Proactive GPU Health Monitoring}}

    A suite of utilities for monitoring and troubleshooting hardware accelerators (GPUs), these tools collect telemetry on utilization, temperature, memory usage, and errors. They run diagnostic tests to identify failures, bottlenecks, or configuration issues, providing essential data to the autoscaler and GPU Optimizer. This proactive health monitoring ensures system reliability, prevents hardware-related outages, and maintains efficient operation at scale.

\begin{figure}[h]
    \centering
    \includegraphics[width=1.0\linewidth]{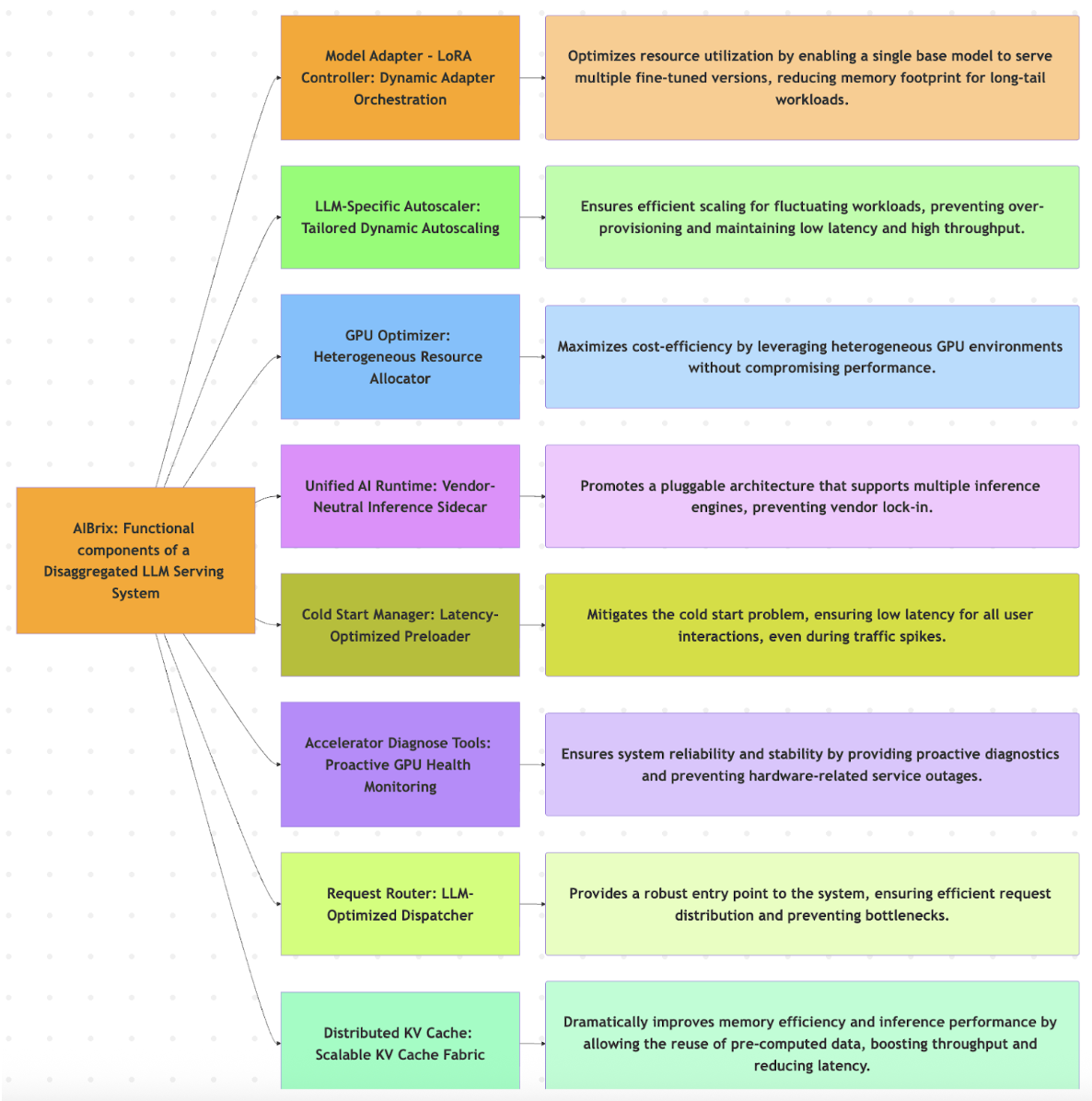}
    \caption{Role in Disaggregated Serving}
    \label{fig:placeholder}
\end{figure}

{\color{blue} \bfseries Data Plane Components}

\item \textbf{Request Router: \emph {\color{blue} LLM-Optimized Dispatcher}}

As the central dispatcher for inference requests, this LLM-aware API gateway routes traffic to appropriate workers across models and replicas. It enforces fairness policies, rate limiting, and workload isolation while using prefix-cache and load-aware strategies to optimize latency and throughput. By efficiently distributing requests, it prevents bottlenecks and provides a robust, reliable entry point for LLM inference.

\item \textbf{Distributed KV Cache: \emph {\color{blue} Scalable KV Cache Fabric}}

This framework enables Key-Value (KV) cache distribution and sharing across nodes and inference engines, providing high-capacity, low-latency network access. It allows reuse of pre-computed KV cache data for long-context or repeated prompts, reducing redundant computation through a scan-resistant eviction policy. By dramatically improving memory efficiency and inference performance, it boosts throughput and reduces latency—particularly for long-context and prefill-heavy workloads.
    
\end{enumerate}

\subsubsection{NVIDIA Dynamo : \emph {\color{blue} Hardware-Accelerated Disaggregation}}

{\bfseries  Core Concept }

NVIDIA Dynamo is a modular, enterprise-grade platform that leverages tight hardware-software co-design to achieve high-performance disaggregated inference. It is optimized end-to-end for NVIDIA's full technology stack (GPUs, NVLink, NVSwitch).

{\bfseries  Key Mechanism }

Its architecture is built around specialized modules—a Planner, Smart Router, and multi-tiered KV Cache Manager—orchestrated by the NVIDIA Inference Transfer Library (NIXL). This library provides ultra-fast, hardware-accelerated data transfer between disaggregated components, minimizing communication overhead.

{\bfseries  Primary Value }

Dynamo delivers unmatched performance and scalability for massive models (e.g., 671B parameters) by turning hardware integration into its greatest advantage. It provides a full-stack solution that ensures enterprise-grade reliability, advanced monitoring, and seamless integration with popular inference engines like TensorRT-LLM and vLLM.

{\bfseries  Archetype }

It represents the full-stack, internally optimized archetype, where the disaggregation logic is fundamentally designed to maximize the potential of the hardware’s proprietary interconnects and memory hierarchies.

{\bfseries System Components}

\begin{enumerate}
    \item \textbf{Dynamo Planner:: \emph {\color{blue} Dynamic Resource Optimizer}}

    An event-driven resource allocator that continuously monitors system performance to dynamically adjust GPU allocation between prefill and decode phases. It ingests real-time metrics—including request rates, sequence lengths, and GPU utilization—and applies both load-based and SLO-aware strategies to optimize throughput while meeting performance guarantees. The Planner automatically scales prefill workers during surges in long input sequences and supports zero-downtime scaling via Kubernetes or Circus process management. By adapting to fluctuating demand and minimizing over-provisioning, it ensures high GPU utilization across large-scale deployments.

    \item \textbf{Smart Router: \emph {\color{blue} KV Cache-Optimized Dispatcher}}

    A KV cache-aware request dispatcher that routes user queries to workers with the highest cache hit rates while maintaining cluster load balance. It uses a global radix tree to track KV block locations and combines KV overlap scoring with utilization metrics in its routing algorithms. Supporting multiple modes—random, round-robin, and KV-aware—the router also enables replica synchronization via NATS for fault tolerance. Its intelligent routing slashes Time to First Token (TTFT) by reducing recomputation, achieving up to 3× improvements in latency.

    \item \textbf{NIXL: \emph {\color{blue} High-Performance Data Transfer Fabric}}

    Working alongside the autoscaler, this profiler-based optimizer maximizes cost-efficiency in heterogeneous GPU environments. It analyzes workload patterns and GPU profiles to determine the most cost-effective GPU mix—such as leveraging older or less expensive GPUs for lighter workloads—while still meeting Service Level Objectives (SLOs). This significantly reduces operational costs without compromising performance.

    \item \textbf{KV Cache Block Manager: \emph {\color{blue} Hierarchical Cache Manager}}
    
    A hierarchical, multi-tiered cache system that manages KV data across GPU memory, CPU DRAM, SSD, and object storage to optimize memory use and cost. It offers asynchronous GET/PUT APIs that overlap with inference computation and supports pluggable eviction policies like LRU and cost-aware strategies. Integrated with NIXL for efficient data movement, it allows organizations to cost-effectively scale cache capacity to petabytes while maintaining high hit rates, especially beneficial for long-context workloads.

    \begin{figure}[h]
        \centering
        \includegraphics[width=1.0\linewidth]{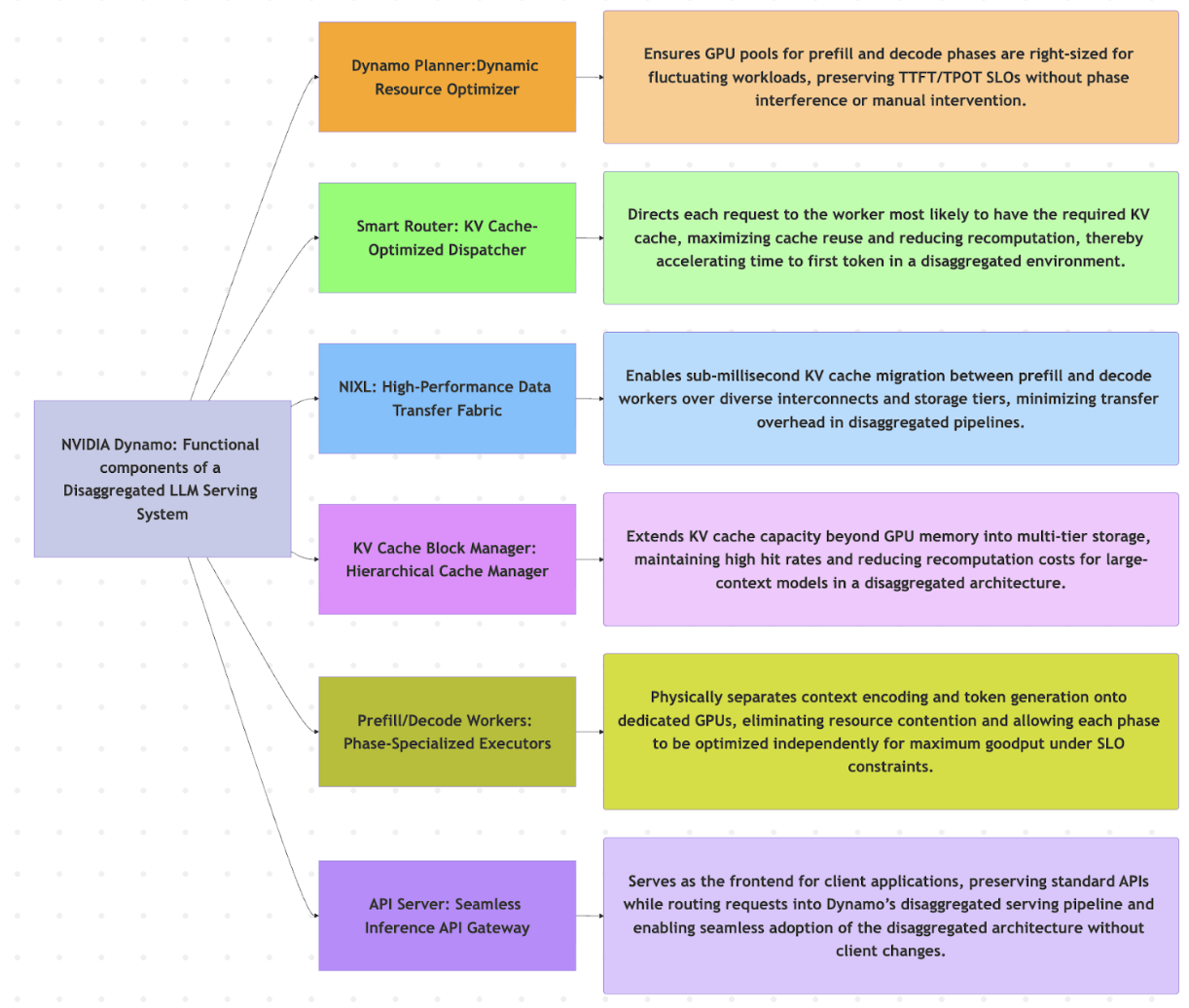}
        \caption{Role in Disaggregated Serving}
        \label{fig:placeholder}
    \end{figure}

    \item \textbf{Prefill/Decode Workers: \emph {\color{blue} Phase-Specialized Executors}}
    
    Specialized GPU units independently optimized for distinct inference phases: prefill workers handle compute-intensive context processing, while decode workers manage memory-bound token generation. Each supports phase-specific parallelism strategies and uses disaggregated routing logic to balance local and remote execution. With nonblocking KV transfers via NIXL, these workers eliminate inter-phase interference, enabling finer resource control, improved scalability, and higher overall performance.

    \item \textbf{API Server: \emph {\color{blue} Seamless Inference API Gateway}}
    
    A user-facing HTTP server providing OpenAI-compatible REST endpoints for seamless integration with existing applications. It translates external requests into internal Dynamo protocols, handles authentication, validation, and response streaming, and integrates with Dynamo’s runtime for service discovery. By maintaining standard API compliance while leveraging distributed inference optimizations, it allows users to benefit from Dynamo’s performance without modifying their client applications.

\end{enumerate}

\subsection{Resource Management and Scheduling}

\subsubsection{DistServe : \emph {\color{blue} Bandwidth-Aware Optimization}}

DistServe optimizes large language model serving by strategically disaggregating the prefill and decoding phases of LLM inference. This system employs a bandwidth-aware algorithm to intelligently allocate resources and place computational tasks, alongside a First-Come-First-Served (FCFS) scheduling policy that includes dynamic replanning for shifting workloads. A key innovation is its focus on maximizing "goodput" requests completed within defined Time-to-First-Token (TTFT) and Total-Processing-Over-Time (TPOT) bounds rather than solely targeting peak throughput, enabling independent parallelism settings for tensor and pipeline splits in prefill versus decode to meet specific latency goals.

\begin{figure}[h]
    \centering
    \includegraphics[width=1.0\linewidth]{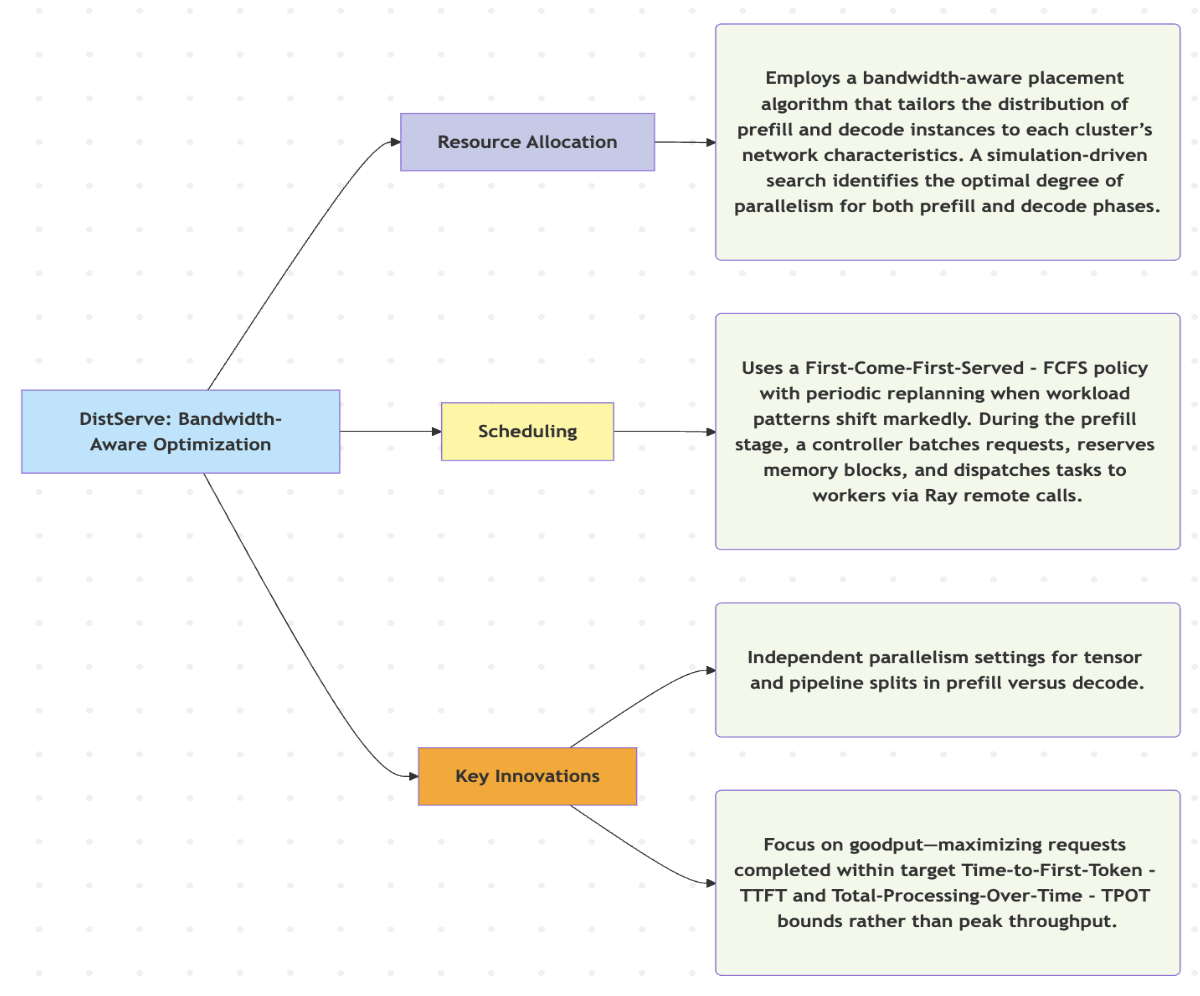}
    \caption{Bandwidth-Aware Optimization}
    \label{fig:placeholder}
\end{figure}

\subsubsection{AIBrix : \emph {\color{blue} Intelligent Multi-Grain Orchestration}}

AlBrix intelligently orchestrates Large Language Model (LLM) inference by seamlessly blending Kubernetes for high-level resource allocation with Ray for precise execution control. This system ensures efficient, low-latency scaling through LLM-specific autoscaling driven by dynamic sliding-window metric aggregation, while its innovative features, including standardized AI Runtime Metrics and SLO-driven optimization, guarantee a balance between cost-efficiency and performance objectives.

\begin{figure}[h]
    \centering
    \includegraphics[width=1.0\linewidth]{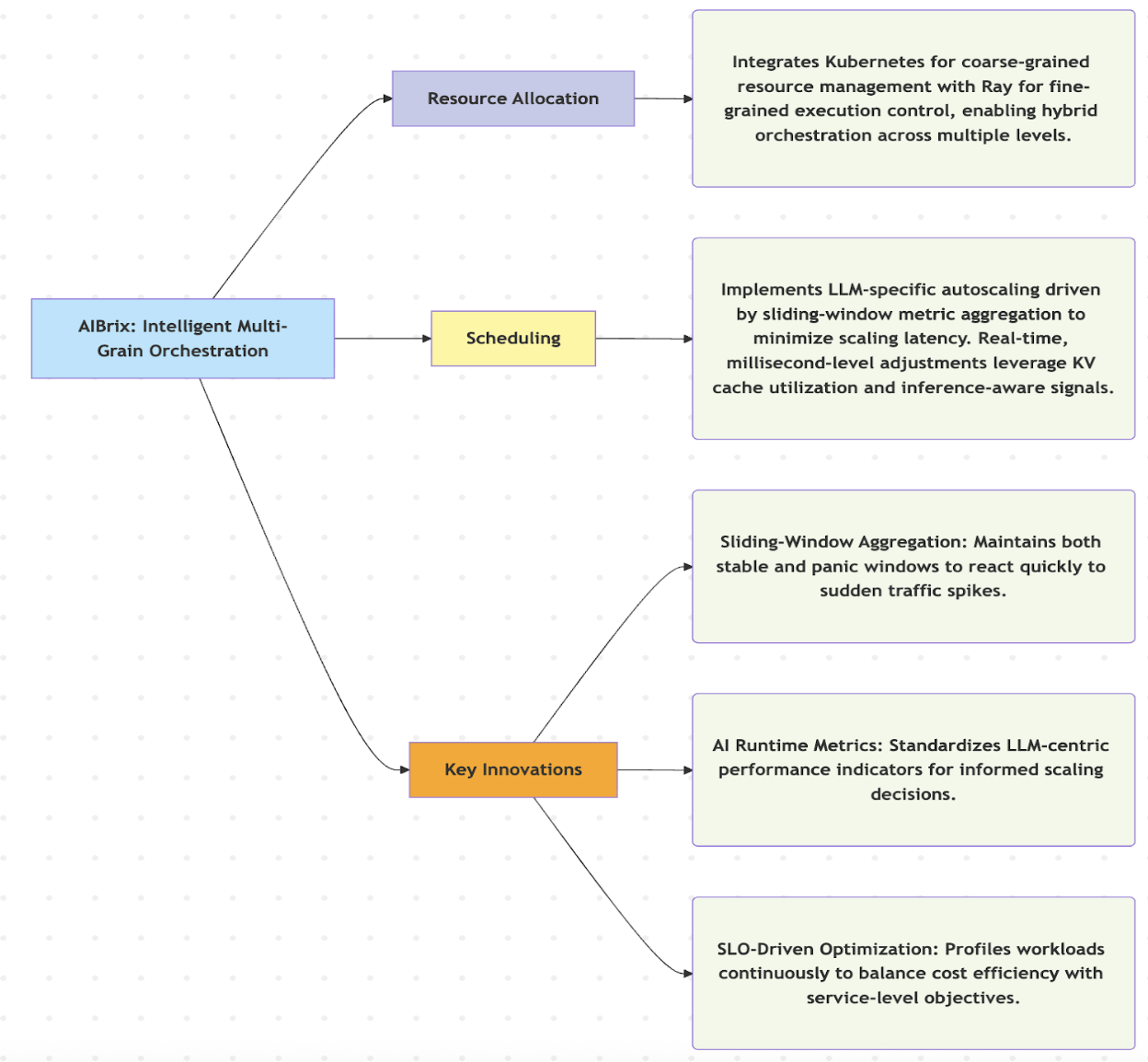}
    \caption{Intelligent Multi-Grain Orchestration}
    \label{fig:placeholder}
\end{figure}

\subsubsection{NVIDIA Dynamo : \emph {\color{blue} Event-Driven Dynamic Allocation}}

NVIDIA Dynamo is an event-driven framework designed for dynamic allocation in distributed environments, focusing on optimizing GPU resource allocation, efficient scheduling, and enabling seamless scaling for AI workloads. Key innovations include dynamic GPU scheduling through an event-driven Planner that adapts resource allocation to real-time demand, zero-downtime scaling of workers for flexible deployment and load management via an event plane, and intelligent scaling to automatically manage prefill workers based on request volumes and input sequence lengths. This modular design facilitates high-throughput, low-latency inference serving of generative AI and reasoning models by ensuring efficient resource utilization and adaptive scaling without manual intervention.

\begin{figure}[h]
    \centering
    \includegraphics[width=1.0\linewidth]{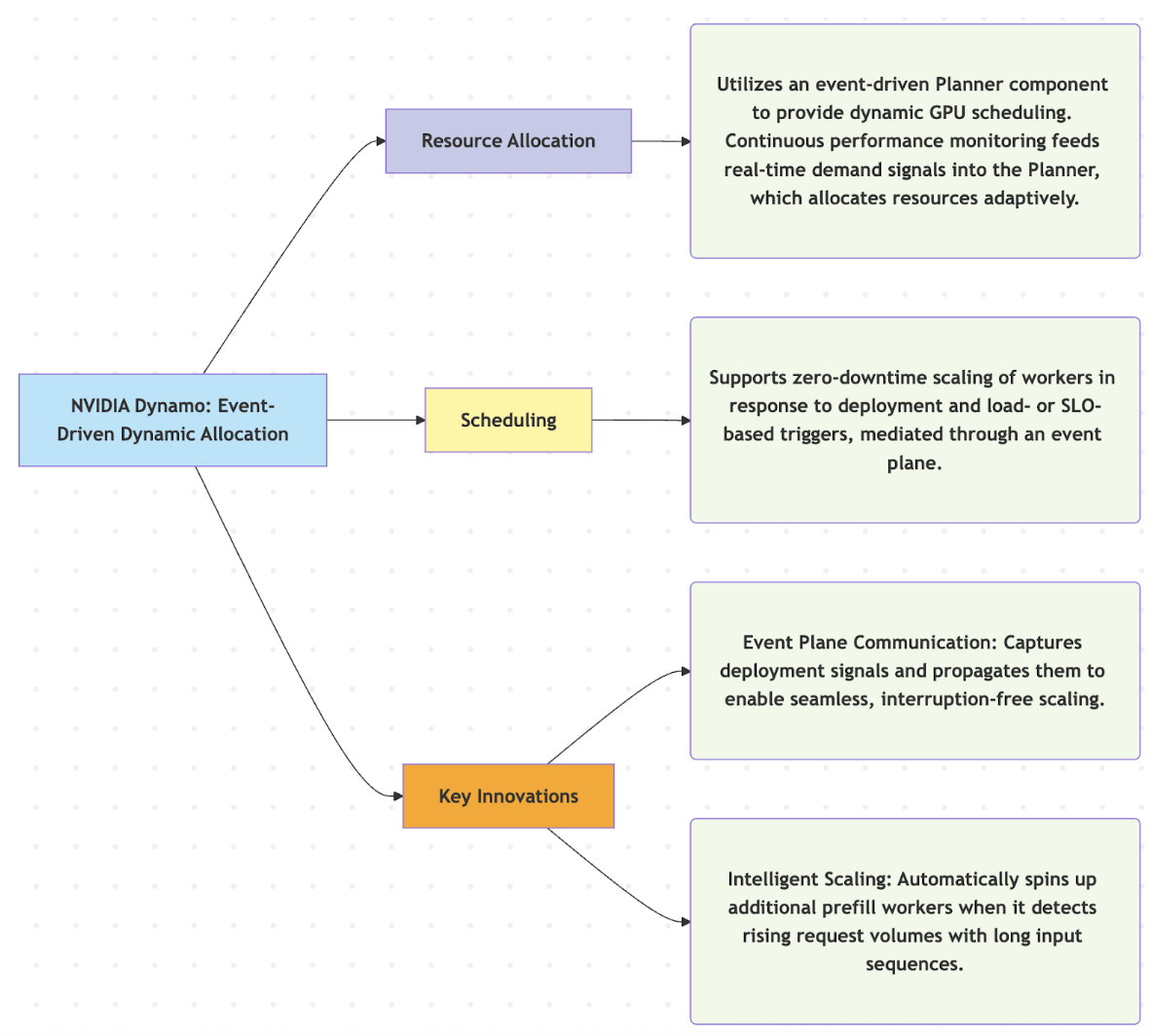}
    \caption{Event-Driven Dynamic Allocation}
    \label{fig:placeholder}
\end{figure}

\subsection{Communication and Data Transfer}

\subsubsection{DistServe : \emph {\color{blue} High-Bandwidth Intra-Node Communication}}

DistServe's High-Bandwidth Intra-Node Communication, a strategy optimized for single-node GPU deployments leveraging high-bandwidth NVLink for efficient GPU-to-GPU communication with negligible overhead. Its data transfer mechanism utilizes Ray remote calls for batch dispatching and cudaMemcpyAsync within the SwiftTransformer engine for asynchronous KV cache block transfers. While effective within a single node, DistServe's reliance on NVLink limits its scalability to single-node setups, as multi-node extensions would incur significant communication overhead. 

\begin{figure}
    \centering
    \includegraphics[width=1.0\linewidth]{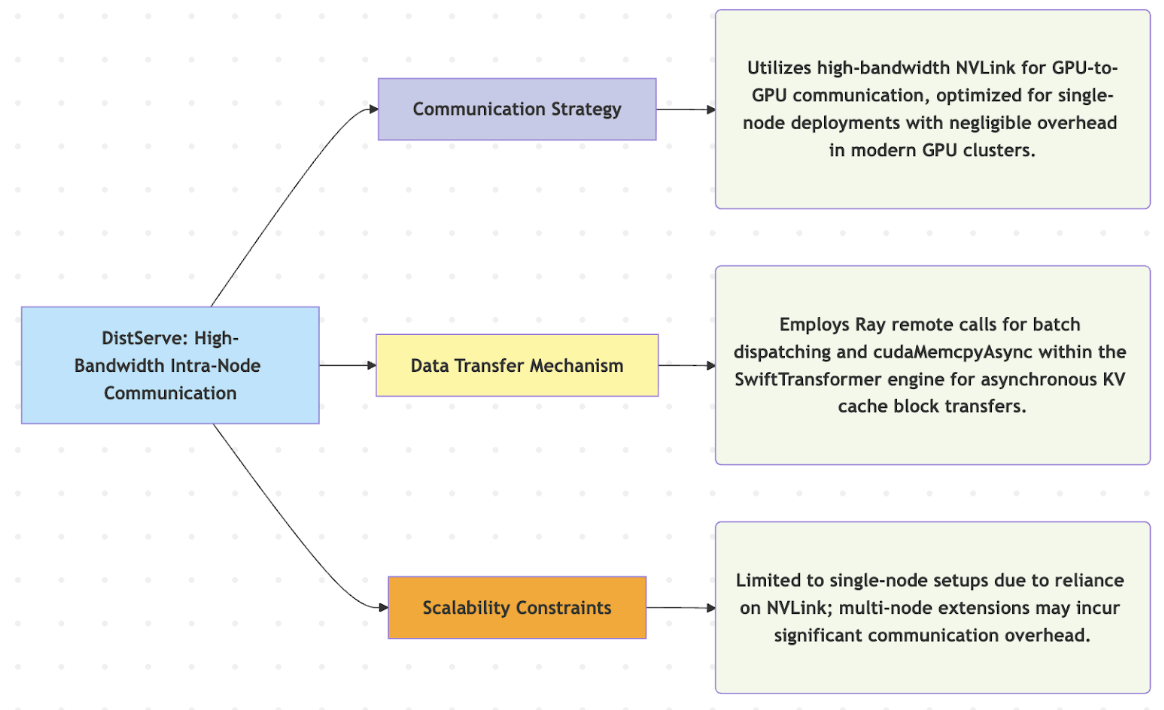}
    \caption{High-Bandwidth Intra-Node Communication}
    \label{fig:placeholder}
\end{figure}

\subsubsection{AIBrix : \emph {\color{blue} Distributed Cache-Aware Communication}}

AlBrix Distributed Cache-Aware Communication employs a sophisticated approach to optimize distributed systems, particularly for large language models. It leverages a communication strategy featuring a distributed Key-Value (KV) cache with cache-aware routing and shared-memory optimization to enhance network efficiency and reduce redundant data transfers. Furthermore, the system utilizes asynchronous metadata updates and shared-memory-based data exchange, with prefix-aware routing in the API Gateway to maximize KV cache locality during data transfer. Finally, AlBrix ensures multi-node scalability through the use of RayClusterFleet controllers, facilitating efficient coordination across distributed Ray clusters.

\begin{figure}
    \centering
    \includegraphics[width=1.0\linewidth]{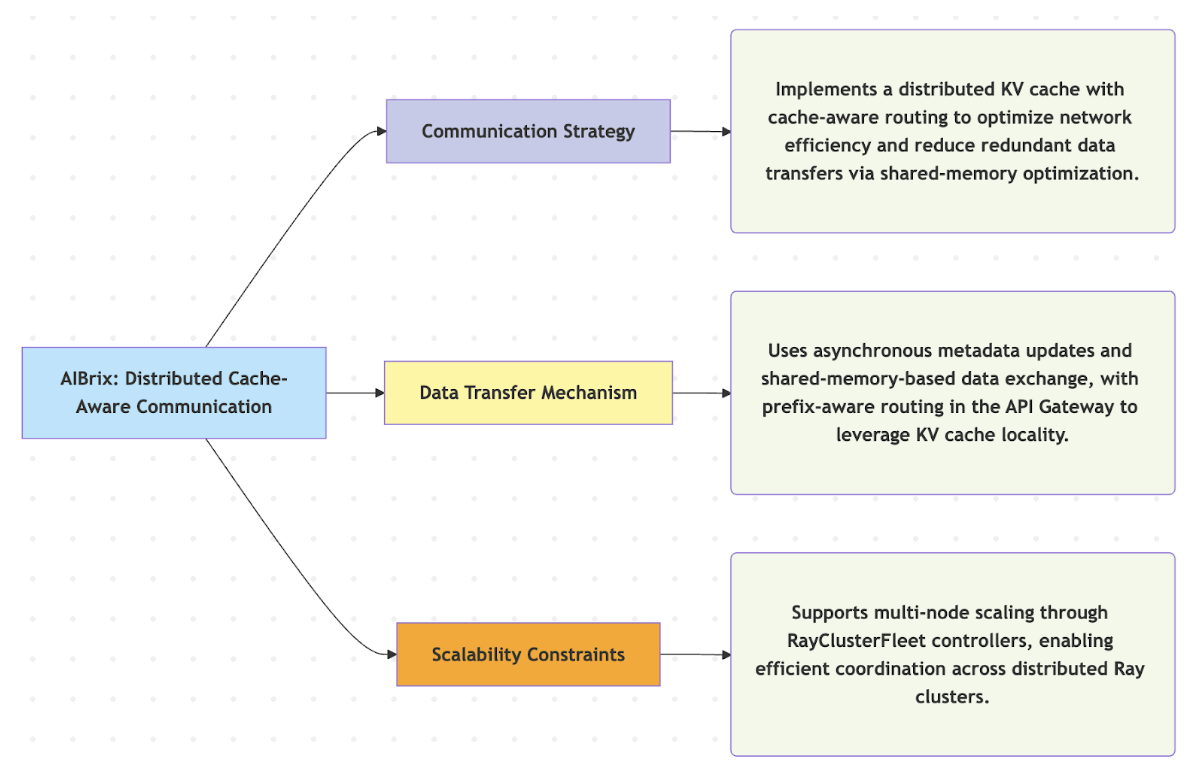}
    \caption{Distributed Cache-Aware Communication}
    \label{fig:placeholder}
\end{figure}

\subsubsection{NVIDIA Dynamo : \emph {\color{blue} Advanced Transfer Library}}

NVIDIA Dynamo is an advanced transfer library designed to optimize AI inference, particularly in large-scale, distributed GPU environments. It achieves this through a Communication Strategy that leverages NIXL for efficient GPU-to-GPU data exchange across heterogeneous devices, a Data Transfer Mechanism featuring reduced synchronization and intelligent batching via NIXL to minimize latency, and a focus on Scalability Constraints by built-in optimizations for coordinating distributed GPU fleets in enterprise deployments.

\begin{figure}
    \centering
    \includegraphics[width=1.0\linewidth]{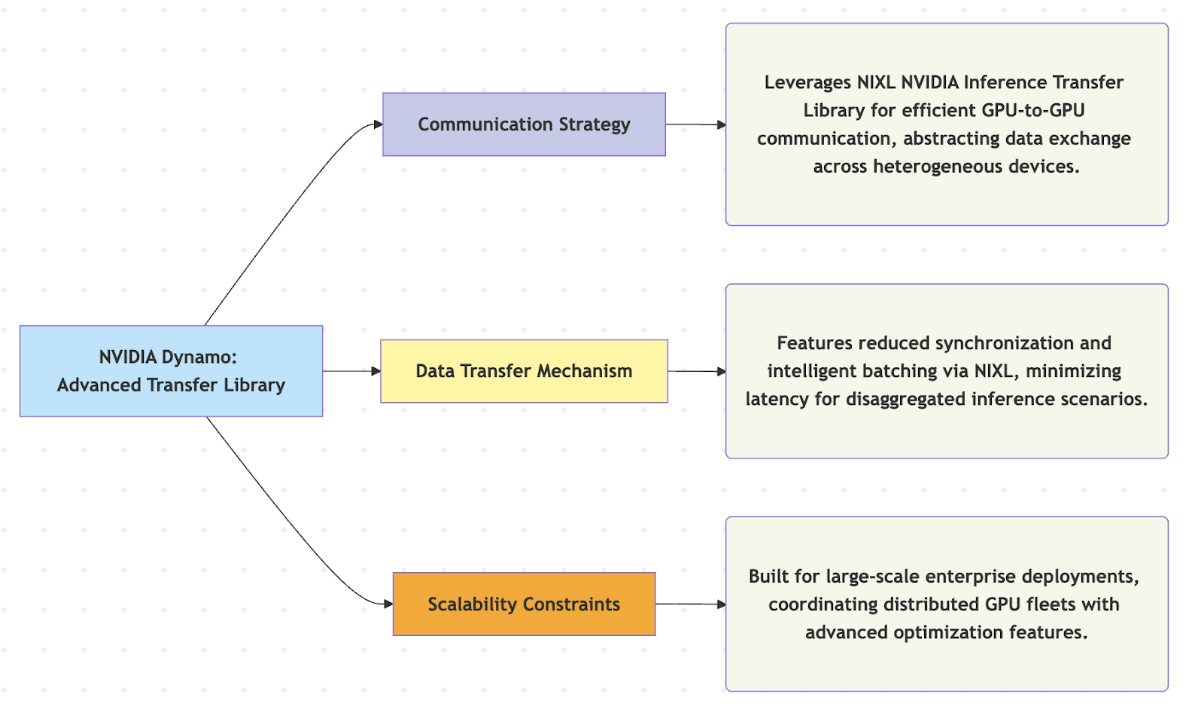}
    \caption{Advanced Transfer Library}
    \label{fig:placeholder}
\end{figure}

\subsection{Core Technical Architecture Specifications}

\subsubsection{DistServe : \emph {\color{blue} SwiftTransformer Integration}}

DistServe integrates with SwiftTransformer to boost Large Language Model (LLM) inference by utilizing a high-performance C++ backend that includes features like model/pipeline parallelism, FlashAttention, and paged attention for efficient serving. This integration also optimizes memory transfers by employing cudaMemcpyAsync for non-blocking KV cache operations, thereby reducing synchronization overhead and maintaining data consistency across prefill and decode stages.

\begin{figure}
    \centering
    \includegraphics[width=1.0\linewidth]{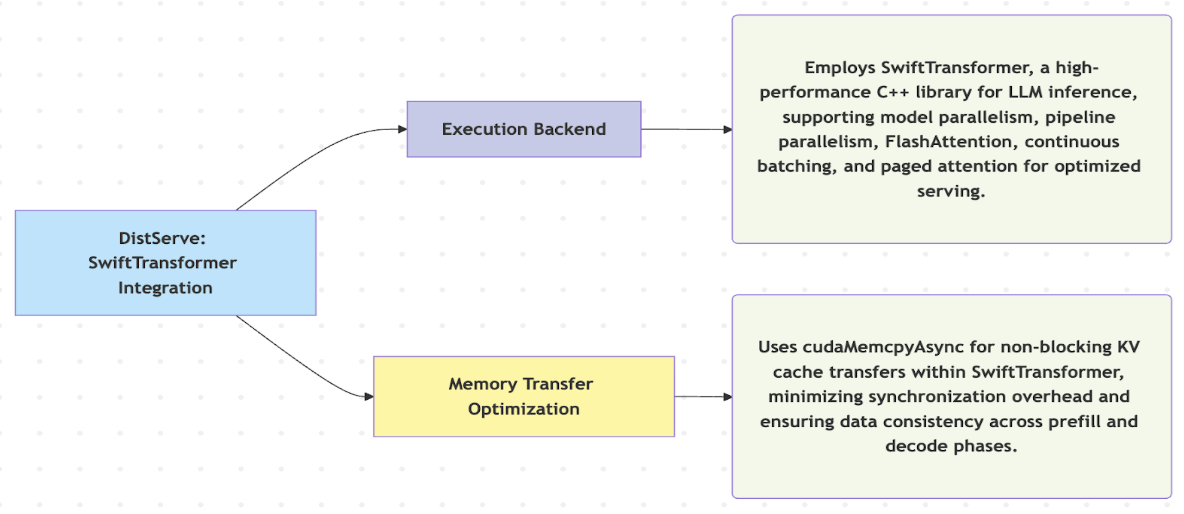}
    \caption{SwiftTransformer Integration}
    \label{fig:placeholder}
\end{figure}

\subsubsection{AIBrix : \emph {\color{blue} Advanced Autoscaling Architecture}}

AIBrix's Advanced Autoscaling Architecture is designed for efficient LLM serving, featuring Sliding Window Aggregation for smoothed metric collection and responsive autoscaling. It also includes Cold Start Optimization to minimize latency by prefetching model artifacts, alongside Multi-LoRA Management for dynamic adapter scheduling, which together optimize GPU memory and enhance overall scalability. This integrated approach ensures scalable, cost-effective, and high-performance LLM inference.

\begin{figure}
    \centering
    \includegraphics[width=1.0\linewidth]{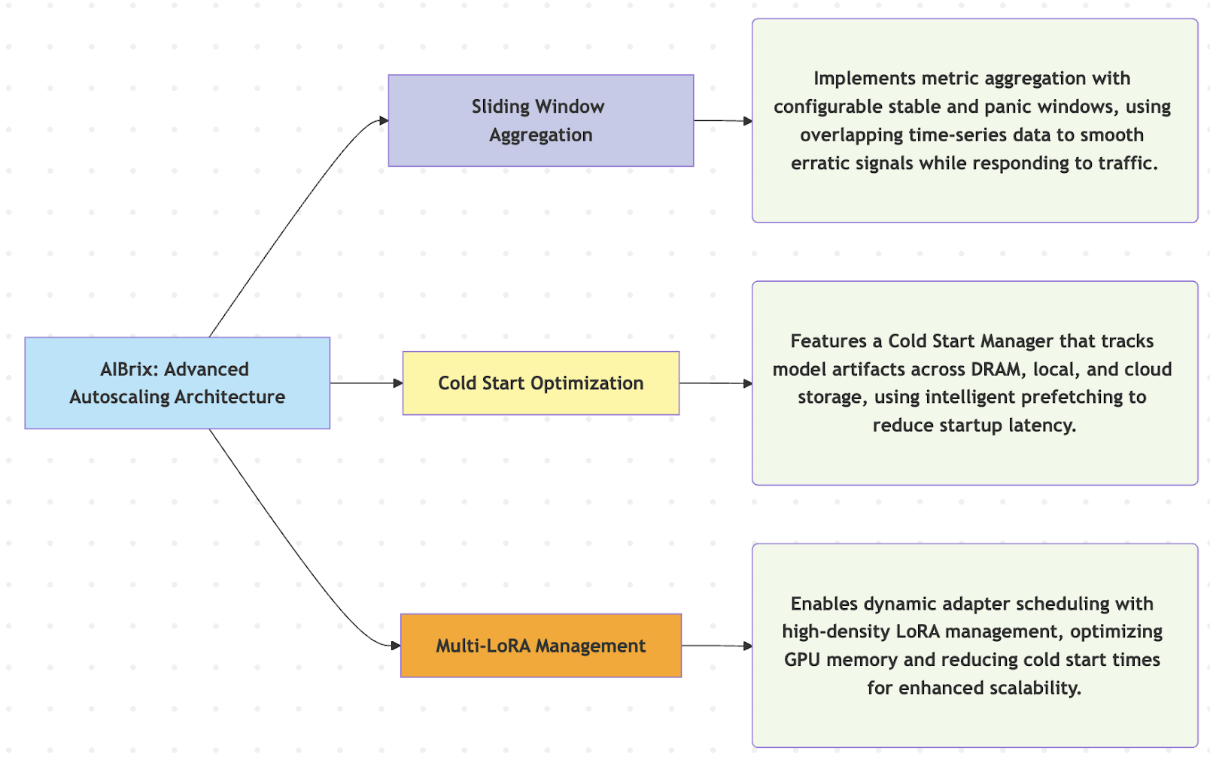}
    \caption{Advanced Autoscaling Architecture}
    \label{fig:placeholder}
\end{figure}

\subsubsection{NVIDIA Dynamo : \emph {\color{blue} NIXL Communication Library}}

The NVIDIA Dynamo NIXL Communication Library serves as the foundation, enabling both the NIXL Architecture and an Event-Driven Planner. The NIXL Architecture uses the NVIDIA Inference Transfer Library for efficient, unified data movement across diverse memory and storage systems, while the Event-Driven Planner intelligently allocates resources for local and Kubernetes environments, continuously optimizing to meet service level agreements.

\begin{figure}
    \centering
    \includegraphics[width=1.0\linewidth]{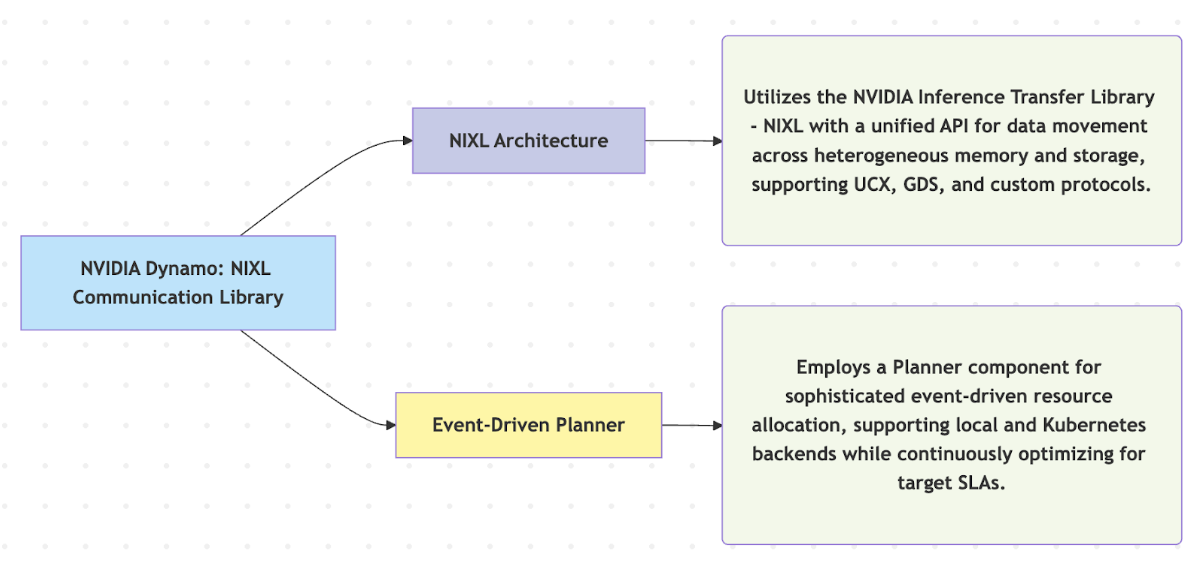}
    \caption{NIXL Communication Library}
    \label{fig:placeholder}
\end{figure}

\subsection{Deployment and Integration Complexity}

\subsubsection{DistServe : \emph {\color{blue} Research-Focused Deployment}}

DistServe is a research-centric deployment system that mandates a Ray cluster setup and specialized knowledge in optimizing tensor/pipeline parallelism, primarily designed for academic and research contexts. Although its single-node focus eases initial configuration, its integration capabilities are restricted to vLLM and Ray-based environments, demanding custom solutions for production-level integrations. Ray's distributed task management serves to address and alleviate any multi-node scaling complexities that may arise.

\begin{figure}
    \centering
    \includegraphics[width=1.0\linewidth]{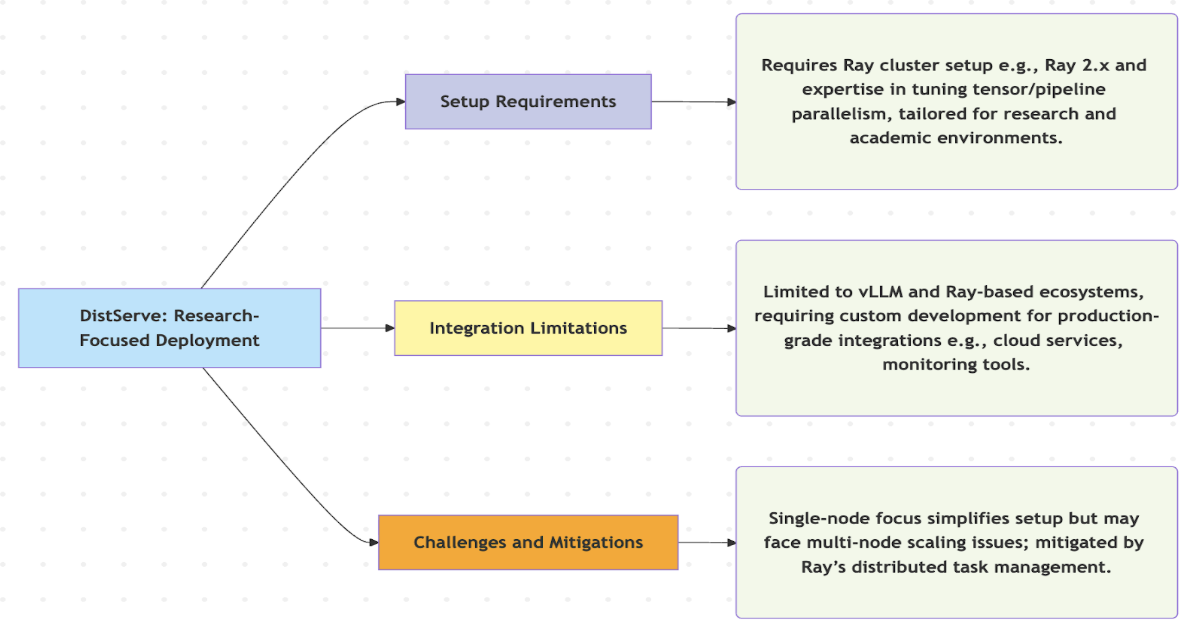}
    \caption{Research-Focused Deployment}
    \label{fig:placeholder}
\end{figure}

\subsubsection{AIBrix : \emph {\color{blue} Production-Ready Kubernetes Integration}}

AlBrix provides production-ready Kubernetes integration, characterized by its demanding enterprise deployment requiring expertise in cloud-native orchestration and components like GPU optimizers. It offers extensive cloud-native features through integrations with service meshes (e.g., Istio), monitoring tools (e.g., Prometheus), and cloud services (e.g., AWS EKS) via YAML configurations. While setup can be complex, AlBrix mitigates these challenges with automated deployment scripts and compatibility with major cloud providers for scalable, fault-tolerant operations.

\begin{figure}
    \centering
    \includegraphics[width=1.0\linewidth]{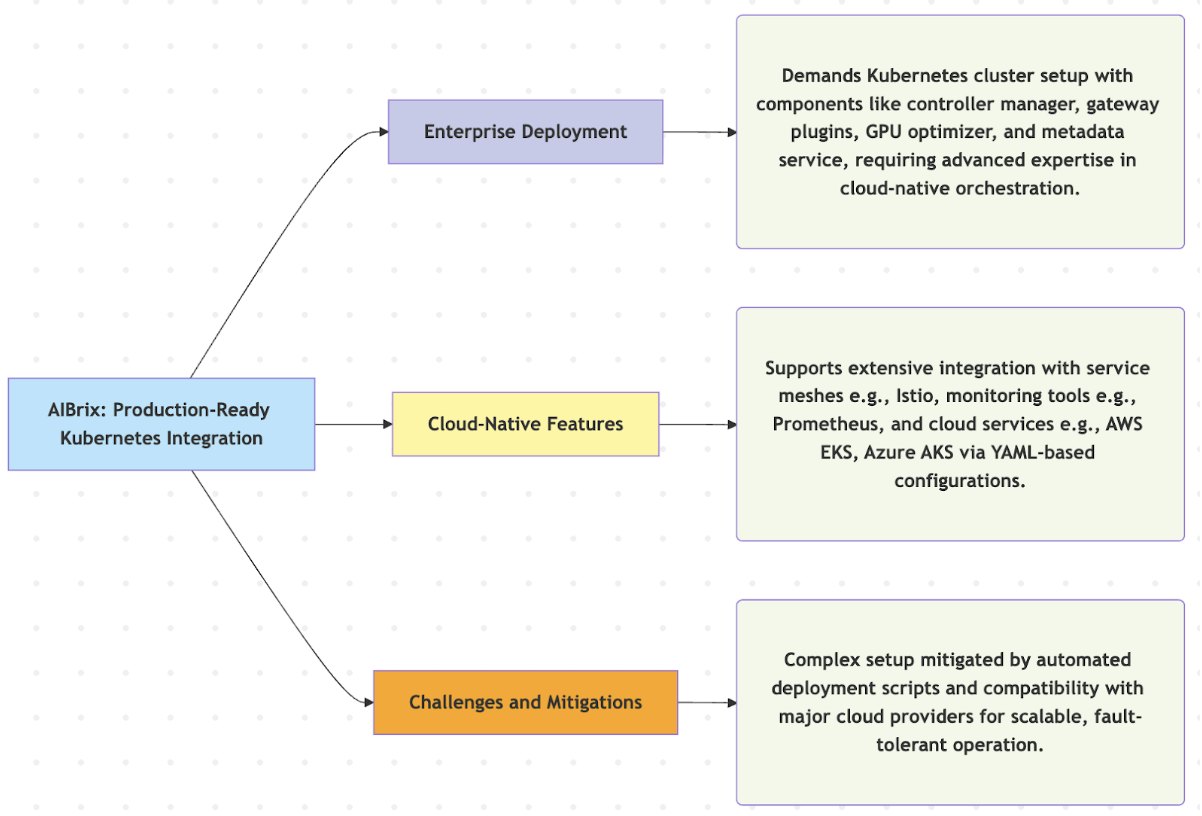}
    \caption{Production-Ready Kubernetes Integration}
    \label{fig:placeholder}
\end{figure}

\subsubsection{NVIDIA Dynamo : \emph {\color{blue} Modular Enterprise Framework}}

The NVIDIA Dynamo Modular Enterprise Framework offers a highly adaptable and efficient solution for deploying AI models at scale. It boasts a modular architecture, enabling selective use of components like NIXL for optimized data transfer and a Planner for dynamic resource allocation, enhancing flexibility for diverse enterprise requirements. Furthermore, Dynamo supports flexible deployment across on-premises and cloud environments using technologies like Docker and Kubernetes, and mitigates potential version conflicts in heterogeneous setups through comprehensive compatibility testing and modular design.

\begin{figure}
    \centering
    \includegraphics[width=1.0\linewidth]{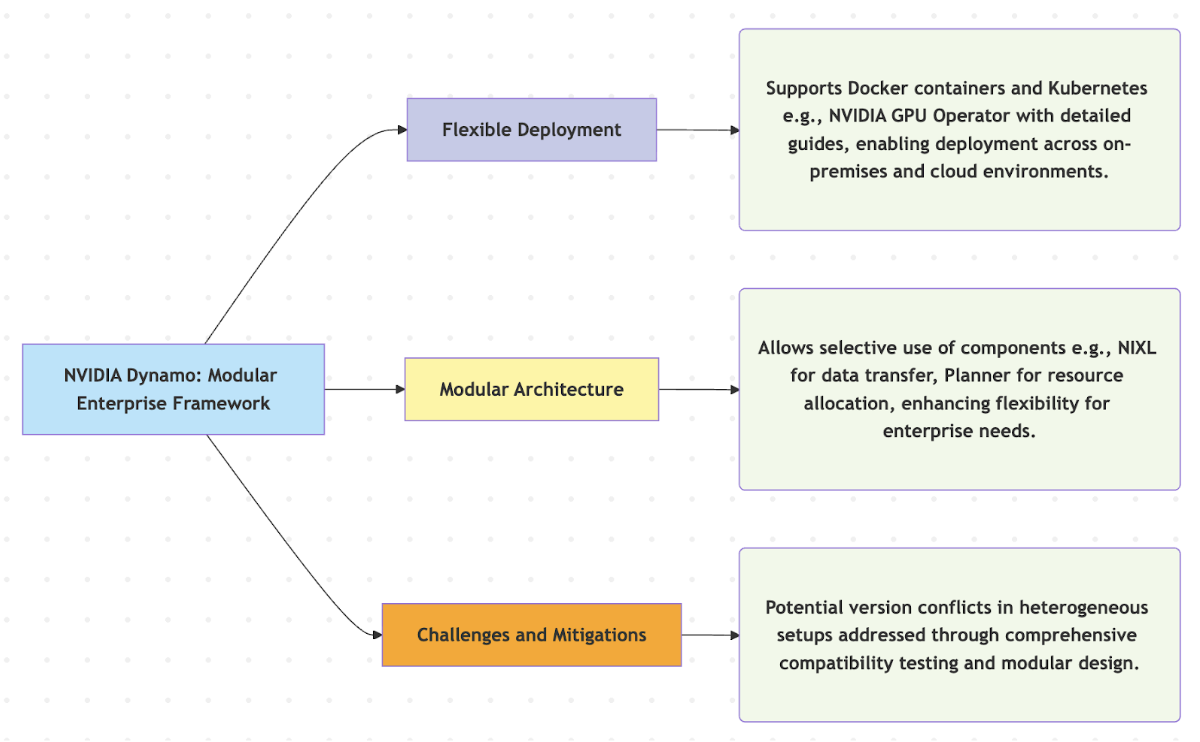}
    \caption{Modular Enterprise Framework}
    \label{fig:placeholder}
\end{figure}

\subsection{Performance Characteristics and Benchmarks}

\subsubsection{DistServe : \emph {\color{blue} Goodput-Optimized Performance}}

DistServe demonstrates "Goodput-Optimized Performance" for LLM serving by achieving substantial performance gains, including a 7.4x increase in requests and 12.6x tighter SLOs compared to vLLM, with over $90\%$ of requests meeting latency targets. This system sustains significantly higher request rates than both vLLM and DeepSpeed-MII across various LLMs and workloads, including Llama-70B and GPT-2 on 8x NVIDIA A100 GPUs. Its technical optimization prioritizes goodput to minimize cost per query by intelligently optimizing GPU utilization while adhering to Time To First Token (TTFT) and Time Per Output Token (TPOT) constraints.

\begin{figure}
    \centering
    \includegraphics[width=1.0\linewidth]{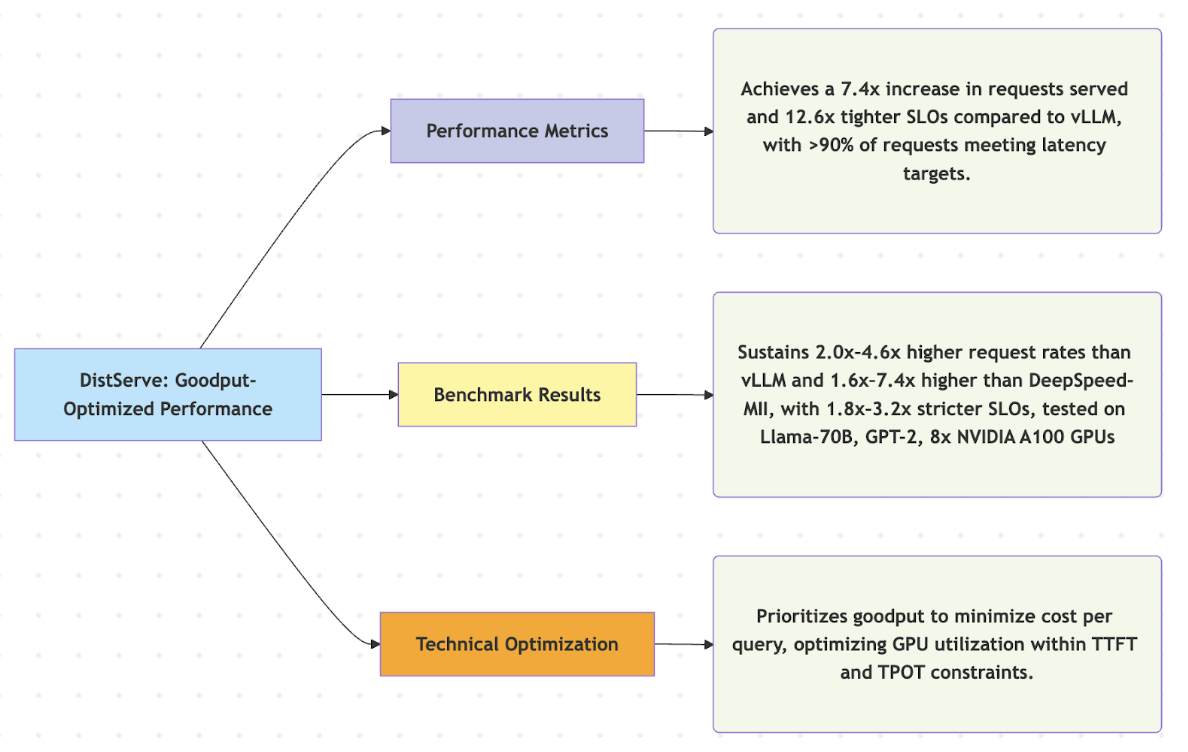}
    \caption{Goodput-Optimized Performance}
    \label{fig:placeholder}
\end{figure}

\subsubsection{AIBrix : \emph {\color{blue} Cost-Effective Scalability} }

AIBrix delivers cost-effective scalability for LLM inference by significantly enhancing performance and reducing costs. It achieves this through distributed KV cache optimization, leading to $50\%$ higher throughput and $70\%$ lower inference latency, coupled with a 4.7x cost reduction in low-traffic scenarios, particularly for LoRA-based models like Llama-13B on heterogeneous clusters, and by optimizing GPU usage through intelligent resource allocation and system-wide optimizations tested on AWS EKS with NVIDIA T4 GPUs.

\begin{figure}
    \centering
    \includegraphics[width=1.0\linewidth]{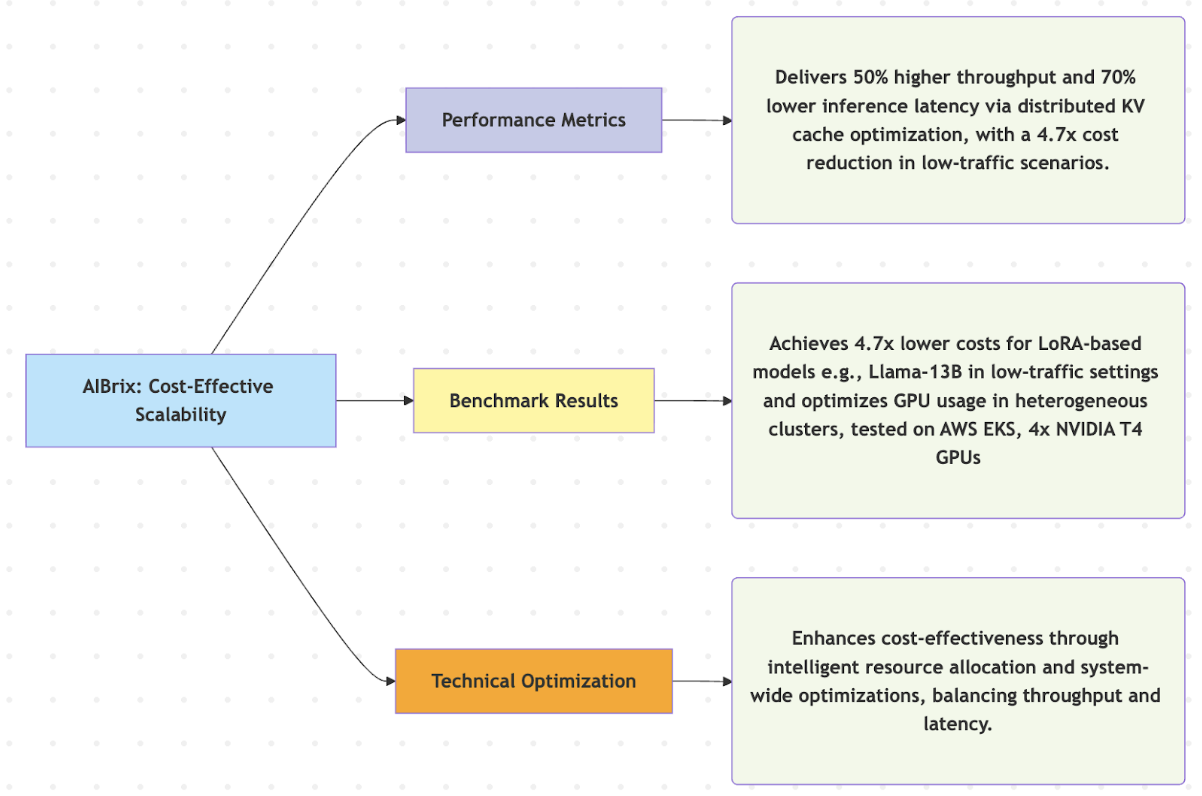}
    \caption{Cost-Effective Scalability}
    \label{fig:placeholder}
\end{figure}

\subsubsection{NVIDIA Dynamo : \emph {\color{blue} Enterprise-Scale Performance}}

The NVIDIA Dynamo platform provides enterprise-scale performance for AI inference, delivering significant performance enhancements for large language models (LLMs). It achieves this through benchmark results like supporting up to 30x more requests for large models on NVIDIA GB200 NVL72 and doubling throughput for Llama-70B on NVIDIA Hopper GPUs, driven by dynamic, disaggregated serving. Technically, Dynamo targets enterprise scalability via multi-framework compatibility and efficient resource management for large-scale AI deployments.

\begin{figure}
    \centering
    \includegraphics[width=1.0\linewidth]{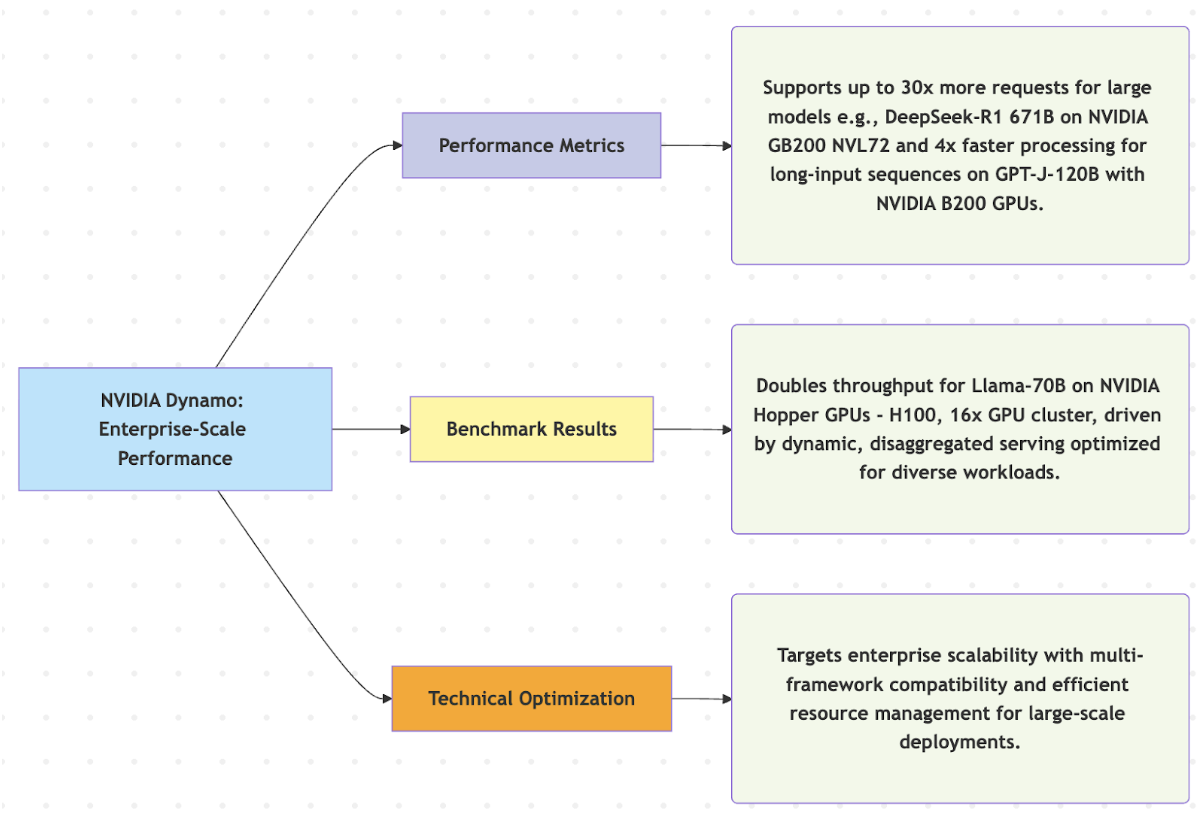}
    \caption{Enterprise-Scale Performance}
    \label{fig:placeholder}
\end{figure}

\section{Conclusion}

This comprehensive technical analysis reveals the distinct positioning and capabilities of each disaggregated inference framework. DistServe establishes the academic foundation with strong theoretical contributions and goodput optimization, making it ideal for research environments exploring disaggregated serving concepts. AIBrix provides comprehensive cloud-native orchestration optimized for production environments, emphasizing cost-effectiveness and enterprise-grade reliability. NVIDIA Dynamo delivers enterprise-grade performance with extensive framework support and a modular architecture, targeting large-scale distributed deployments.
The choice between frameworks depends on specific deployment requirements, technical expertise, target scale, and integration needs within existing infrastructure ecosystems. Organizations should consider their primary objectives: research and experimentation (DistServe), production cloud-native deployment (AIBrix), or enterprise-scale multi-framework support (NVIDIA Dynamo)


\begin{thebibliography}{99} 

\bibitem{Vaswani2017Transformer} Ashish Vaswani et al. 2017. Attention Is All You Need (Transformer architecture). \url{https://arxiv.org/abs/1706.03762}
\bibitem{Radford2018GPT1} Alec Radford et al. 2018. Improving Language Understanding by Generative Pre-Training (GPT-1). \url{https://cdn.openai.com/research-covers/language-unsupervised/language_understanding_paper.pdf}
\bibitem{Radford2019GPT2} Alec Radford et al. 2019. Language Models are Unsupervised Multitask Learners (GPT-2). \url{https://cdn.openai.com/better-language-models/language_models_are_unsupervised_multitask_learners.pdf}
\bibitem{Brown2020GPT3} Tom B. Brown et al. 2020. Language Models are Few-Shot Learners (GPT-3). \url{https://arxiv.org/abs/2005.14165}
\bibitem{OpenAI2023GPT4} OpenAI. 2023. GPT-4 Technical Report (GPT-4). \url{https://arxiv.org/abs/2303.08774}
\bibitem{Shoeybi2019MegatronLM} Mohammad Shoeybi et al. 2019. Megatron-LM: Training Multi-Billion Parameter Language Models Using Model Parallelism (KV Cache introduction). \url{https://arxiv.org/abs/1909.08053}
\bibitem{Megatron2020KV} 2020. Key-Value (KV) Cache: Megatron-LM Paper (write-once, read-many caching). \url{https://arxiv.org/abs/1909.08053}
\bibitem{Dao2022FlashAttention1} Tri Dao et al. 2022. Flash Attention v1: Fast and Memory-Efficient Exact Attention with IO-Awareness. \url{https://arxiv.org/abs/2205.14135}
\bibitem{Dao2023FlashAttention2} Tri Dao et al. 2023. Flash Attention v2: Faster Attention with Better Parallelism and Work Partitioning. \url{https://arxiv.org/abs/2307.08691}
\bibitem{Dao2024FlashAttention3} Tri Dao et al. 2024. Flash Attention v3: Fast and Accurate Attention with Asynchrony and Low-precision. \url{https://arxiv.org/abs/2407.08608}
\bibitem{Leviathan2022Speculative} Y. Leviathan et al. 2022. Speculative Decoding: Fast Inference from Transformers via Speculative Decoding. \url{https://arxiv.org/abs/2211.17192}
\bibitem{Anyscale2023ContinuousBatching} Anyscale. 2023. Continuous Batching: Efficient Memory Management for Large Language Model Serving with Orca. \url{https://www.anyscale.com/blog/continuous-batching-llm-inference}
\bibitem{Kwon2023PagedAttention} Woosuk Kwon et al. 2023. Paged Attention: Efficient Memory Management for Large Language Model Serving with PagedAttention (vLLM). \url{https://arxiv.org/abs/2309.06180}
\bibitem{SGLang2024Radix} 2024. Radix Attention: SGLang: Efficient Execution of Structured Language Model Programs. \url{https://arxiv.org/abs/2312.07104}
\bibitem{Zhong2024DistServeOSDI} Yinmin Zhong et al. 2024. DistServe: Disaggregating Prefill and Decoding for Goodput-optimized LLM Serving (OSDI'24). \url{https://arxiv.org/abs/2401.09670}
\bibitem{Zhong2024DistServePDF} Yinmin Zhong et al. 2024. DistServe: USENIX OSDI'24 Full Paper PDF. \url{https://www.usenix.org/system/files/osdi24-zhong-yinmin.pdf}
\bibitem{DistServeGitHub} 2024. DistServe GitHub Repository. \url{https://github.com/LLMServe/DistServe}
\bibitem{AIBrix2025} 2025. AIBrix: Towards Scalable, Cost-Effective Large Language Model Serving. \url{https://arxiv.org/abs/2504.03648}
\bibitem{NVIDIADynamo} NVIDIA. NVIDIA Dynamo: A Low-Latency Distributed Inference Framework. \url{https://developer.nvidia.com/dynamo}
\bibitem{NVIDIADynamoArch} NVIDIA. NVIDIA Dynamo Architecture: Disaggregation Guide. \url{https://docs.nvidia.com/dynamo/latest/architecture/disagg_serving.html}
\bibitem{NVIDIADynamoBlog} NVIDIA. Introducing NVIDIA Dynamo: Scaling Reasoning AI Models. \url{https://developer.nvidia.com/blog/introducing-nvidia-dynamo-a-low-latency-distributed-inference-framework-for-scaling-reasoning-ai-models/}
\bibitem{PDServe2024} 2024. P/D-Serve: Serving Disaggregated Large Language Model at Scale. \url{https://arxiv.org/abs/2408.08147}
\bibitem{NVIDIAInferenceOpt} NVIDIA. Mastering LLM Techniques: Inference Optimization. \url{https://developer.nvidia.com/blog/mastering-llm-techniques-inference-optimization/}
\bibitem{SurveyLLMInference2024} 2024. Inference Optimizations for Large Language Models: A Survey. \url{https://arxiv.org/abs/2408.03130}
\bibitem{SurveyEfficientLLM2025} 2025. Taming the Titans: A Survey of Efficient LLM Inference Serving. \url{https://arxiv.org/abs/2504.19720}

\end{thebibliography}
\end{document}